\documentclass[%
 reprint,
 twocolumn,
superscriptaddress,
 amsmath,amssymb,
 aps,
 bbm,
]{revtex4-2}

\usepackage{graphicx}% Include figure files
\usepackage{dcolumn}% Align table columns on decimal point
\usepackage{bm}% bold math
\usepackage{bbm}

\usepackage{graphicx}
\usepackage[colorlinks=true]{hyperref}
\usepackage{comment} 
\usepackage{color}
\usepackage{mathtools}
\usepackage{amsmath}
\usepackage{amssymb}
\usepackage{amsfonts}
\usepackage{empheq}
\usepackage{bbm}
\usepackage{MnSymbol}
\usepackage[dvipsnames]{xcolor}
\usepackage[normalem]{ulem}

\begin{document}

\preprint{APS/123-QED}

\title{Theory of gating in recurrent neural networks}% Force line breaks with \\

\author{Kamesh Krishnamurthy}
\altaffiliation{lead author}%: kameshk@princeton.edu}
\affiliation{Joseph Henry Laboratories of Physics and PNI, Princeton University, Princeton, NJ}%Lines break automatically or can be forced with 
%\affiliation{lead author}

\author{Tankut Can}
\altaffiliation{corresponding authors: tankut.can@gmail.com,  kameshk@ princeton.edu}
\affiliation{
Institute for Advanced Study, Princeton, NJ
}%

\author{David J. Schwab}
\affiliation{
Initiative for Theoretical Sciences, Graduate Center, CUNY, New York
}%

\date{\today}% It is always \today, today,

\begin{abstract}

Recurrent neural networks (RNNs) are powerful dynamical models, widely used in machine learning (ML)  and  neuroscience. Prior theoretical work  has focused on RNNs  with additive interactions. However, gating – i.e. multiplicative – interactions are ubiquitous in real neurons and  also {\it the} central feature of the best-performing RNNs in ML.  Here, we show that gating offers flexible control of two salient features of the collective dynamics: i) timescales and ii) dimensionality. The gate controlling timescales leads to a novel, marginally stable state, where the network functions as a flexible integrator. Unlike previous approaches, gating permits this important function without parameter fine-tuning or special symmetries.
Gates also provide a flexible, context-dependent mechanism to reset the memory trace, thus complementing the memory function. 
The gate modulating the dimensionality can induce a novel, discontinuous chaotic transition, where inputs push a stable system to strong chaotic activity, in contrast to the typically stabilizing effect of inputs. At this transition, unlike additive RNNs, the proliferation of critical points (topological complexity) is decoupled from the appearance of chaotic dynamics (dynamical complexity).
 The rich dynamics are summarized in  phase diagrams, thus providing a map for principled parameter initialization choices to ML practitioners.

\end{abstract}

\maketitle

%\tableofcontents

%************************ INTRO ***********************

\section{Introduction}

Recurrent neural networks (RNNs) are powerful dynamical systems that can represent a rich repertoire of trajectories and are popular models in neuroscience and machine learning.  In modern machine learning, RNNs are used to learn complex dynamics from data with rich sequential/temporal structure such as speech \cite{graves2013speech,pascanu2013construct},  turbulent flows \cite{pathak2018model,vlachas2020backpropagation,guastoni2020use} or text sequences \cite{jozefowicz2015empirical}.
RNNs are also influential in neuroscience as models to study the collective behavior of a large network of neurons \cite{vogels2005neural} (and references therein). For instance, they have been used to explain the dynamics and temporally-irregular fluctuations observed in cortical networks \cite{ahmadian2019dynamical,kadmon2015transition}, and how the motor-cortex network generates movement sequences \cite{sussillo2009generating,laje2013robust}.

Classical RNN models typically involve units that interact with each other in an additive fashion -- i.e. each unit integrates a weighted sum of the output of the rest of the network. However, researchers in machine learning have empirically found that RNNs with {\it gating} -- a form of multiplicative interaction -- can be trained to perform
significantly more complex tasks than classical RNNs  \cite{hochreiter1997long,jozefowicz2015empirical}. 
% This is attributed to the superior ability of gated RNNs to learn long-time dependencies in data \cite{hochreiter1997long,hochreiter2001gradient}.
Gating interactions are also ubiquitous in real neurons due to mechanisms such as shunting  inhibition \cite{mitchell2003shunting}. Moreover, when single-neuron models are endowed with more realistic conductance dynamics, the effective interactions at the network level have gating effects, which confer robustness to time-warped inputs \cite{gutig2009time}. Thus, RNNs with gating interactions not only have superior information processing capabilities, but they also embody a prominent feature found in real neurons.

Prior theoretical work on understanding the dynamics and functional capabilities  of RNNs has mostly focused on RNNs with additive interactions. The original work by Crisanti et al. \cite{sompolinsky1988chaos} identified a phase transition in the autonomous dynamics of randomly connected RNNs from stability to chaos.
% when the variance of the coupling weights exceeded a critical value. 
Subsequent work has extended this analysis to cases where the random connectivity additionally has correlations \cite{marti2018correlations}, a low-rank structured component \cite{schuessler2020dynamics,mastrogiuseppe2018linking}, strong self-interaction \cite{PhysRevE.90.062710} and heterogeneous variance across blocks \cite{aljadeff2015transition}. The role of sparse connectivity and the single-neuron nonlinearity was studied in \cite{kadmon2015transition}. The effect of a Gaussian noise input was analyzed in \cite{schuecker2018optimal}.

In this work, we study the consequences of gating interactions on the dynamics of RNNs. We introduce a gated RNN model that naturally extends a classical RNN by augmenting it with two kinds of gating interactions: i) an {\it update} gate that acts like an adaptive time-constant and ii) an {\it output} gate which modulates the output of a neuron. The choice of these forms for gates are motivated by biophysical considerations (e.g.  \cite{gutig2009time,brette2006exact}) and retain the most functionally important aspects of the gated RNNs in machine learning.  Our gated RNN reduces to the classical RNN (\cite{sompolinsky1988chaos,amari1972characteristics}) when the gates are open, and  is closely related to the state-of-the-art gated RNNs in machine learning when the dynamics are discretized \cite{cho2014learning}.  We further elaborate on this connection in the Discussion.

We develop a theory  for the gated RNN based on non-Hermitian random matrix techniques \cite{chalker1998eigenvector,feinberg1997non} and the Martin-Siggia-Rose-De Dominicis-Jansen (MSRDJ) formalism \cite{martin1973statistical,de1978dynamics,hertz2016path,janssen1976lagrangean,schuecker2018optimal,crisanti2018path,helias2019statistical}, and  use the theory to map out, in a phase diagram, the rich, functionally significant dynamical phenomena produced by gating.

 We show that the update gate produces slow modes and a  marginally stable critical state. Marginally stable systems are of special interest in the context of biological information processing (e.g. \cite{mora2011biological}).
%  and making a system marginally stable often requires parameter fine-tuning. We show that gating allows this behaviour over a wide range of parameters. 
 Moreover, the network in this marginally stable state can function as a robust integrator --  a function that is critical for memory capabilities in biological systems \cite{seung1998continuous,seung1996brain,seung2000stability,machens2005flexible}, but has been hard to achieve without parameter fine-tuning and hand-crafted symmetries \cite{chaudhuri2016computational}. Gating permits the network to serve this function without any symmetries or fine-tuning. For a detailed discussion of these issues, we refer the reader to \cite{bialek2012biophysics} pp. 329-250;  \cite{goldman2009memory,chaudhuri2016computational}. Integrator-like dynamics are also empirically observed in  gated ML RNNs successfully trained on complex sequential tasks \cite{maheswaranathan2019reverse}; our theory shows how gates allow for this robustly.

 The output gate allows fine control over the dimensionality of the network activity; control of the dimensionality can be  useful during learning tasks \cite{farrell2019recurrent}.    In certain regimes, this gate can mediate an input-driven chaotic transition, where static inputs  can push a stable system abruptly to a chaotic state. This behavior with gating is in stark contrast to the typically stabilizing effect of inputs in high-dimensional systems \cite{molgedey1992suppressing,rajan2010stimulus,schuecker2018optimal}. In this novel discontinuous chaotic transition, the proliferation of critical points (a static property) is decoupled from the appearance of  chaotic transients (a dynamical property); this is in contrast to the tight link between the two properties in additive RNNs as shown by Wainrib \& Touboul \cite{wainrib2013topological}. This transition is also characterized by a non-trivial state where a stable fixed-point coexists with long chaotic transients. Gates also provide a flexible, context-dependent way to reset the state, thus 
 providing a way to selectively erase the memory trace of past inputs.

 We summarize these functionally significant phenomena in phase diagrams, which are also practically useful for ML practitioners -- indeed,  the choice of parameter initialization is known to be one of the most important factors deciding the success of training \cite{sutskever2013importance}, with best outcomes occurring near critical lines \cite{legenstein2007edge,jaeger2004harnessing,toyoizumi2011beyond, sussillo2009generating}.  Phase diagrams thus allow a principled and exhaustive exploration of dynamically distinct initializations.

% *************** MODEL DEFINITION ***************

\section{A recurrent neural network model to study gating}
\label{sec:gRNN_model}

We study an extension of a classical RNN \cite{amari1972characteristics,sompolinsky1988chaos} by augmenting it with multiplicative {\it gating} interactions. 
Specifically, we consider two gates: (i) an {\it update} (or $z-$) gate which controls the rate of integration, and (ii) an {\it output} (or $r-$) gate which modulates the strength of the output. The equations describing the gated RNN are given by:

\begin{eqnarray} \label{eq:gatedRNN-eom1}
 \dot{h}_i(t) = \sigma_z(z_i)  \Big[-h_i(t) + R_{i}(t) \Big] + I_{i}^{h}(t)
\end{eqnarray}
where $h_i$ represents the internal state of the $i^{th}$ unit, 
 and $\sigma_{(\cdot)}(x) = [1+\exp(-\alpha_{(\cdot)} x + \beta_{(\cdot)})]^{-1}$ are sigmoidal gating functions. The recurrent input to a neuron is $R_{i}(t) = \sum_{j=1}^{N}J^h_{ij}\phi(h_j(t)) \cdot \sigma_r(r_{j}(t))$, where $J^h_{ij}$ are the coupling strengths between the units, and $\phi(x) = \tanh(g_hx + \beta_h)$ is the activation function. $\phi,\sigma_{z,r}$ are  parametrized by gain parameters  ($g_h,\alpha_{z,r}$) and biases ($\beta_{h,z,r}$), which constitute the parameters of the gated RNN. Finally, $I^{h}$ represents external input to the network. The gating variables $z_i(t),r_i(t)$ evolve according to dynamics driven by the output $\phi(h(t))$ of the network:
\begin{eqnarray} \label{eq:gatedRNN-eom2}
\tau_x\dot{x}_i(t) = && -x_i(t) +\sum_{j=1}^{N}J^x_{ij}\phi(h_j(t))  + I_{i}^{x},% 
\end{eqnarray}
where $x\in \{z,r\}$. 
Note that the coupling matrices $J^{z,r}$ for $z,r$ are distinct from $J^h$.  We also allow for different inputs $I^{r}$ and $I^{z}$ being fed to the gates. For instance, they might be zero, or they might be equal up to a scaling factor to $I^{h}$.

The value of $\sigma_z(z_i)$ can be viewed as a dynamical time-constant for the $i^{th}$ unit, while the output gate, $\sigma_r(r_i)$,  modulates the output strength of unit $i$. In the presence of external input, the r-gate can control the relative strengths of the internal (recurrent) activity and the external input $I^{h}$. In the limit $\sigma_z, \sigma_r \rightarrow 1$, we recover the dynamics of the classical RNN.

We choose the coupling weights from a Gaussian distribution with variance scaled such that the input to each unit will remain $O(1)$. Specifically,
$J^{h,z,r}_{ij} \sim \mathcal{N}\left(0,N^{-1} \right)
$. 
This choice of couplings is a popular initialization scheme for RNNs in machine learning \cite{jozefowicz2015empirical,sutskever2013importance}, and also  in models of cortical neural circuits \cite{aljadeff2015transition, sompolinsky1988chaos}. If the gating variables were purely internal, then $(J^{z,r})$ will be diagonal; however, we do not consider this case below. 
In the rest of the paper, we analyze the various dynamical regimes the gated RNN exhibits and their functional significance.

% ************ JACOBIAN SPECTRUM ****************

\section{How the gates shape the linearised dynamics}
We first study the linearised dynamics of the gated RNN through the lens of the  instantaneous Jacobian, and
describe how these dynamics are shaped by the gates. 
  The instantaneous Jacobian describes the  linearized dynamics about an operating point, and the eigenvalues of the Jacobian inform us about the  timescales of growth/decay of perturbations and  the local stability of the dynamics.
 As we show below, the  spectral density of the Jacobian depends on {\it equal-time} correlation functions, which are the order parameters in the mean-field picture of the dynamics, developed in Appendix \ref{app:DMFT_derivation}. 
 We study how the gates shape the support and the density of Jacobian eigenvalues in the steady state, through their influence on the correlation functions.

The linearized  dynamics in the tangent space at an operating point $\mathbf{x} = (\mathbf{h},\mathbf{z},\mathbf{r})$ is given by 
\begin{eqnarray} \label{eq:tangent-eom}
\delta \dot{\mathbf{x}} = \mathcal{D}(t) \delta \mathbf{x}
\end{eqnarray}
where $\mathcal{D}$ is the $3N \times 3N$ dimensional instantaneous Jacobian of the full network equations. Linearization of eqs. (\ref{eq:gatedRNN-eom1}) and (\ref{eq:gatedRNN-eom2})
yields
\begin{align} \label{eq:Jacobian_1}
\mathcal{D} = \begin{pmatrix}
\left[\sigma_z\right]\left( -\mathbbm{1} + J^h \left[\phi' \sigma _{r}\right]\right) 
& \mathbb{D}  & \left[\sigma_{z}\right] J^h \left[\phi \sigma_{r}^{\prime}\right] \\
\tau_z^{-1}J^z \left[ \phi'\right]  & -\tau_z^{-1} \mathbbm{1} & 0 \\
\tau_r^{-1}J^r \left[ \phi' \right] & 0 & -\tau_r^{-1} \mathbbm{1} 
\end{pmatrix}    
\end{align}
where $\left[x\right]$ denotes a diagonal matrix with the diagonal entries given by the vector $x$. The term $\mathbb{D}_{ij} = \delta_{ij}\sigma_{z}'(z_{i})  \left( - h_{i} + \sum_{j}J_{ij}^{h} \phi(h_j) \sigma_{r}(r_{j}) \right)= \left[ - \sigma_{z}'(z) h\right] + \left[\sigma_{z}'\odot J^{h} ( \phi \odot \sigma_{r}) \right]$ %= %-\mathbb{D}_{h G_{z}^{\prime}} + \mathbb{D}_{G_{z}^{\prime}\cdot (J^h[\phi \odot \sigma_r])} $ 
arises when we linearize about a time-varying state and will be zero for fixed-points. We introduce the additional shorthand $\phi'(t) = \phi'(h(t))$ and  $\sigma_{r/z}' = \sigma_{r/z}'(r/z(t))$. 

The Jacobian is a block-structured matrix involving random elements ($J^{z,h,r}$) and functions of various state variables. We will need additional tools from non-Hermitian random matrix theory (RMT) \cite{feinberg1997non} and dynamical mean-field theory (DMFT)\cite{sompolinsky1988chaos} to  analyze the spectrum of the Jacobian $\mathcal{D}$. We provide a detailed, self-contained derivation of the calculations in Appendix \ref{app:DMFT_derivation} (DMFT) and Appendix \ref{app:RMT_spectral_curve} (RMT). Here, we only state the main results derived from these formalisms. 

One of the main results is an analytical expression for the spectral curve, which describes the boundary of the  Jacobian spectrum, in terms of the moments of the state variables. The most general expression for the spectral curve (Appendix \ref{app:RMT_spectral_curve}, Eq. \ref{eq:RMT_spectral_curve_general}) involves empirical averages over the $3N$ dimensional state variables. However, for large $N$, we can appeal to a concentration of measure argument to replace these discrete sums with averages over the steady-state distribution from the DMFT (c.f. Appendix \ref{app:DMFT_derivation}) -- i.e. we can replace empirical averages of any function of the state variables  $\frac{1}{N} \sum_{i} F(h_{i}, z_{i}, r_{i})$ with
$ \langle F(h(t), z(t), r(t))\rangle$, where the brackets indicate average over the steady-state distribution. The DMFT + RMT prediction for the spectral curve for a generic steady-state point is given in Appendix \ref{app:RMT_spectral_curve} eq. \ref{eq:RMT_MFT_spectral_curve}. 
 Strictly speaking, the analysis of the DMFT around a generic time-dependent steady state is complicated by the fact that the distribution for $h$ will not be Gaussian (while $r$ and $z$ {\it are} Gaussian). For fixed-points, however, the distributions of $h,z,r$ are all Gaussian, and the expression for the spectral curve reduces simplifies. It is given by the set of $\lambda \in \mathbbm{C}$ which satisfy:

\begin{align}\label{eq:RMT_FP_spectral_curve}
\langle \phi'^{\,2} \rangle \left( \langle\sigma_{r}^{2} \rangle  + \frac{\langle \phi^{2} \rangle \langle \sigma_{r}'^{\,2}\rangle}{|1 + \tau_{r} \lambda|^{2}}   \right) \left\langle \frac{\sigma_{z}^{2} }{| \lambda + \sigma_{z}|^{2}} \right\rangle_{z} = 1.
\end{align}
Here, the averages are taken over the Gaussian fixed-point distributions $(h, z, r) \sim \mathcal{N}(0, \Delta_{h,z,r})$ which follow from the MFT Eq.(\ref{eq:MFT_FP_variance_1}). For example, $\langle \phi'^{2} \rangle = \mathbbm{E}_{h \sim \mathcal{N}(0, \Delta_{h})}[ \phi'(h)^{2}]$.  

We would like to make two comments on the Jacobian of a time-varying state: i) one might wonder if any useful information can be gleaned by studying the Jacobian at a time-varying state where the Hartman-Grobman theorem is not valid. Indeed, as we will see below, the limiting form of the Jacobian in steady-state crucially informs us about  the suppression of unstable directions and the emergence of slow dynamics due to {\it pinching} and marginal stability in certain parameter regimes (also see 
\footnote{  Since the Jacobian spectral density depends on correlation functions (see Appendix \ref{app:RMT_spectral_curve}), in the dynamical steady-state the spectral density becomes time-translation invariant. In other words, the spectral density also reaches a steady state distribution. As a result, a snapshot of the spectral density at any given time will have the same form. Instability then implies that the eigenvectors must evolve over time in order to keep the dynamics bounded.
The timescale involved in the evolution of the eigenvectors should correspond roughly with the correlation time implied by the DMFT. Within this window, the spectral analysis of the Jacobian in the steady state gives a meaningful description of the range of timescales involved. Furthermore, we see empirically that the local structure appears very informative of the true dynamics, in particular with understanding the emergence of continuous attractors and marginal stability, as we discuss in Sec. \ref{sec:marginal_stability}}).  In other words, the instantaneous Jacobian charts the approach to marginal stability, and provides a quantitative justification for the approximate integrator functionality exhibited in Sec. \ref{subsec:marginal_stability_line_attractors}; 
ii) interestingly, the spectral curve calculated using the MFT (Eq.( \ref{eq:RMT_FP_spectral_curve})) for a {\it time-varying} steady-state not deep in the chaotic regime is a very good approximation for the true spectral curve (see Fig. \ref{fig:Jac_spec_full_v_MFT}).

% ---------------- FIGURE JACOBIAN SPECTRUM -------------

\begin{figure}
\begin{centering}
\includegraphics[scale=0.58]{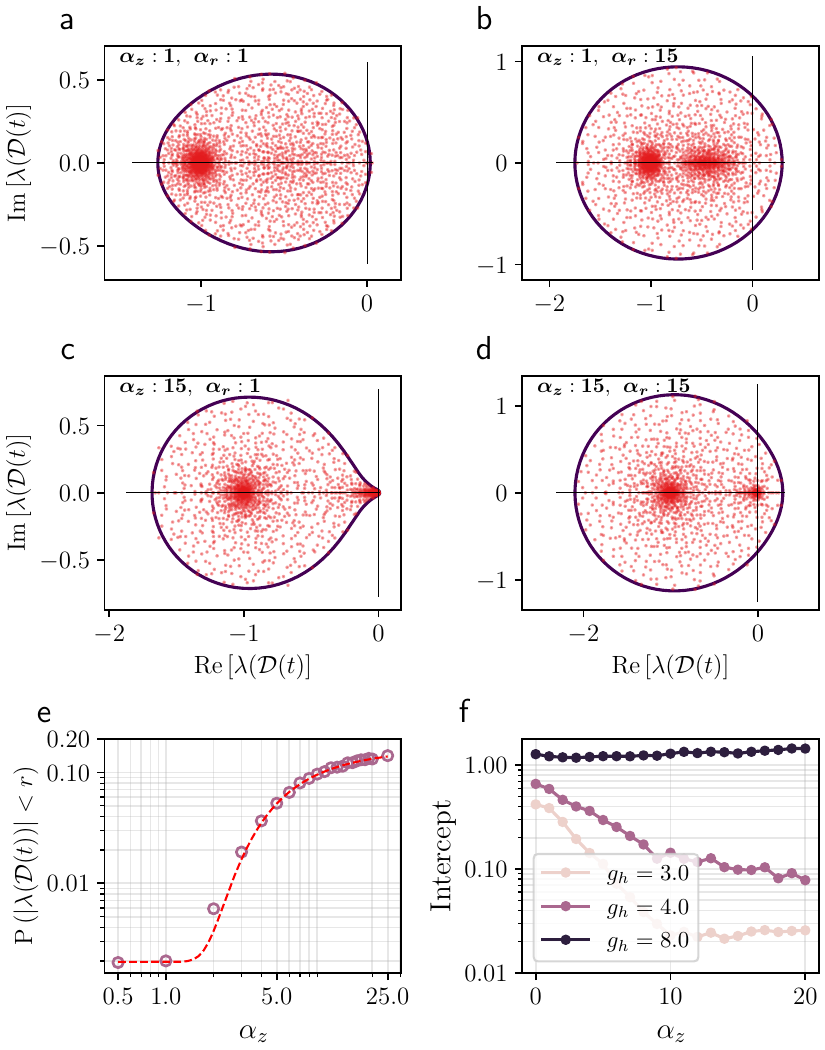}
\par\end{centering}
\caption{\label{fig:Jac_spectrum_RMT} {\it How gates shape the Jacobian spectrum:}  (a-d) Jacobian eigenvalues (red dots) of the gated RNN in (time-varying) steady-state. The dark outline is the spectral support curve predicted by eq. \ref{eq:RMT_FP_spectral_curve}. Bottom row corresponds to larger $\alpha_z$ and right column corresponds to large $\alpha_r$. (e) Cumulative distribution function of  Jacobian eigenvalues in a disk of radius $r=0.05$ centered at the origin plotted against $\alpha_z$. Circles are numerical density calculated from the true network Jacobian (averaged over 10 instances), and the dashed line is a fit from eq. \ref{eq:ev_zero_density}.  (f) Intercept of the spectral curve on the imaginary axis, plotted against $\alpha_z$ for three different values of $g_h$ $(\alpha_r=0)$. For network simulations $N=2000$, $g_h=3, \tau_r=\tau_z=1$ unless otherwise stated, and all biases are zero.
}
\end{figure}

 Fig. \ref{fig:Jac_spectrum_RMT} [a-d] show that the RMT prediction of the spectral support (dark outline) agrees well with the numerically calculated spectrum (red dots) in different dynamical regimes. As a consequence of eq. \ref{eq:RMT_FP_spectral_curve}, we get a condition for the stability of the zero fixed-point. The leading edge of the spectral curve for the zero fixed-point will cross the origin when 
  $g_h < 1 + e^{-\beta_r}$. So, in the absence of biases, $g_h>2$ will make the zero FP unstable.  More generally,  the leading 
 edge of the spectrum crossing the origin gives us the condition for the FP to become unstable:
 \begin{align}\label{eq:transition-to-chaos}
& \left\langle \phi^{\prime 2}\right\rangle \Big(\left\langle\phi^{2}\right\rangle
\left\langle \sigma_{r}'^{2}\right\rangle
+ 
\left\langle\sigma_{r}^{2}\right\rangle\Big)
 > \: \:
1 \quad  \Rightarrow {\rm unstable \, \, FP }
\end{align}

 We will see later on that the time-varying state corresponding to this regime is chaotic.  We now proceed to analyze how the two gates shape the Jacobian spectrum via the equation for the spectral curve.

% ----------------  CLUMPING AND PINCHING ------------

\subsection{{\it Update} gate facilitates slow modes and
{\it Output} gate causes instability}
\label{subsec:spectral_shaping}

To understand how each gate shapes the local dynamics,  we study their effect on the density of Jacobian eigenvalues and the shape of the spectral support curve -- the eigenvalues tell us about the rate of growth/decay of small perturbations, and thus timescales in the local dynamics, and the spectral curve informs us about stability. For ease of exposition, we consider the case without biases in the main text ($\beta_{r,z,h}=0$); we discuss the role of biases in Appendix \ref{app:role_of_biases}. 

Fig. \ref{fig:Jac_spectrum_RMT} shows how the gain parameters of the update and output gates -- $\alpha_z$ and $\alpha_r$ respectively -- shape the Jacobian spectrum. In Fig. \ref{fig:Jac_spectrum_RMT} [a-d],  we see that $\alpha_z$ has two salient effects on the spectrum: increasing $\alpha_{z}$ leads to i) an accumulation of eigenvalues near zero; and ii) a {\it pinching} of the spectral curve for certain values of $g_h$ wherein the intercept on the imaginary axis gets smaller (Fig. \ref{fig:Jac_spectrum_RMT}f; also see Sec. \ref{sec:marginal_stability}A ). In Fig. \ref{fig:Jac_spectrum_RMT} [a-d], we also see that increasing the value of $\alpha_r$ leads to an increase in the spectral  radius, thus pushing the leading edge  (${\rm max}\, {\rm Re} \lambda_{i}$) to the right and thereby increasing the local dimensionality of the unstable manifold. This behavior of the linearized dynamics is also reflected in the nonlinear dynamics, where, as we show in Sec.\ref{sec:output-gate}, $\alpha_r$ has the effect of controlling the {\it dimensionality} of full phase-space dynamics.

The accumulation of eigenvalues near zero with increasing $\alpha_z$ suggests the emergence of a wide spectrum of timescales in the local dynamics. To understand this accumulation quantitatively, it is helpful to consider the scenario where $\alpha_z$ is large and we  replace the $\tanh$ activation functions with a piece-wise linear approximation. In this limit,  the density of eigenvalues within a radius $\delta$ of the origin is well approximated by the following functional form (details in Appendix \ref{app:clumping_and_pinching}):
\begin{align} \label{eq:ev_zero_density}
    P\left( | \lambda(\mathcal{D}_x) | < \delta \right)
    \sim c_0 \textrm{erf}\left( \frac{c_1}{\alpha_z} \right)
\end{align}
where $c_0,c_1$ are constants that, in general, depend on $a_r, \delta$, and $g_h$. 
 Fig. \ref{fig:Jac_spectrum_RMT}e, shows this scaling for a specific value of $\delta$: the dashed line shows the predicted curve and the circles indicate the actual eigenvalue density calculated using the full Jacobian. In the limit of $\alpha_z \rightarrow \infty$ we get an extensive number of eigenvalues at zero, and the eigenvalue density converges to (see Appendix \ref{app:clumping_and_pinching}):
\begin{align*}
     \mu(\lambda)=(1-f_z) \delta(\lambda)+ f_z(1-f_h) \delta(\lambda+1)  + \frac{4}{\pi g_h^2} 
     \mathbb{I}_{\{ |\lambda| \leq g_h^2/4 \} },
 \end{align*}
 where $f_z = \langle \sigma_{z}(z) \rangle$ is the fraction of update gates which are non-zero, and $f_h$ is the fraction of unsaturated activation functions $\phi(h)$. For other choices of 
 saturating non-linearities, the extensive number of eigenvalues at zero remains; however, the expressions are more complicated.
  Analogous phenomena were observed for discrete-time gated RNNs in \cite{can2020gating}, using a similar combination of analytical and numerical techniques \footnote{The continuous-time gated RNN we study in this paper is most closely related to the Gated Recurrent Unit (GRU) architecture studied in \cite{can2020gating}.}

In Sec.\ref{sec:long-time}, we show that the slow modes, as seen from linearization, persist asymptotically (i.e. in the nonlinear regime). This can be seen from the Lyapunov spectrum in Fig(\ref{fig:Lyapunov_analyses}a), which for large $\alpha_{z}$ exhibits an analogous accumulation of Lyapunov exponents near zero.  

In the next section, we study the profound {\it functional} consequences of the combination of spectral pinching and accumulation of eigenvalues near zero.

 %************** MARGINAL STABILITY ********************

\section{Marginal stability and its consequences} \label{sec:marginal_stability}

As the update gate becomes more switch-like (higher $\alpha_z$), we see an accumulation of slow modes and a pinching of the spectral curve which drastically suppresses the unstable directions. In the limit $\alpha_z \rightarrow \infty$, this can make previously unstable points  marginally stable by pinning  the leading edge of the spectral curve exactly at zero. Marginally stable systems are of significant interest because of the potential benefits in information processing  -- for instance, they can generate long timescales in their collective modes \cite{mora2011biological,bialek2012biophysics}. Moreover, achieving marginal stability often requires fine-tuning parameters close to a bifurcation point. As we will see, gating allows us to achieve a marginally-stable critical state over a wide range of parameters; this has been typically highly non-trivial to achieve (e.g. \cite{bialek2012biophysics} pp. 329-350,\cite{mora2011biological}). We first investigate the conditions under which marginal stability arises, then we touch on one of its important functional consequences : appearance of {\it ``line attractors''} which allow the system to be used as a robust integrator.

\subsection{Condition for marginal stability}
\label{subsec:cond_marg_stability}
Marginal stability is a consequence of   pinching of the spectral curve with increasing $\alpha_z$, wherein the (positive) leading edge of the spectrum and the intercept of the spectral curve on the imaginary axis both shrink with $\alpha_z$ (e.g. Fig.  \ref{fig:Jac_spectrum_RMT}f and compare a,c). However, we see in Fig. \ref{fig:Jac_spectrum_RMT}f (via the intercept) that pinching does not happen if $g_h$ is sufficiently large (even as $\alpha_z \to \infty$).
Here, we provide the conditions when pinching can occur and thus marginal stability can emerge.   For simplicity, let us consider the case where $\tau_r=1$ and there are no biases.

Marginal stability strictly only exists for $\alpha_{z} = \infty$. We first examine the conditions under which the system can become marginally stable in this limit, and then we explain the {\it route} to marginal stability for large but finite $\alpha_z$,  i.e. how a time-varying state ends up as a marginally stable fixed-point. For $\alpha_z = \infty$, the spectral density has an extensive number $N(1- \langle \sigma_{z}(z) \rangle)$ of zero eigenvalues, and the remaining eigenvalues are distributed in a disc centered at $\lambda = -1$ with radius $\rho$. If $\rho < 1$, then the spectral density has two topologically  disconnected configurations (the disc and the zero modes) and the system is marginally stable. If $\rho > 1$  the zero modes get absorbed in the interior of the disc and the system will be unstable with fast, chaotic dynamics. The radius $\rho$ is given by $\rho^{2} = \frac{1}{2} a + \frac{1}{2} \sqrt{4 b + a^{2}}<1$, where $a = \langle \phi'^{2}\rangle \langle \sigma_{z}\rangle \langle \sigma_{r}^{2}\rangle$ and $b = \langle \phi'^{2}\rangle \langle \sigma_{z}\rangle\langle \phi^{2}\rangle \langle \sigma_{r}'^{2}\rangle$.
This follows from Eq. \ref{eq:RMT_FP_spectral_curve} by evaluating the $z$-expectation value assuming $\sigma_{z}$ is a binary variable. Thus the system will be marginally stable in the limit $\alpha_z=\infty$ as long as 
\begin{align}\label{eq:condn-marg-stab-general}
\left\langle \phi'^{\, 2}\right\rangle
\Big(
\left\langle \phi^{2} \right\rangle
\left\langle \sigma_{r}'^{\,2}\right\rangle
+ 
\left\langle \sigma_{r}^2\right\rangle
\Big)
<
\langle \sigma_{z}\rangle ^{-1}
\end{align}

The crucial difference between this expression and Eq. (\ref{eq:transition-to-chaos}) is that the RHS now has a factor of $\langle \sigma_{z}\rangle^{-1}$ which can be  greater than unity, thus pushing the transition to chaos further out along the $g_{h}$ and $\alpha_{r}$ directions, as depicted in the phase diagram Fig. (\ref{fig:phaseDiag_combined}). For concreteness, we report here how the transition changes at $\alpha_{r} = 0$. In this setting, the transition to chaos moves from $g_{h} = 2$ to $g_{h} \lessapprox 6.2$, and the system is marginally stable for $2 < g_{h} \lessapprox 6.2$. 

Having identified the region in the phase diagram that can be made marginally stable for $\alpha_{z} = \infty$, we can now discuss the route to marginal stability for large but finite $\alpha_{z}$.  In other words, how does an unstable chaotic state become marginally stable with increasing $\alpha_{z}$? Since the marginally stable region is characterized by a disconnected spectral density, evidently increasing $\alpha_{z}$ must lead to singular behavior in the spectral curve. This takes the form of a {\it pinching} at the origin. We show that for values of $g_h$ supporting marginal stability, the leading edge $\lambda_{e}$ of the spectrum for  the time-varying state gets pinched exponentially fast with $\alpha_z$ as $ \lambda_{e} \sim   e^{-c \alpha_{z} \sqrt{\Delta_{h}}}$ (see Appendix \ref{app:clumping_and_pinching}). This accounts for the fact that already for $\alpha_{z} = 15$, we observe the pinching in Fig. (\ref{fig:Jac_spectrum_RMT} c). In contrast, the parameters in Fig. (\ref{fig:Jac_spectrum_RMT}d) lie outside the marginally stable region, and thus there is no pinching, since the zero modes are asymptotically (in $\alpha_{z}$) buried in the bulk of the spectrum.

\begin{comment}
 Thus, marginal stability is possible for all values of $g_h$ for which
 \begin{align} \label{eq:condn-marg-stab-noBias}
     \left\langle\phi^{\prime\: 2}\right\rangle < 8, \quad \textrm{where }  \left\langle\phi'^{\,2}\right\rangle \equiv \int Dx \phi'(\sqrt{\Delta_{h}} x)^{2} 
 \end{align}
which approximately corresponds to $g_h \lessapprox 6.2$. It is easy to see from eq.(\ref{eq:RMT_FP_spectral_curve}) that unstable points satisfy 
 $\left\langle\phi^{\prime\: 2}\right\rangle > 4$, i.e. $g_h > 2$, so we see that unstable points corresponding to $2 < g_h \lessapprox 6.2$ can be made marginally stable.
 More generally, for finite $\alpha_r$, in the limit $\alpha_z \rightarrow \infty$, the condition for marginal stability is:

\begin{align}\label{eq:condn-marg-stab-general}
\left\langle \phi'^{\, 2}\right\rangle
\Big(
\left\langle \phi^{2} \right\rangle
\left\langle \sigma_{r}'^{\,2}\right\rangle
+ 
\left\langle \sigma_{r}^2\right\rangle
\Big)
<
\langle \sigma_{z}\rangle ^{-1}
\end{align}

\end{comment}

 To summarize, as $\alpha_{z} \to \infty$ the Jacobian spectrum undergoes a topological transition from a single simply connected domain to two domains, both containing an extensive number of eigenvalues. A finite fraction of eigenvalues will end up sitting exactly at zero, while the rest occupy a finite circular region. If the leading edge of the circular region crosses zero in this limit, then the state remains unstable; otherwise, the state becomes marginally stable. The latter case is achieved through a gradual  pinching of the spectrum near zero; there is no pinching in the former case.

We emphasize that marginal stability requires more than just an  accumulation of eigenvalues near zero. Indeed, this will happen even when $g_h$ is outside the range supporting marginal stability as $\alpha_z \to \infty$, but there will be no pinching and the system remains unstable (e.g. see Fig. \ref{fig:Jac_spectrum_RMT}d). 
We will return to this when we describe the phase diagram for the gated RNN (Sec. \ref{sec:phase_diagrams}). There we will see that the marginally 
stable region occupies a macroscopic volume in the parameter space 
adjoining the critical lines on one side.

%------------------- LINE ATTRACTOR FIGURE -------------

\begin{figure}
\begin{centering}
\includegraphics[scale=0.4
]{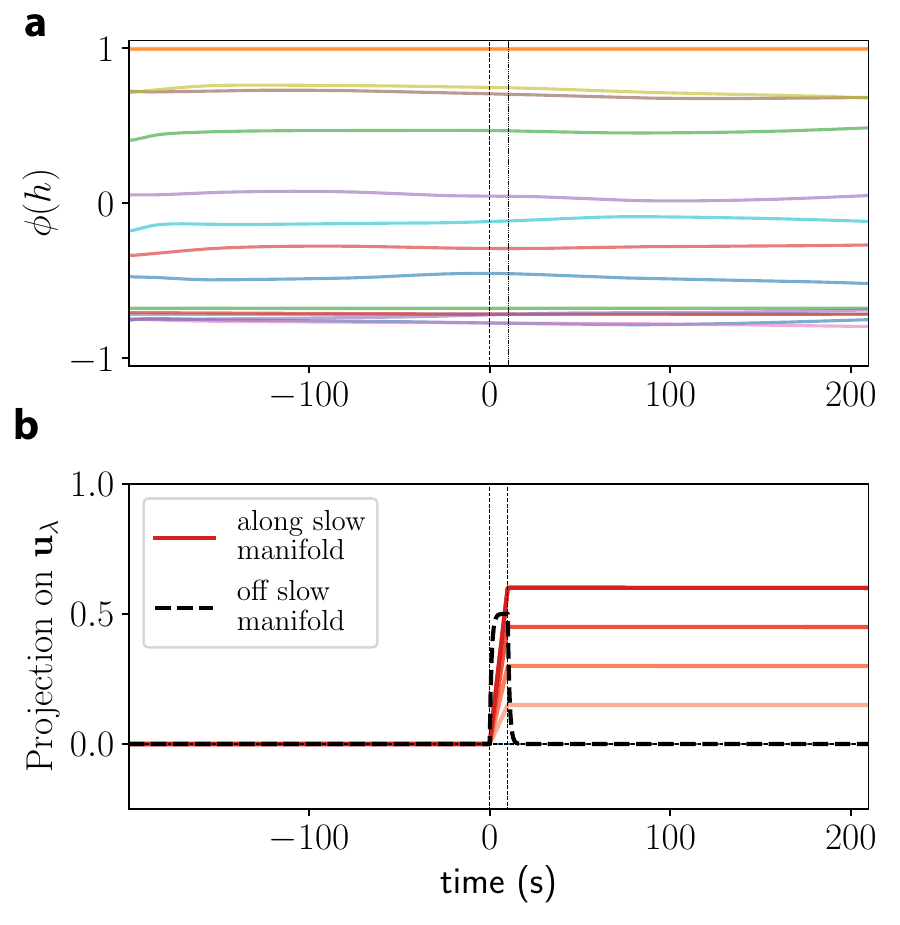}
\par\end{centering}
\caption{\label{fig:line-attractor-1} {\it Network in the marginally-stable state functions as an integrator:}
 a) sample traces from a network with switch-like update gates $(\alpha_z=30, g_h=3)$ show slow evolution (time on $x-$axis is relative to $\tau_h$). b) An input is applied to the same network in (a) from $t=0$ till $t=10$, either aligned with a slow eigenvector $\mathbf{u}_{\lambda}$ (red traces) or unaligned with slow modes (black dashed trace).
Plot shows the excess projection of the network state on the left eigenvector $\mathbf{u}_{\lambda}$.
Different  shades of red correspond to different input strengths.
If the input is along the slow manifold, the trace of the input is retained for a long time after the cessation of input. (the traces in (a) are for the network with an input along the manifold)
}
\end{figure}

\subsection{Functional consequences of marginal stability}\label{subsec:marginal_stability_line_attractors}

The marginally-stable critical state produced by gating can subserve the function of a robust integrator. This integrator-like function  is crucial for a variety of computational functions such as motor control \cite{seung1998continuous,seung1996brain,seung2000stability}, decision making \cite{machens2005flexible} and auditory processing \cite{eguiluz2000essential}. However, achieving this function  has typically required fine-tuning or special hand-crafted architectures\cite{chaudhuri2016computational}, but gating permits the integrator function over a range of parameters and without any specific symmetries in $J^{h,z,r}$. Specifically, 
% the clumping of eigenvalues near zero and importantly the pinching, lead to a significant reduction in the number of unstable directions, and a manifold of slow directions emerges. Thus, 
for large $\alpha_z$, any perturbation in the span of the eigenvectors corresponding to the eigenvalues with magnitude close to zero will be integrated by the network, and once the input perturbation ceases the memory trace of the input will be retained for a duration much longer than the intrinsic time-constant of the neurons; perturbations along other directions, however, will relax with a spectrum of timescales dictated by the inverse of (the real part of) their eigenvalues. Thus the manifold of slow directions will form an approximate continuous attractor on which input can effortlessly move the state vector around. These approximate continuous attractor dynamics are illustrated in Fig. \ref{fig:line-attractor-1}. At time $t = 0$, an  input $I^{h}$ (with $I^{r} = I^{z} = 0$) is applied till $t=10$ (between dashed vertical lines) along an eigenvector of the Jacobian with an eigenvalue close to zero. Inputs along this slow manifold with varying strengths (different shades of red) are integrated by the network as evidenced by the excess projection of the network activity on the left-eigenvector $\mathbf{u}_{\lambda}$  corresponding to the slow mode; on the other hand, inputs not aligned with the slow modes decay away quickly (dashed black line).  Recall that the intrinsic time-constant of the neurons here is set to 1 unit. The exponentially fast (in $\alpha_{z}$) pinching of the spectral curve (discussed above in sec. \ref{subsec:cond_marg_stability}) suggests this slow-manifold behavior should also hold for moderately large $\alpha_z$ (as in Fig. \ref{fig:line-attractor-1}).  Therefore, even though the state is technically unstable, the local structure of the Jacobian is responsible for giving rise to extremely long timescales, and allows the network to operate as an approximate integrator within relatively time windows of time, as demonstrated in Fig.(\ref{fig:line-attractor-1}).

 Of course, after sufficiently long times, the instability will cause the state to evolve and the memory will be lost. Exactly how long the memory will last depends on the asymptotic stability of the network, which is revealed by the Lyapunov spectrum, discussed below in Sec.(\ref{sec:long-time}).

 %**************** OUTPUT GATE and LONG TIME BEHAVIOUR *******
 
 \section{ {\it output} gate controls dimensionality and leads to a novel chaotic transition} \label{sec:output-gate}
 
 We have thus far used insights from local dynamics to study the functional consequences of the gates.   To study the salient features of the output gate, it is useful to analyze the effect of inputs and the long-time behavior of the network through the lens of Lyapunov spectra. We will see that the output gate controls the {\it dimensionality} of the dynamics in the phase space; dimensionality is a salient aspect of the dynamics for task function \cite{farrell2019recurrent}. 
 The output gate also gives rise to a novel discontinuous chaotic transition, near which inputs (even static ones) can abruptly push a stable system into strongly chaotic behavior -- contrary to the typically stabilizing effect of inputs. Below, we begin with the Lyapunov analyses of the dynamics, and then proceed to study the chaotic transition.

 \subsection{Long-time behavior of the network}\label{sec:long-time}
 
\begin{figure}
\begin{centering}
\includegraphics[scale=0.37
]{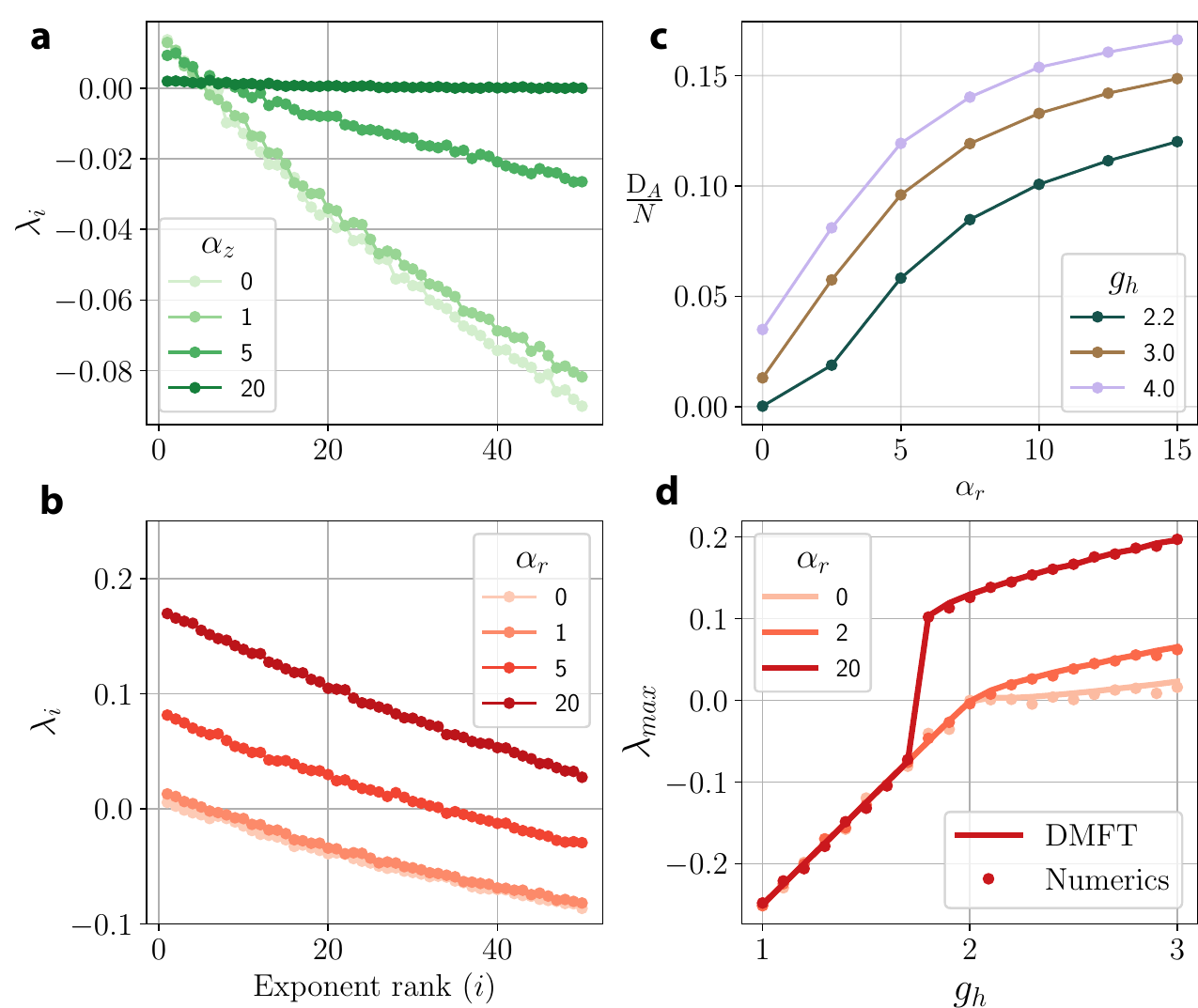}
\par\end{centering}
\caption{\label{fig:Lyapunov_analyses} {\it Lyapunov spectra and dimensionality of the gated RNN: }
(a,b): The first 50 ordered Lyapunov exponents for a gated RNN $(N=2000)$ as a function of varying (a) $\alpha_z$  and (b) $\alpha_r$. The Lyapunov spectrum was calculated as described in Appendix \ref{app:Lyapunov_numerics} (c) The Kaplan-Yorke dimensionality of the dynamics as a function of $\alpha_r$ (d) The maximal Lyapunov exponent $\lambda_{max}$ predicted by the DMFT  (solving eqns.  \ref{eq:Lyapunov-DMFT-eigenvalue_1}-\ref{eq:Lyapunov-DMFT-eigenvalue_2}; solid line) and obtained numerically using the $QR-$method (circles; $N=2000,\alpha_z=0$).
Note that the transition for $\alpha_r=20$ is sharp; also c.f. Fig. \ref{fig:DMFT_bifurcation}c. $\tau_z=\tau_r=2.0$ here.
}
\end{figure}

 We study the asymptotic behavior of the network and the nature of the time-varying state through the lens of its Lyapunov spectra.  In this section, where we study the effects of $\alpha_{z}$, our results are numerical except in cases where $\alpha_{z} = 0$ (e.g. in Fig.\ref{fig:Lyapunov_analyses}d). Lyapunov exponents specify how infinitesimal perturbations $\delta \mathbf{x}(t)$ grow or shrink along
the trajectories of the dynamics -- in particular if the growth/decay is exponentially fast, then the rate will  be dictated by
the maximal Lyapunov exponent defined as \cite{eckmann1985ergodic}:
$\lambda_{max} :=  \lim_{T \rightarrow \infty}
T^{-1} \: \:
\lim_{\| \delta \mathbf{x}(0) \| \rightarrow 0 }
\ln (\| \delta \mathbf{x}(T) \| /
\| \delta \mathbf{x}(0) \|)$. More generally, the set of all Lyapunov exponents -- the Lyapunov spectrum --  yields the 
rates at which perturbations along different directions shrink or diverge, and thus provide a fuller characterization of asymptotic behavior.  We first numerically study how the gates shape the full Lyapunov spectrum (details in Appendix \ref{app:Lyapunov_numerics}), and derive an analytical prediction for the maximum Lyapunov exponent using the DMFT (Sec. \ref{subsec:DMFT_lambdaMax}) \footnote{For reference, we also supply a bound on the maximal Lyapunov exponent in Appendix \ref{app:RMT_Lyapunov}, showing that the relaxation time of the dynamics enters into an upper bound on $\lambda_{max}$.}

 %******************** LYAPUNOV SPECTRA & DIMENSIONALITY FIGURE ****

 Fig. \ref{fig:Lyapunov_analyses}a,b shows how the update $(z-)$ and output $(r-)$ gates shape
the Lyapunov spectrum. We see that as the update gets more 
sensitive (larger $\alpha_z$), the Lyapunov spectrum flattens
pushing more exponents closer to zero, generating long timescales. As the output gate becomes more sensitive 
 (larger $\alpha_r$) all Lypunov exponents increase, thus increasing the rate of growth in unstable directions. 
 
We can estimate the 
 dimensionality of the activity in the chaotic state by calculating an upper-bound $D_A$ on the dimension according to a conjecture by Kaplan \& Yorke \cite{eckmann1985ergodic}. 
The Kaplan-Yorke upper bound for the attractor dimension $D_A$ is given by 
\begin{align} \label{eq:Kaplan_Yorke}
  &  D_A = M +\frac{\sum_{i=1}^{M} \lambda_{i}}{\left|\lambda_{M+1}\right|} \quad \textrm{ where }
  M = \max_j \left\{ \sum_{i=1}^{j} \lambda_{i} \geqslant 0 \right\},
\end{align}
where $\lambda_i$ are the rank-ordered Lyapunov exponents.
We see in Fig. \ref{fig:Lyapunov_analyses}c, that the sensitivity of the output gate ($\alpha_r$) can shape the dimensionality of the dynamics -- a more sensitive output gate  leads to higher dimensionality. As we will see below, this effect of the output gate is different from how the gain $g_h$ shapes dimensionality, and can lead to a novel chaotic transition. Even more directly, if the $r-$gate for neurons $i_1 \ldots i_K$ are set to zero, then the activity is constrained to evolve in a $N-K$ dimensional subspace; however, the $r-$gate allows the possibility -- i.e. the ``inductive bias'' -- of doing this {\it dynamically}.

 \subsubsection{DMFT prediction for $\lambda_{max}$}
 \label{subsec:DMFT_lambdaMax}

We would also like to study the chaotic nature of the time-varying phase by means of the  maximal Lyapunov exponent, and characterize when the transition to chaos occurs. We extend the DMFT for the gated RNN to calculate the maximum Lyapunov exponent, and to do this, we make use of a technique suggested by \cite{derrida1986random,cessac1995increase} and clearly elucidated in
\cite{schuecker2018optimal}. The details are provided in Appendix \ref{app:Lyapunov_DMFT}, and the end result of the calculation is the DMFT prediction for $\lambda_{max}$ as the solution to a  generalized eigenvalue problem for $\kappa$ involving the correlation functions of the state variables:

\begin{align}
  & \big[
  \left(\left\langle \sigma_{z}\right\rangle+ \kappa \right)^{2}-\partial_{\tau}^{2}+ C_{\sigma_{z}}(\tau) -\left\langle \sigma_{z}\right\rangle^{2}
  \big] 
  \chi_{h}\left(\tau \right) = \qquad \nonumber \\
& \qquad \qquad C_{\sigma_{z}'}(\tau) \big[
 C_{\phi_{} \cdot \sigma_{r}}(\tau) - C_{h}(\tau)
\big] 
\chi_{z}\left(\tau \right)
 \nonumber \\
& \qquad \qquad \qquad + \: C_{\sigma_{z}}\left(\tau\right) \frac{\partial C_{\phi_{} \cdot \sigma_{r}}\left(\tau\right)}{\partial C_{h}} \chi_{h}\left(\tau\right), \label{eq:Lyapunov-DMFT-eigenvalue_1}
\\
& \big[\left(1+\tau_{z/r} \kappa \right)^{2}-\tau_{z/r}^{2} \partial_{\tau}^{2}\big] \chi_{z/r}\left(\tau \right) 
=
\frac{\partial C_{\phi_{}}\left(\tau\right)}{\partial C_{h}} \chi_{h}\left(\tau\right)  \label{eq:Lyapunov-DMFT-eigenvalue_2} 
% \\
% & \big[\left(1+\tau_{r} \kappa \right)^{2}-\tau_{r}^{2} \partial_{\tau}^{2}\big] \chi_{r}\left(\tau \right) 
% =
% \frac{\partial C_{\phi_{}}\left(\tau\right)}{\partial C_{h}}) \chi_{h}\left(\tau\right). \label{eq:Lyapunov-DMFT-eigenvalue_3}
\end{align}
 where we have denoted the two-time correlation function $C_{x}(t, t') \equiv \langle x(t) x(t') \rangle$ for different (functions of) state variables $x(t)$ (see Eq.\ref{eq:corr_fn} for more context). The largest eigenvalue solution to this problem is the required 
maximal Lyapunov exponent  \footnote{  One might worry that the $h$ and $\sigma(z)$ correlators are not separable in general. However, this issue only arises for large $\alpha_z$. For moderate $\alpha_z$, the separability assumption is valid.}. Note, this is the analogue of the 
Schr{\"o}dinger equation for the maximal Lyapunov exponent in the vanilla RNN.  When $\alpha_z=0$ (or small), the $h$ field is Gaussian and we can use Price's theorem for Gaussian integrals to replace the variational derivatives on the r.h.s of eq. \ref{eq:Lyapunov-DMFT-eigenvalue_1}-\ref{eq:Lyapunov-DMFT-eigenvalue_2} by simple correlation functions, for instance:  $\partial C_{\phi_{}}\left(\tau\right)/ \partial C_{h}(\tau) =
C_{\phi_{}^{\prime}}\left(\tau\right)$.
In this limit, we see good agreement  between the numerically 
calculated maximal Lyapunov exponent (Fig. \ref{fig:Lyapunov_analyses}c dots) compared to the DMFT prediction (Fig. \ref{fig:Lyapunov_analyses}c solid line) obtained by solving the eigenvalue problem (eq. \ref{eq:Lyapunov-DMFT-eigenvalue_1}-\ref{eq:Lyapunov-DMFT-eigenvalue_2})  
For large values of $\alpha_z$ we see quantitative deviations between the DMFT prediction and the true $\lambda_{max}$. Indeed, for large $\alpha_{z}$ the distribution of $h$ is strongly non-Gaussian, and there is no reason to expect that variational formulas given by Price's theorem will be even approximately correct.  For more on this point, see the discussion toward the end of Appendix \ref{app:DMFT_derivation}.

% ------------------- condition for transition to chaos

\subsubsection{Condition for continuous transition to chaos}
 The value of $\alpha_z$ affects the precise value 
of the maximal Lyapunov exponent $\lambda_{max}$; however, numerics suggest that across a continuous transition to chaos, the point at which $\lambda_{max}$ becomes positive is not dependent on $\alpha_z$ (data not shown). 
We can see this more clearly by calculating the transition to chaos when the leading edge of the spectral curve (for a FP) crosses zero. 
This condition is given by eq. \ref{eq:transition-to-chaos}, and we see that it has no dependence on $\alpha_z$ or the update gate. 
We stress that this condition, eq. \ref{eq:transition-to-chaos}, for the transition to chaos -- when the stable fixed-point becomes unstable -- is valid when the chaotic attractor emerges continuously from the fixed-point (Fig. \ref{fig:Lyapunov_analyses}c, $\alpha_r=0,2$). However, in the gated RNN, there is another discontinuous transition to chaos (Fig.\ref{fig:Lyapunov_analyses}c, $\alpha_r=20$): for sufficiently large $\alpha_r$, the transition to chaos is discontinuous and occurs at a value of $g_h$  where the zero FP is still stable ($g_h < 2$ with no biases).  To our knowledge, this is a novel type of transition which is not present in the vanilla RNN, and not visible from an analysis that only considers the stability of fixed points. We characterize this phenomenon in detail below.

% --------------- DMFT Bifrucation figure ----------------------

\begin{figure*}
\begin{centering}
\includegraphics[scale=0.43]{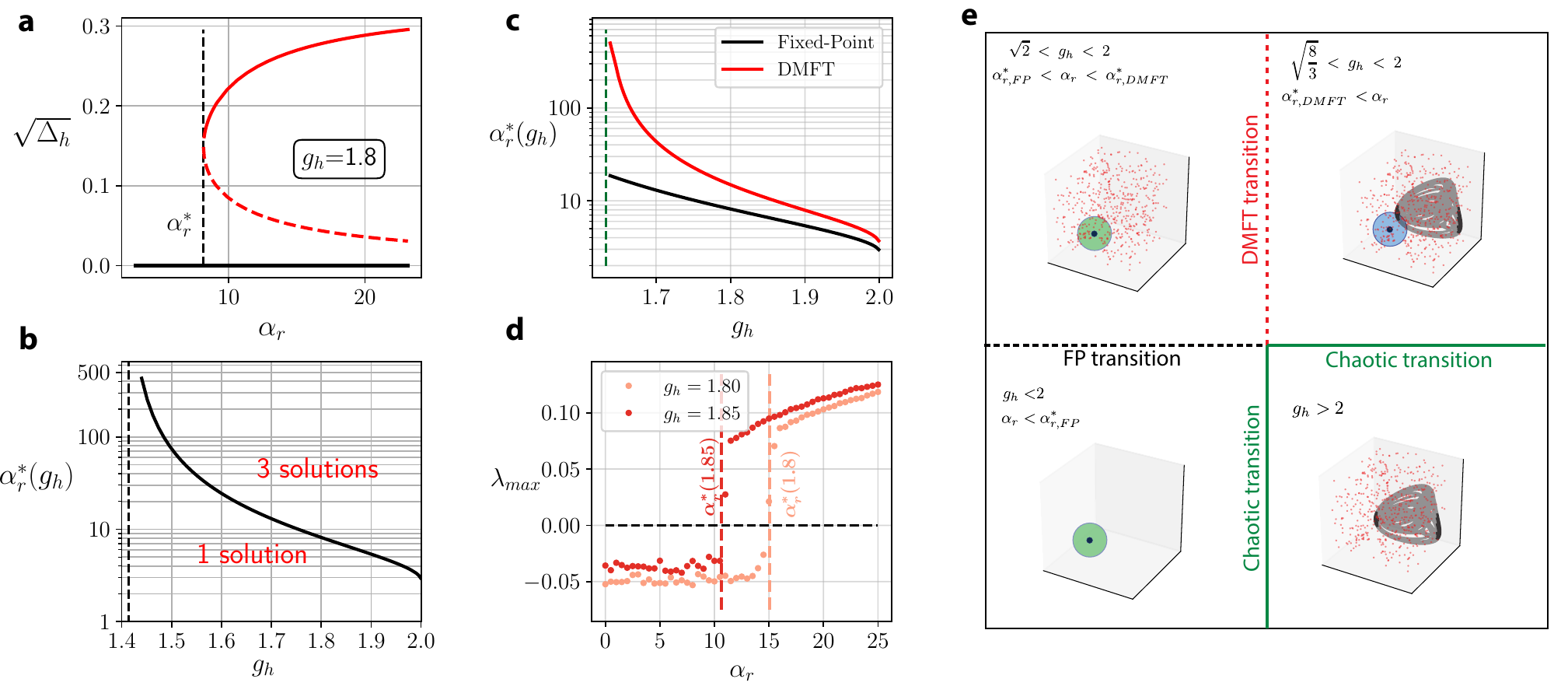}
\par\end{centering}
\caption{\label{fig:DMFT_bifurcation} {\it The discontinuous 
dynamical transition : }
(a) spontaneous appearance of non-zero solutions (dashed/solid red lines) to the FP equations once $\alpha_r$ crosses a critical value $\alpha^{*}_{r,FP}(g_h)$ at fixed $g_h$.  (b) The critical $\alpha^{*}_{r,FP}(g_h)$ as a function of $g_h$. The vertical dashed line represents left critical value $g_c=\sqrt{2}$ below which a bifurcation is not possible.
(c) The critical DMFT transition curve $\alpha_{r,DMFT}^{*}(g_h)$ (red curve) calculated using eqns. \ref{eq:DMFT_BF_eqn_1}, \ref{eq:DMFT_BF_eqn_2}. The FP transition curve from (b) is shown in black. Green dashed line corresponds to $g_c = \sqrt{8/3}$ below which
the dynamical transition is not possible. 
% (d) The corresponding
% DMFT solution to the variance (eq. \ref{eq:DMFT_BF_root}) for $\alpha_r=\alpha_{r,DMFT}^{*}(g_h)$.
(d) Numerically calculated maximum Lypunov exponent $\lambda_{max}$ as a function of $\alpha_r$ for two different values of $g_h$. The dashed lines correspond to the DMFT prediction for the discontinuous transition from (c).  
% (f) Attractor dimension $D_A$  (eq. \ref{eq:Kaplan_Yorke})  for $g_h=1.85$.  (Inset) The first 450 ordered Lyapunov exponents for $\alpha_r > \alpha_{r,DMFT}^{*}(g_h)$ (green) and $\alpha_r < \alpha_{r,DMFT}^{*}(g_h)$ (red). $N=2000; \alpha_z=0$ for the numerical simulations. 
(e) Schematic of the bifurcation transition:
for $g_h < 2$ and $\alpha_r< \alpha_{r,FP}^{*}$ the zero FP is
the only (stable) solution (bottom left box); for $\sqrt{2} < g_h < 2$  and $ \alpha_{r,FP}^{*} < \alpha_r < \alpha_{r,DMFT}^{*}$ the zero FP is still stable, but there is a proliferation of unstable FPs without any  obvious dynamical signature (top left); for $\sqrt{8/3} < g_h < 2$ and $\alpha_r >  \alpha_{r,DMFT}^{*}$  chaotic dynamics coexist with the stable FP and this transition is discontinuous (top right); finally for $g_h>2.0$ the stable FP becomes unstable, and only the chaotic attractor remains; this transition is continuous (bottom right).
}
\end{figure*}
%----------------------------------------------------------

 \subsection{{\it output} gate induces a novel chaotic transition}
 \label{sec:BF_transition}
Here, we describe  a novel phase, 
characterized by a proliferation of unstable fixed-points, and the coexistence of a stable fixed-point with chaotic dynamics.  It is the appearance of this state that gives rise to the discontinuous transition observed in  Fig. \ref{fig:Lyapunov_analyses}c. The appearance of this state is mediated by the output gate becoming more switch-like (i.e. increasing 
$\alpha_r$) in the quiescent region for $g_h$. To our knowledge no such
comparable phenomenon exists in RNNs with additive interactions.  The full details of the calculations for this transition are provided in Appendix \ref{app:DMFT_BF}. Here we simply state and describe the salient features. For ease of presentation, the rest of the section will assume that all biases are zero.  The results in this section are strictly only valid for $\alpha_{z} = 0$. In Appendix \ref{app:z-gate-BF-line}, we argue that they should also hold for moderate $\alpha_{z}$. 

This discontinuous transition is characterized by a few noteworthy features: %has a few {\red no} aspects: 

i) {\it Spontaneous emergence of fixed-points}

When $g_h < 2.0$ the zero fixed-point is stable. Moreover, for $\sqrt{2} < g_{h} < 2$, when $\alpha_r$ crosses a threshold value 
  $\alpha_{r,FP}^{*}(g_h)$, unstable fixed-points spontaneously appear in the phase space.  The only dynamical signature of these unstable FPs are short-lived transients which do not scale with system size (see Fig.\ref{fig:BF_transient_times}). Thus we have a
  \begin{align}\label{eq:FP_BF_condn}
 \textrm{ {\it condition for fixed-point transition}:}
 \nonumber \\
 \sqrt{2} < g_h \leq 2 
 \quad \textrm{and} \quad \alpha_r > \alpha_{r,FP}^{*}(g_h),
\end{align}
 These unstable fixed-points correspond to the emergence of non-trivial solutions to the time-independent MFT. Fig. (\ref{fig:DMFT_bifurcation}a) shows  the appearance of fixed-point MFT solutions for a fixed $g_h$, and Fig. (\ref{fig:DMFT_bifurcation}b) shows the critical $\alpha_{r,FP}^{*}(g_h)$ as a function of $g_h$. As $g_h \to 2^-$, we see that $\alpha_{r,FP}^{*} \to \sqrt{8}$.

These spontaneous MFT fixed-point solutions are unstable according to the criterion Eq. (\ref{eq:transition-to-chaos}) derived from RMT. Moreover, in Appendix \ref{app:Kac-Rice}, using a Kac-Rice analysis, we show that in this region the full $3N$-dimensional system does indeed have a number of unstable fixed-points that grows exponentially fast with $N$.  Thus, this transition line $\alpha_{r, FP}^{*}$ represents a topological trivialization transition as conceived by, e.g. \cite{fyodorov2004complexity,fyodorov2014topology}.  Our analysis shows that instability is intimately connected to the proliferation of fixed-points. Remarkably, however, a time-dependent solution to the DMFT does {\it not} emerge across this transition, and the microscopic dynamics are insensitive to the transition in topological complexity, bringing us to the next point:

ii) {\it A delayed dynamical transition that shows a decoupling between topological and dynamical complexity}

On increasing $\alpha_r$ beyond $\alpha_{r,FP}^{*}$, there is a second transition when $\alpha_r$ crosses a critical value $\alpha_{r,DMFT}^{*}$.  This happens when we satisfy the
\begin{align}\label{eq:DMFT_BF_condn}
 \textrm{ {\it condition for dynamical transition}:}
 \nonumber \\
 \sqrt{\frac{8}{3}} < g_h \leq 2 
 \quad \textrm{and} \quad \alpha_r > \alpha_{r,DMFT}^{*}(g_h),
\end{align}
derived in Appendix \ref{app:subsec:DMFT_transition}. Fig.(\ref{fig:DMFT_bifurcation}c) shows how $\alpha_{r,DMFT}^{*}(g_h)$ varies with $g_h$. As $g_h \to 2^-$, we see that $\alpha_{r,DMFT}^{*} \to \sqrt{12}$. 
Across this transition, a dynamical state spontaneously emerges, and the maximum Lyapunov exponent jumps from a negative value to a positive value (Fig. \ref{fig:DMFT_bifurcation}d). This state
 exhibits chaotic dynamics that coexist with the stable zero fixed-point. The presence of the stable FP means that the dynamical state is not strictly a chaotic attractor but rather a spontaneously appearing ``chaotic set''.  
 On increasing $g_h$ beyond $2.0$, for large but fixed $\alpha_r$, the stable fixed-point disappears and the state smoothly transitions into a full chaotic attractor that was characterized above. This picture is summarized in the schematic in Fig. \ref{fig:DMFT_bifurcation}e. This gap between the proliferation of unstable fixed-points and the appearance of the chaotic dynamics differs from the result of  Wainrib \& Touboul \cite{wainrib2013topological} for purely additive RNNs, where the proliferation  (topological complexity) is tightly linked to the chaotic dynamics (dynamical complexity). Thus, for gated RNNs, there appears to be another distinct mechanism for the transition to chaos, and the accompanying transition is a discontinuous one.

 iii) {\it Long chaotic transients}

 For finite systems, across the transition the dynamics will eventually flow into the zero FP after chaotic transients.   Moreover,  we expect this transient time to scale with the system size,  and in the infinite system size limit, the transient time should diverge in spite of the fact that the stable fixed-point still exists. This is because the relative volume of the basin
of attraction of the fixed-point  will vanish as $N \rightarrow \infty$. In Appendix \ref{app:DMFT_BF} Fig. (\ref{fig:BF_transient_times}c,d) we do indeed see that the transient time for a fixed $g_h$ scales with system size (Fig. \ref{fig:BF_transient_times}c) once $\alpha_r$ is above the second transition (dashed line), and not otherwise (see Fig. (\ref{fig:BF_transient_times}a,e) dashed lines).

 \subsubsection{An input-induced chaotic transition}
 
\begin{figure}
\begin{centering}
\includegraphics[scale=0.35]{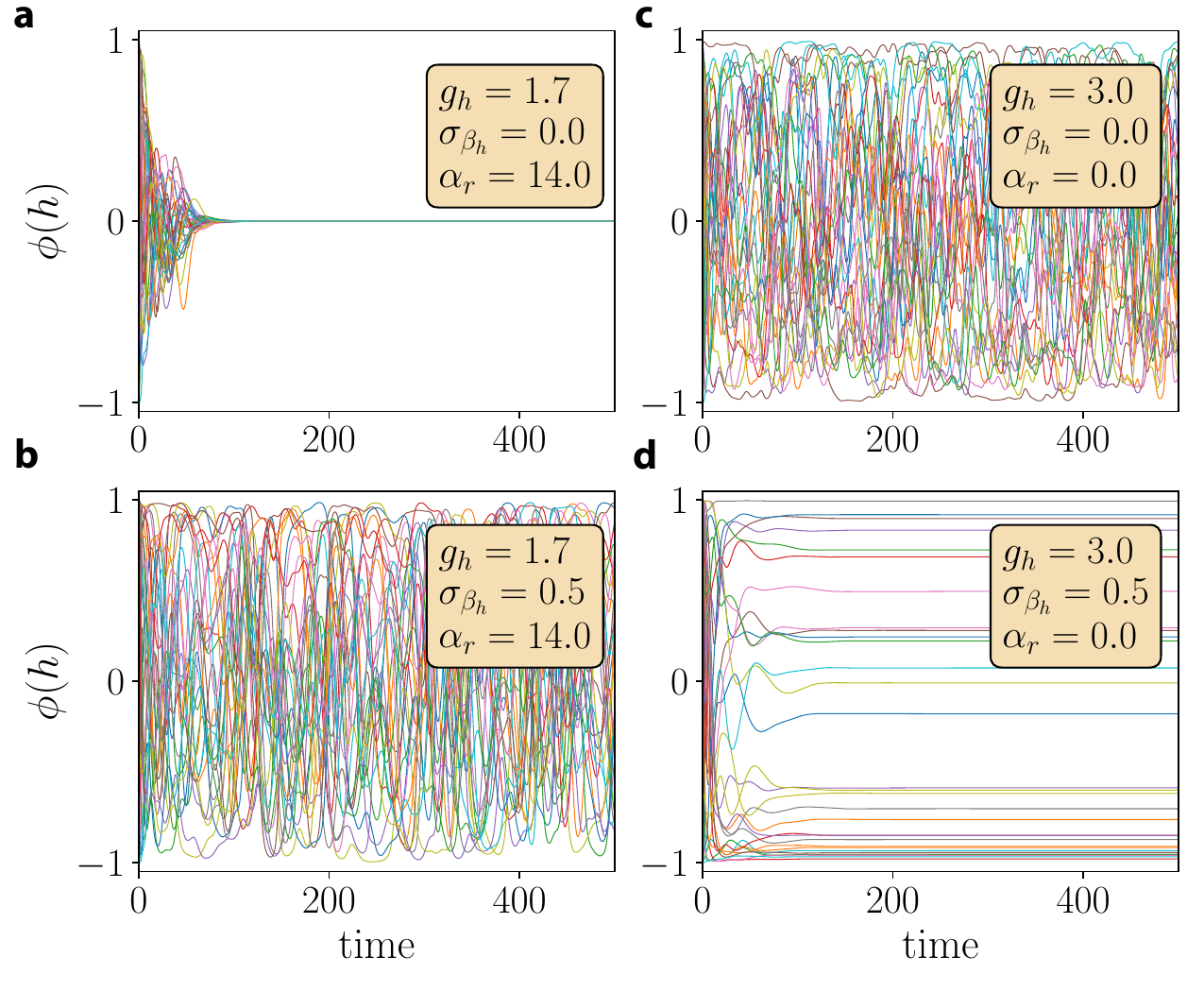}
\par\end{centering}
\caption{\label{fig:input_driven_chaos} {\it Input driven chaos} (a,b): Near the discontinuous chaotic transition  (in region 2 of Fig.(\ref{fig:phaseDiag_combined}))}, static input  $I^{h}$ (with $I^{r} = I^{z} = 0$) can push a stable system (a) to chaotic activity (b); (c,d): In the purely chaotic state (c, $g_h=3.0$), input has the familiar effect of stabilising the dynamics (d). The elements of the input vector $I^{h}$ are random Gaussian variables with zero mean and variance $\sigma_{\beta_{h}}^{2}$.

\end{figure}

The discontinuous chaotic transition has a functional consequence : near the transition, static inputs  can push a stable system to strong chaotic activity. This is in contrast to the typically stabilizing effects of inputs on the activity of random additive RNNs \cite{rajan2010stimulus,molgedey1992suppressing,schuecker2018optimal}. In Fig. \ref{fig:input_driven_chaos}a,b we see that when static input with variance $\sigma_{\beta_h}$ is applied to a  stable system (a) near the discontinuous chaotic transition (in region 2 in Fig. \ref{fig:phaseDiag_combined}), it  induces chaotic activity (b); however, the same input when applied to the system in the chaotic state (Fig. \ref{fig:input_driven_chaos}c) , the dynamics are stabilized (d) as has been reported before. 
% As we will see below this input-induced chaos can be leveraged to implement a reset mechanism for the memory trace.
 
This phenomena for static inputs can be understood using the phase diagram with nonzero biases, discussed in Sec.(\ref{sec:phase_diagrams}). There, we see how the transition curves move when a random bias $\beta_{h}$ is included. Near the classic chaotic transition ($\alpha_{r} = 0$), the bias moves the curve towards larger $g_{h}$, thus suppressing chaos. Near the discontinuous chaotic transition $\alpha_{r,DMFT}^{*}$, the bias pulls the curve toward smaller values of $\alpha_{r}$, thus {\it promoting} chaos. Thus, inputs can have opposite effects of inducing/stabilizing chaos within the same model in different parameter regimes. This phenomenon could in principle be leveraged for shaping the interaction between inputs and internal dynamics.

 %*************** r-gate reset mechanism ***********

  \section{ Gates provide a flexible reset mechanism}

Here we discuss how the gates provide another critical function -- a mechanism to flexibly reset the memory trace depending on external input or the internal state. This function complements the memory function; a memory that can't be erased when needed is not very useful. To build intuition, let us consider a linear network $\dot{\mathbf{h}} = - \mathbf{h} + J\mathbf{h}$, where the matrix $-\mathbbm{1} + J$ has a few eigenvalues that are zero, while the rest have negative real part. The slow modes are good for memory function, however that fact also makes it hard to forget memory traces along the slow modes.  This trade-off was pointed out in \cite{pereira2015tradeoff}. To be functionally useful, it is critical that the memory trace can be erased flexibly in a context-dependent manner. The $r-$gate allows this function naturally. Consider the same net, but now augmented with a $r-$gate:  $\dot{\mathbf{h}} = - \mathbf{h} + J\mathbf{h} \odot \pmb{\sigma}_r$. If the gate is turned off ($\sigma_{r} = 0$) for a short duration, the state $\mathbf{h}$ will be reset to zero. One can actually be more specific: 
% if we choose $\pmb{\sigma}_{r,i} = \left[ \mathbb{I} - \mathbf{v}^{(1)}_i\mathbf{v}^{(1)T}\phi(\mathbf{h}) \right]^+ $, then we can selectively increase the decay rate for the projection along the slow mode $\mathbf{v}^{(1)}$. This basic insight also suggests that one can 
we may choose a $J^r = -\mathbbm{1}\mathbf{u}^T$ with $\pmb{\sigma}_r = \sigma(J^r \phi(\mathbf{h}))$, such that the $r-$gate turns off whenever $\phi(\mathbf{h})$ gets aligned with $\mathbf{u}$, thus providing an internal-context dependent reset. 

\begin{figure}
\begin{centering}
\includegraphics[scale=0.5]{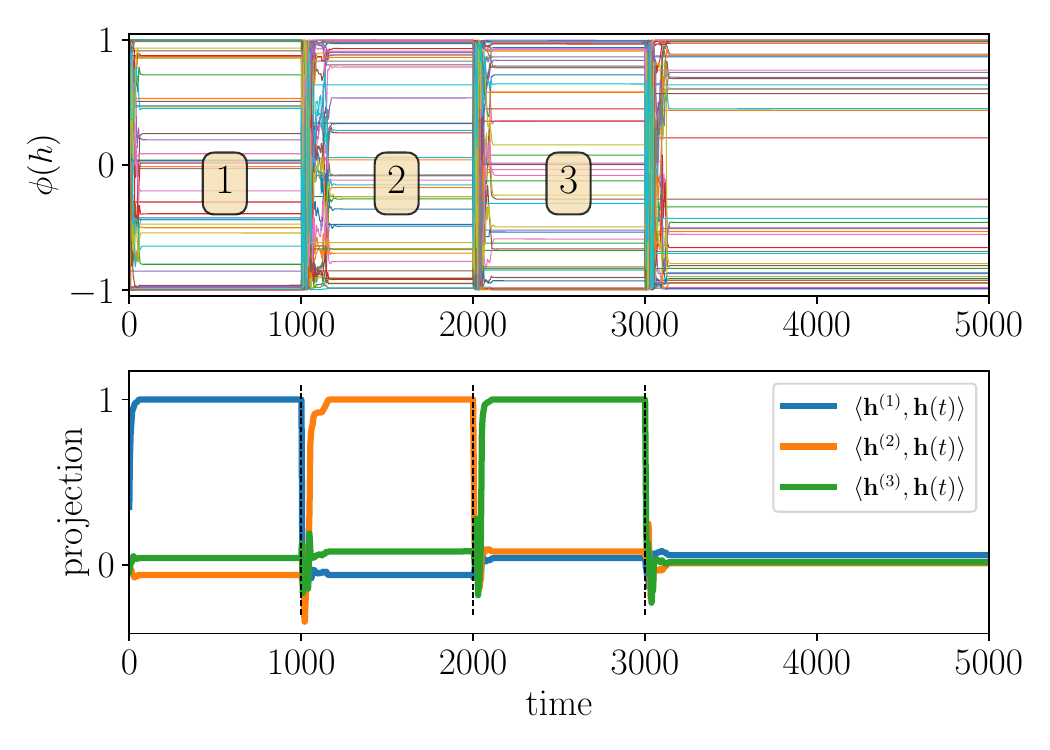}
\par\end{centering}
\caption{\label{fig:zGate_reset} {\it Gates provide a reset mechanism:}  Positive static inputs  are applied to the $z$-gate when the RNN is in the marginally stable state ($g_h=3.0, \alpha_r=2.5, \alpha_z=\infty)$ for 20 time units at times indicated by dashed lines. The input induces chaos which rapidly scrambles the network state thus erasing the trace of the previous memory; bottom panel shows the normalized projection of the state $\mathbf{h}(t)$ on the directions $\mathbf{h}^{(1,2,3)}$ aligned with the state in regions 1,2,3.
}
\end{figure}

Apart from resetting to zero, the $z-$gate also allows the possibility of rapidly scrambling the state to a random value by means of the input-induced chaos. This phenomenon is illustrated in Fig. \ref{fig:zGate_reset}, where the network in the marginally stable state normally functions as  a memory (retains traces for long times, as in Fig. \ref{fig:line-attractor-1}), but positive inputs $I^{z}$ (with $I^{h} = I^{r} = 0$) to the $z-$gate above a threshold strength even for a short duration can induce chaos, thereby scrambling the state and erasing the previous memory state (Fig. \ref{fig:zGate_reset} bottom panel). The mechanism for this scrambling can be understood by appealing to Eq.(\ref{eq:condn-marg-stab-general}). A finite input $I^{z}$ with nonzero mean is able to change $\langle \sigma(z)\rangle$, and thus push the critical line for marginal stability in one way or the other. For instance, if $\langle I^{z}\rangle > 0$, $\langle \sigma(z) \rangle >1/2$, which (for $\alpha_{r}= 0$) moves the transition to marginal stability to a smaller value of $g_{h}$. This implies that a marginally stable state can be made chaotic in the presence of $I^{z}$ with finite mean.  This mechanism for input-induced chaos actually appears to be different from that explored in the previous section, which occurs across the discontinuous chaotic transition. We discuss this more in Sec.(\ref{sec:phase_diagrams}).

 In summary, gating imbues the RNN with the capacity to flexibly reset memory traces, providing an ``inductive bias'' for context-dependent reset.
%Thus, the gates offer the possibility of both flexibly resetting memory and flexibly both resetting the memory trace with appropriate choices of $J^{z,r}$ and how inputs couple to the internal dynamics -- i.e. they provide the gated RNN with an ``inductive bias'' for context-dependent reset. 
The specific method of reset will depend on the task/function and this can be selected, e.g., by gradient-based training. This inductive bias for resetting has been found to be critical for performance in ML tasks \cite{greff2016lstm}.

  %*************** PHASE DIAGRAMS ********************

 \section{Phase diagrams for the gated network}
 \label{sec:phase_diagrams}

 Here, we 
summarize the rich dynamical phases of the gated RNN
and the critical lines separating them. The key parameters determining the critical lines and the 
phase diagram are the activation and output-gate gains
and the associated biases: $(g_h,\beta_h,\alpha_r,\beta_r)$.  The update gate does not play a role in determining continuous or critical chaotic transitions. On the other hand, it will influence the discontinuous transition to chaos for sufficiently large values of $\alpha_{z}$ (see \ref{app:z-gate-BF-line} for discussion). Furthermore, the update gate has a strong effect on the dynamical aspects of the states near the critical lines.  There are macroscopic regions of the parameter space adjacent to the critical lines  where the states can be made marginally stable in the limit of $\alpha_z \rightarrow \infty$. The shape of this marginal stability region is influenced by $\beta_{z}$ and $I^{z}$.

 %----------------------- PHASE DIAGRAM combined -------

\begin{figure}
\begin{centering}
\includegraphics[scale=0.5]{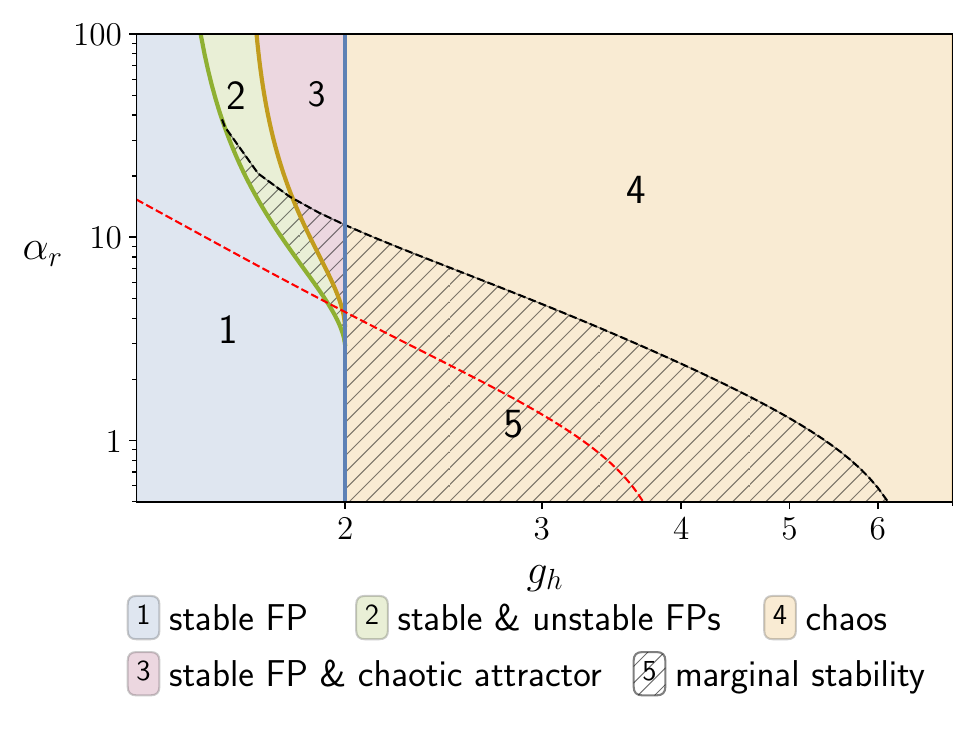}%fig_phaseDiag_combined}
\par\end{centering}
\caption{\label{fig:phaseDiag_combined} {\it Phase diagram for the gated RNN}   a): (no biases) In regions 1 \& 2 the zero FP is the global attractor of dynamics; however in region 2 there is a proliferation of unstable FPs without any  asymptotic dynamical signatures. In region 3, the (stable) zero FP coexists with chaotic dynamics.  Note that the plotted curve separating regions 2 and 3 is computed for $\alpha_{z} = 0$, and remains valid for sufficiently small values of $\alpha_{z}$. In region 4 the zero FP is unstable, and dynamics are chaotic. For all parameter values in region 5, a previously unstable/chaotic state can be made marginally stable when $\alpha_z=\infty$. For any given parameter values in region 5, there are infinitely many marginally stable points in the phase space to which the dynamics will converge. The red dashed line indicates the critical transition between a stable fixed-point (below the line) and chaos (above the line) in the presence of static random input (to the $h-$variable) with standard deviation $\sigma_{h} = 0.5$. Note that while chaos is suppressed for small $\alpha_{r}$ along the $g_{h}$ axis, for larger $\alpha_{r}$ there are regions of stable FPs that become chaotic with finite input. This leads to the phenomenon of input-induced chaos. 
}
\end{figure}

 Fig. \ref{fig:phaseDiag_combined}a shows the dynamical phases for the network with no biases in the $(g_h,\alpha_r)$ plane. 
 When $g_h$ is below $2.0$ and $\alpha_r < \alpha_{r,FP}^{*}$, the zero-fixed point is the only solution ({\bf region 1}). As discussed in Sec.(\ref{sec:BF_transition}), on crossing the fixed-point bifurcation line (green line, Fig. \ref{fig:phaseDiag_combined}a), there is a spontaneous proliferation of unstable fixed-points in the phase space ({\bf region 2}). This can occur only when $g_h>\sqrt{2}$. The proliferation of fixed-points is not accompanied by any obvious dynamical signatures. However, if $\sqrt{8/3} < g_h < 2$, we can increase $\alpha_r$ further to cross a second discontinuous transition where a dynamical state spontaneously appears featuring the coexistence of chaotic activity and a stable fixed-point ({\bf region 3}).  When $g_h$ is increased beyond the critical value of $2.0$, the stable zero fixed-point becomes unstable for all $\alpha_r$, and we  get a chaotic attractor ({\bf region 4}). All the critical lines are determined by $g_h$ and $\alpha_r$, and $\alpha_z$ has no explicit role; however, for large $\alpha_z$ there is a large region of the parameter space on the chaotic side of the chaotic transition that can be made marginally stable  (thatched {\bf region 5} in 
Fig. \ref{fig:phaseDiag_combined}a). 

\subsubsection{Role of biases and static inputs} 
\label{subsec:role_of_biases}
 Biases have the effect of generating non-trivial fixed points and controlling stability by moving the edge of the spectral curve. Another key feature of biases is the suppression of the discontinuous bifurcation transition observed without biases. This is explained in detail in Appendix \ref{app:role_of_biases}. 
A particularly illuminating illustration of the effects of a bias can be inferred from the critical line (red-dashed) for finite bias shown in Fig. \ref{fig:phaseDiag_combined}. This curve, computed using the FP stability criterion (\ref{eq:transition-to-chaos}) combined with the MFT equations (\ref{eq:mft_input1}-\ref{eq:mft_input3}), represents the transition between stability and chaos for finite bias with zero mean and non-zero variance. Equivalently, we may think of this as the critical line for a network with static input $I_{i}^{h} \sim \mathcal{N}(0, \sigma_{h}^{2})$ (with $I_{r} = I_{z} = 0$). Along the $g_{h}$ axis, we can observe the well-documented phenomena whereby an input will suppress chaos. This corresponds to the region $g_{h}>2$ which lies to the left of the red dashed critical line, which is chaotic in the absence of input, and flows to a stable fixed point in the presence of input. However, this behavior is reversed for $g_{h}<2$. Here we see a significant swath of phase space which is stable in the absence of input, but which becomes chaotic when input is present. Thus, the stability-to-chaos phase boundary in the presence of biases (or inputs) reveals that the output (r-) gate can facilitate an input-induced transition to chaos. %for we show the effect of the bias $\beta_h$ on the stability-to-chaos transition. We see that larger values of $\beta_h$ require higher activation gains to transition to chaos, and as in the case without biases, on the chaotic side of the critical line, there is a large region of the parameter space that can become marginally stable for $\alpha_z \rightarrow \infty$.

% ********************* DISCUSSION *******************

\section{Discussion}

Gating is a form of multiplicative interaction that is a central feature of the best performing RNNs in machine learning, and it is also a prominent feature of  biological neurons. Prior theoretical work on RNNs has only considered RNNs with additive interactions. Here, we present the first detailed study on the consequences of gating for RNNs and show that gating can produce dramatically richer behavior that have significant functional benefits.

The continuous-time gated RNN (gRNN) we study resembles a popular model used in machine learning applications, the Gated Recurrent Unit (GRU) (see note below Eq. (\ref{eq:MFT_FP_variance_2})). Previous work \cite{can2020gating} looked at the instantaneous Jacobian spectrum for the discrete-time GRU using RMT methods similar to those presented in Appendix \ref{app:RMT_spectral_curve}; however, this work did not go beyond time-independent MFT, and presented a phase diagram showing only boundaries across which the MFT fixed-points become unstable \footnote{In fact, the fixed-point phase diagrams for the current model and the GRU are in one-to-one correspondence. What this {\it static} phase diagram importantly lacks is region 3 in Fig.(\ref{fig:phaseDiag_combined})} In the present manuscript, we have illuminated the full {\it dynamical} phase diagram for our gated RNN, uncovering much richer structure.  Both the GRU and our gRNN have a gating function which dynamically scales the time-constant, which in both cases leads to a marginally stable phase in the limit of a binary gate. However, the dynamical phase diagram presented in Fig.(\ref{fig:phaseDiag_combined}) reveals a novel discontinuous transition to chaos. We conjecture that such a phase transition should also be present in the GRU. Also, \cite{can2020gating} lacked any discussion of the influence of inputs or biases. The present paper includes discussion of the functional significance of the gates in the presence of {\it inputs}. We believe these results, combined with the refined {\it dynamical} phase diagram, can shed some light on the role of analogous gates in the GRU and other gated ML architectures. We review the significance of the gates in more detail below.

{\it Significance of the update gate:}
The update gate modulates the rate of integration. In single neuron models, such a modulation has been shown to make the neuron's responses robust to time-warped inputs \cite{gutig2009time}. Furthermore, normative approaches, requiring time reparameterization invariance
% {\red time-warp invariance [this is weird jargon. we should write ``time reparameterization invariance". Also this sentence is confusing. Like, what is ``normative approaches" doing stuck between commas?]}
in ML RNNs naturally imply the existence of a mechanism that modulates the integration rate  \cite{tallec2018can}. We show that for a wide range of parameters, a more sensitive (or switch-like) update gate leads to marginal stability.
Marginally stable models of biological function have long been of interest 
with regard to their benefits for information processing (c.f. \cite{mora2011biological} and references therein). In the gated RNN, a functional consequence of the marginally stable state is the use of the network as a robust integrator -- such integrator-like function  has been shown to be beneficial for a variety of computational functions such as motor control \cite{seung1998continuous,seung1996brain,seung2000stability}, decision making \cite{machens2005flexible} and auditory processing \cite{eguiluz2000essential}. However, previous  models of these integrators have often required hand-crafted symmetries and fine-tuning \cite{chaudhuri2016computational}. We show that gating allows this function without fine-tuning.  Signatures of integrator-like behaviour are also empirically observed in successfully trained gated ML RNNs on complex tasks \cite{maheswaranathan2019reverse}. We provide a theoretical basis for  how gating produces these.  The update gate facilitates accumulation of slow modes and a pinching of the spectral curve which leads to a suppression of unstable directions and overall slowing of the dynamics over a range of parameters. This is a manifestly self-organised slowing down.  Other mechanisms for slowing down dynamics of have been proposed where the slow timescales of the network dynamics are inherited from other slow internal processes such as synaptic filtering \cite{muscinelli2019single,beiran2019contrasting}; however, such mechanisms differ from the slowing due to gating; they do not seem to display the pinching and clumping,  and they also  do not rely on self-organised behaviour.

{\it Significance of the output gate:}
The output gate dynamically modulates the outputs of individual neurons. Similar shunting mechanisms are widely observed in real neurons and are crucial for performance in ML tasks \cite{greff2016lstm}. We show that the output gate offers fine control over the {\it dimensionality} of the dynamics in phase space, and this ability has been implicated in task performance in ML RNNs \cite{farrell2019recurrent}.  This gate also gives rise to a novel discontinuous chaotic transition where inputs can abruptly push stable systems to strongly chaotic activity; this is in contrast to the typically stabilising role of inputs in additive RNNs. In this transition, there is a decoupling between {\it topological} and {\it dynamical} complexity. The chaotic state across this transition is also characterised by the coexistence of a stable fixed-point with chaotic dynamics -- in finite size systems this manifests as long transients that scale with the system size.  We note that there are other systems displaying either a discontinuous chaotic transition, or the existence of  fixed-points overlapping with chaotic (pseudo-)attractors \cite{PhysRevE.90.062710} or apparent chaotic attractors with finite alignment with particular directions\cite{pereira2018attractor}, however, we are not aware of a transition in large RNNs where, static inputs can induce strong chaos or the topological and dynamical complexity are decoupled across the transition. In this regard the chaotic transition mediated by the output gated seems to be fundamentally different.  More generally, the output gate is likely to have a  significant role in controlling the influence of external inputs on the intrinsic dynamics. 

We also show how the gates complement the memory function of the update gate by providing an important, context and input dependent {\it reset} mechanism. The ability to erase a memory trace flexibly is critical for function \cite{greff2016lstm}.  Gates also provide a mechanism to avoid the accuracy-flexibility trade-off noted for purely additive networks - where the stability of a memory comes at the cost of the ease with which its rewritten \cite{pereira2015tradeoff}. 

We summarize the rich behaviour of the gated RNN via phase diagrams indicating the critical surfaces and regions of marginal stability. From a practical perspective, the phase diagram is useful in light of the observation that it is often easier to train RNNs initialized in the chaotic regime but close to the critical points. This is often referred to as the ``edge of chaos'' hypothesis \cite{bertschinger2004real,legenstein2007makes,boedecker2012information}. 
% This could be a result of the long timescales present in the dynamics at these operating points.
Thus the phase diagrams provide ML practitioners with a map
for principled parameter initialisation -- one of the most critical choices deciding training success.

 %**************** END MAIN SECTION *******************

\vspace{20pt}

\uline{Author contributions:} KK \& DJS conceived the overall project. KK \& TC developed the theory and did all the analyses. KK \& TC wrote the paper. All authors discussed the results and the paper.

\vspace{20pt}

\begin{acknowledgments}
% We are most grateful to William Bialek, Giulio Biroli, Jonathan Cohen, Andrea Crisanti, Rainer Engelken, Moritz Helias, Jonathan Kadmon, Louis Kang, Jimmy Kim, Itamar Landau, Wave Ngampruetikorn, Jeffrey Pennington, Katherine Quinn, Friedrich Schuessler, James Sethna, Julia Steinberg and Merav Stern for fruitful discussions.
KK is supported by a C.V. Starr Fellowship and a CPBF Fellowship  (through NSF PHY-1734030). TC is supported by a grant from the Simons Foundation (891851, TC). DJS was supported by the NSF through the CPBF (PHY-1734030) and by a Simons Foundation fellowship for the MMLS. This work was partially supported by the NIH under award number R01EB026943. KK \& DJS thank the Simons Institute for the Theory of Computing at U.C. Berkeley, where part of the research was conducted. TC gratefully acknowledges the support of the Initiative for the Theoretical Sciences at the Graduate Center, CUNY, where most of this work was completed.
\end{acknowledgments}

\pagebreak
\newpage

% ***************** APPENDIX ***********************

\appendix

%******************** RMT APPENDIX *******************

\section{Details of random matrix theory for spectrum of the Jacobian}
\label{app:RMT_spectral_curve}

In this section, we provide details of the calculation of the bounding curve for the Jacobian spectrum for both fixed-points and time-varying states. Our approach to the problem utilizes the method of Hermitian reduction \cite{feinberg1997non,chalker1998eigenvector} to deal with non-Hermitian random matrices.The analysis here resembles that in \cite{can2020gating}, which also considered Jacobians that were highly structured random matrices arising from discrete-time gated RNNs.

The Jacobian $\mathcal{D}$ is a block-structured matrix constructed from the random coupling matrices $J^{h,z,r}$ and diagonal matrices of the state variables. In the limit of large $N$, we expect the spectrum to be self-averaging -- i.e. the distribution of eigenvalues for a random instance of the network will approach the ensemble-averaged distribution. 
We can thus gain insight about typical dynamical behavior by studying the ensemble (or disorder) averaged spectrum of the Jacobian. Our starting point is the
disorder-averaged spectral density $\mu(\lambda)$ defined as
\begin{equation}
\mu(\lambda) = \frac{1}{3N}\mathbb{E}\left[\sum_{i=1}^{3N}
\delta(\lambda - \lambda_i)
\right]
\end{equation}
where the $\lambda_i$ are the eigenvalues of $\mathcal{D}$ for a given realization of $J^{h,z,r}$ and the expectation is taken over the distribution of real Ginibre random matrices from which $J^{h,z,r}$ are drawn. Using an alternate representation for the Dirac delta function in the complex plane ($\delta(\lambda) = \pi^{-1} \partial_{\bar{\lambda}} \lambda^{-1}$), we can write the average spectral density as 
\begin{align}\label{eq:RMT_density1}
\mu(\lambda)=\frac{1}{\pi} \frac{\partial}{\partial \bar{\lambda}} \mathbb{E}\left[\frac{1}{3N} \operatorname{Tr}\left[\left(\lambda \mathbbm{1}_{3N}-\mathcal{D} \right)^{-1}\right]\right]   
\end{align}
where $\mathbbm{1}_{3N}$ is the $3N$-dimensional identity matrix.  $\mathcal{D}$ is in general non-Hermitian, so the support of the spectrum is not limited to the real line, and the standard procedure of studying the Green's function $G(\lambda,\bar{\lambda}) = (3N)^{-1}
 \operatorname{Tr} \mathbb{E} \left[\left(\lambda \mathbbm{1}_{3N}-\mathcal{D}\right)^{-1}\right]$
 by analytic continuation is not applicable since it is non-holomorphic on the support. Instead, we use the method of Hermitization \cite{feinberg1997non,chalker1998eigenvector} to analyse the resolvent for an expanded $6N \times 6N$ Hermitian matrix $H$
\begin{align}
\mathcal{G}(\eta,\lambda,\bar{\lambda}) = & \: \mathbb{E}
\left[ (\eta \mathbbm{1}_{6N} - H)^{-1}
\right]  \\ 
H = & \: \left(\begin{array}{cc}
0 & \lambda-\mathcal{D} \\
\bar{\lambda}-\mathcal{D}^{T} & 0 
\end{array}\right) 
\end{align}
and the Green's function for the original problem is obtained by considering the lower-left block of $\mathcal{G}$ 
\begin{equation} \label{eq:RMT_Greens_function}
G(\lambda,\bar{\lambda}) 
= \lim_{\eta \rightarrow i0^+}\, \frac{1}{3N} \operatorname{Tr} \mathcal{G}_{21}(\eta,\lambda,\bar{\lambda})
\end{equation}

To make this problem tractable, we invoke an ansatz called the local chaos hypothesis \cite{cessac1995increase,geman1982chaos}, which posits that for large random networks in steady-state, the state variables are statistically independent of the random coupling matrices $J^{z,h,r}$ (also see \footnote{Strictly speaking, the state variables evolve according to dynamics governed by (and thus dependent on) the $J$'s. However, the local chaos hypothesis states that large random networks approach a steady-state where the state variables are independent of $J$s and are distributed according to their steady-state distribution.}).
This implies that the Jacobian (eq. \ref{eq:Jacobian_1}) only has an explicit linear dependence on $J^{h,z,r}$, and the state variables are governed by their steady-state distribution from the disorder-averaged DMFT (Appendix \ref{app:DMFT_derivation}). 
These assumptions make the random matrix problem tractable, and we can evaluate the Green's function by using the self-consistent Born approximation, which is exact as $N\rightarrow \infty$. We detail this procedure below.

The Jacobian itself can be decomposed into structured $(A,L,R)$ and random parts $(\mathcal{J})$:

\begin{align} \label{eq:JacRMT_linearisation}
 \mathcal{D} = &
\underbrace{
\begin{pmatrix}
-[ \sigma_{z}]& \mathbb{D}  & 0 \\
0 & -\tau_z^{-1} \mathbbm{1} & 0 \\
0 & 0 & -\tau_r^{-1} \mathbbm{1} 
\end{pmatrix}}_{A}  + \underbrace{
\begin{pmatrix}
[ \sigma_{z}] & 0  & 0 \\
0 & \tau_z^{-1} \mathbbm{1} & 0 \\
0 & 0 & \tau_r^{-1} \mathbbm{1} 
\end{pmatrix}}_{L} 
\nonumber \\
& \times \:
\underbrace{
\begin{pmatrix}
 J^h  & 0  & 0 \\
0 & J^z & 0 \\
0 & 0 & J^r
\end{pmatrix}}_{\mathcal{J}}
\underbrace{
\begin{pmatrix}
 \left[\phi_{}^{\prime} \sigma_{r}\right] & 0 &  \left[\phi_{} \sigma_{r}^{\prime} \right] \\
 \left[\phi^{\prime}\right] & 0 & 0 \\
[\phi^{\prime}] & 0 & 0 
\end{pmatrix}}_{R}   .
\end{align}

At this point, we must make a crucial assumption: the structured matrices $A$, $L$, and $R$ are {\it independent} of the random matrices appearing $\mathcal{J}$.   This implies that the dynamics is {\it self-averaging}, and that the state variables reach a steady-state distribution determined by the DMFT, and insensitive to the particular quenched disorder $\mathcal{J}$.This self-averaging assumption leads to theoretical predictions which are in very good agreement with simulations of large networks, as presented in Fig.(\ref{fig:Jac_spectrum_RMT}).

This  independence assumption renders $\mathcal{D}$ a linear function of the random matrix $\mathcal{J}$, whose entries are Gaussian random variables. The next steps would be to develop an asymptotic series in the random components of $H$, compute the resulting moments, and perform a resummation of the series. This is conveniently accomplished by the self-consistent Born approximation (SCBA). The output of the SCBA is a self-consistently determined self-energy functional $\Sigma[\mathcal{G}]$ which succinctly encapsulates the resummation of moments. With this, the Dyson equation for $\mathcal{G}$ is given by 
\begin{align} \label{eq:RMT_dyson1}
    \mathcal{G}^{-1} =   \mathcal{G}^{-1}_0 - \Sigma[\mathcal{G}] ,
\end{align}
where the matrices on the right are defined in terms of $3N \times 3N$ blocks:
\begin{align} \label{eq:RMT_dyson2}
\mathcal{G}^{-1}_0  = & \begin{pmatrix}
\eta \mathbbm{1} & \lambda - A   \\
\bar{\lambda} - A^T & \eta \mathbbm{1}
\end{pmatrix}, \\
     \Sigma[\mathcal{G}] = & \begin{pmatrix}
LQ[R\mathcal{G}_{22}R^T]L & 0   \\
0 & R^TQ[L^T\mathcal{G}_{11}L]R
\end{pmatrix} , \label{eq:RMT_dyson3}
\end{align}

and $Q$ is a superoperator which acts on its argument as follows:
\begin{align}
Q[M]=\left(\begin{array}{ccc}
\frac{1}{N} \operatorname{Tr} M_{11} & 0 & 0 \\
0 & \frac{1}{N} \operatorname{Tr} M_{22} & 0 \\
0 & 0 & \frac{1}{N} \operatorname{Tr} M_{33}
\end{array}\right) .   
\end{align}

Here, we have expressed the self-energy using the $3N\times 3N$ subblocks of the Green's function $\mathcal{G}$
\begin{align}
  \mathcal{G} =   \begin{pmatrix}
\mathcal{G}_{11} & \mathcal{G}_{12} \\
\mathcal{G}_{21} & \mathcal{G}_{22} 
\end{pmatrix}.
\end{align}

At this point, we have presented all of the necessary ingredients for computing the Green's function, and thus determining the spectral properties of the Jacobian. These are the Dyson equation (\ref{eq:RMT_dyson1}), along with the free Green's function (\ref{eq:RMT_dyson2}) and the self-energy (\ref{eq:RMT_dyson3}). Most of what is left is complicated linear algebra. However, in the interest of completeness, we will proceed to unpack these equations and give a detailed derivation of the main equation of interest, the bounding curve of the spectral density.

To proceed further, it is useful to define the following transformed Green's functions, which can be written in terms of $N\times N$ subblocks:
\begin{align}
\tilde{\mathcal{G}}_{11} \equiv & \: L^T\mathcal{G}_{11}L =
\begin{pmatrix}
\tilde{G}_{11} & \tilde{G}_{12}  & \tilde{G}_{13} \\
\tilde{G}_{21} & \tilde{G}_{22} & \tilde{G}_{23} \\
\tilde{G}_{31} & \tilde{G}_{32} & \tilde{G}_{33}
\end{pmatrix} ,\label{eq:app_RMT_self-consistent_1}\\
\tilde{\mathcal{G}}_{22} \equiv & \: R\mathcal{G}_{22}R^T =
\begin{pmatrix}
\tilde{G}_{44} & \tilde{G}_{45}  & \tilde{G}_{46} \\
\tilde{G}_{54} & \tilde{G}_{55} & \tilde{G}_{56} \\
\tilde{G}_{64} & \tilde{G}_{65} & \tilde{G}_{66}
\end{pmatrix}.\label{eq:app_RMT_self-consistent_2}
\end{align}
Denote also the mean trace of these subblocks as
\begin{align}
    \tilde{g}_{ij} = \frac{1}{N}\textrm{Tr}[ \tilde{G}_{ij}]. \label{eq:mean_trace}
\end{align}

Then the the self-energy matrix in eq. \ref{eq:RMT_dyson3} is block diagonal, i.e.  $\Sigma[\mathcal{G}] = {\rm bdiag}\left( \Sigma_{11}, \Sigma_{22}\right)$, with
\begin{align}
\Sigma_{11}  = & \:
\begin{pmatrix}
\left[\sigma_z^{2}\right] \tilde{g}_{44} & 0  & 0 \\
0 & \tau_z^{-2} \tilde{g}_{55} & 0 \\
0 & 0 & \tau_r^{-2} \tilde{g}_{66} 
\end{pmatrix}, \\
\Sigma_{22}  =  & \:
\begin{pmatrix}
\left[\phi_{}^{\prime} \sigma_{r}\right]^{2} \tilde{g}_{11} 
+ \left[\phi_{}^{\prime}\right]^2 (\tilde{g}_{22} 
+  \tilde{g}_{33}) & 0  & 
\left[\phi_{}^{\prime} \sigma_{r}\right] \left[ \phi_{} \sigma_{r}^{\prime}\right] \tilde{g}_{11}
\\
0 & 0 & 0 \\
[\phi_{}^{\prime} \sigma_{r}] [\phi_{} \sigma_{r}^{\prime}] \tilde{g}_{11} & 0 & [\phi_{} \sigma_{r}^{\prime}]^2 \tilde{g}_{11} 
\end{pmatrix}.
\end{align}

%These equations allow us to define self-consistent relations between the various $\tilde{g}_{ii}$. 
With the self-energy in this form, we are able to solve the Dyson equation for the full Green's function $\mathcal{G}$ by direct matrix inversion:
\begin{align} \label{eq:app_RMT_G_1}
\mathcal{G} = \begin{pmatrix}
\eta -\Sigma_{11} & \lambda - A   \\
\bar{\lambda} - A^T & \eta -\Sigma_{22}
\end{pmatrix}^{-1},
\end{align}
which can be carried out easily by symbolic manipulation software. The R.H.S of eq. \ref{eq:app_RMT_G_1} is a function of $\tilde{g}_{ii}$, whereas the L.H.S. is a function of the Green's function before the transformations (\ref{eq:app_RMT_self-consistent_1}) and (\ref{eq:app_RMT_self-consistent_2}). Thus, to get a set of equations we can solve, we apply these same transformations to both sides of eq.(\ref{eq:app_RMT_G_1}) after solving the Dyson equation. The final step is to take the limit $\eta \to 0$, recovering the problem we originally wished to solve. 

The result of these manipulations will be a set of six equations for the mean traces of the transformed Green's function defined in (\ref{eq:mean_trace}). In order to write these down, we introduce some additional notation. The self-consistent equations will be of the form
\begin{align}\label{eq:RMT_self-consistent}
\tilde{g}_{ii} 
= \left\langle \frac{\Gamma_i}{\mathbf{\Gamma}}\right\rangle , 
\end{align}
where we denote $\langle M \rangle \equiv N^{-1} {\rm Tr} M$ for shorthand, and $i$ runs from $1$ to $6$.
% transparent, 
%and then we can read off the diagonal blocks 
%to impose the self consistent relations using
%\begin{align}
%\tilde{\mathcal{G}}_{11} =  L^T\mathcal{G}_{11}L & \quad 
%\tilde{\mathcal{G}}_{22} =  R\mathcal{G}_{22}R^T , \label{eq:app_RMT_self-consistent_1} \\ 
 %\tilde{g}_{ii} = & \: \frac{1}{N}\textrm{Tr}[\tilde{G}_{ii}] \quad i \in {1,\ldots,6}  .  \label{eq:app_RMT_self-consistent_2}
%\end{align}
 %write the blocks of the Greens functions, $\tilde{G}_{ii}$, as the ratio of a numerator and a denominator that is common to all the blocks. 
Denote the state-variable dependent diagonal matrices as
\begin{align}
  p = [ \phi'], \quad q = [ \phi \sigma_{r}'], \quad   r  = [ \phi' \sigma_{r}], 
\end{align}
and, because they appear frequently in the resulting equations, define
\begin{align}
    X &= \tilde{g}_{11} |\lambda \tau_{r} + 1|^{2} r^{2} + (\tilde{g}_{22} + \tilde{g}_{33}) p^{2} Z,\\
Y &= \mathbb{D}^{2} \tilde{g}_{55} +  |\lambda \tau_{z} + 1|^{2} \left[\sigma_{z}^{2}\right]\tilde{g}_{44},\\
Z& = |\lambda \tau_{r} + 1|^{2} - \tilde{g}_{11} \tilde{g}_{66} q^{2}.
\end{align}

The denominator in (\ref{eq:RMT_self-consistent})is then given by
\begin{align}
\mathbf{\Gamma}&=  |\lambda \tau_{z} + 1|^{2} |\lambda + \sigma_{z}|^{2} Z - X Y,
\end{align}

and the numerators $\Gamma_{i}$ are given by:

\begin{align}
\Gamma_{1}& = \sigma_{z}^{2} |\lambda \tau_{z} + 1|^{2} X, \\
\Gamma_{2}  &= \mathbb{D}^{2} X,\\
\Gamma_{3}& = \tilde{g}_{11} |\lambda \tau_{z} + 1|^{2} |\lambda + \sigma_{z}|^{2} q^{2}\nonumber\\
&- \tilde{g}_{11}(\tilde{g}_{22} + \tilde{g}_{33}) p^{2}q^{2}  Y, \\
\Gamma_{4}& = \tilde{g}_{66}|\lambda \tau_{z} + 1|^{2} |\lambda + \sigma_{z}|^{2} q^{2} \nonumber\\
&+ \left( |\lambda \tau_{r} + 1|^{2}  r^{2} - \tilde{g}_{66}(\tilde{g}_{22} + \tilde{g}_{66}) p^{2} q^{2}\right) Y,\\
\Gamma_{5}& = \Gamma_{6} =  p^{2} YZ.
\end{align}

The numerators and denominator are all  diagonal matrices with real entries, which is why we use the simple notation of a ratio when referring to matrix inversion.  

%The self-consistent equations implied by eqns. \ref{eq:app_RMT_self-consistent_1},\ref{eq:app_RMT_self-consistent_2}  for the six scalar variables $\tilde{g}_{ii}$ are then given by
%\begin{align}\label{eq:RMT_self-consistent}
%\tilde{g}_{ii} = \left\langle \frac{\Gamma_i}{\mathbf{\Gamma}}\right\rangle , 
%\end{align}

%where we denote $\langle M \rangle \equiv N^{-1} {\rm Tr} M$ for shorthand, and $i$ runs from $1$ to $6$.
Solving these equations gives us the $\tilde{g}_{ii}$ as implicit functions of $\lambda$. They are in general complicated, and resist exact solution.
However, the situation simplifies considerably when we are looking for the spectral curve. In this case, we are looking for all $\lambda \in \mathbb{C}$ that satisfy the self-consistent equations with $\tilde{g}_{ii} \to 0$. 

We must take this limit carefully, since the ratio of these functions can remain constant. For this reason, it is necessary to define
\begin{align}
    x_{2} = \tilde{g}_{22}/\tilde{g}_{11}, \quad x_{3} = \tilde{g}_{33}/\tilde{g}_{11}.
\end{align}
We may do the same for $\tilde{g}_{44}$, $\tilde{g}_{55}$ and $\tilde{g}_{66}$, but it turns out that $x_{2}$ and $x_{3}$ are sufficient to compute the spectral curve.  Next, divide by $\tilde{g}_{11}$ and send all $\tilde{g}_{ii} \to 0$ keeping the ratios fixed. Applying this to the equation for $\tilde{g}_{11}$ results in
% Therefore, to get the equation for the spectral curve we look at the numerator terms in the limit of small $\tilde{g}_{ii}$ and eliminating the ratios mentioned above to get an implicit equation in the state variables and $\lambda$. 
%We divide the equations for $\tilde{g}_{ii}$ by 
%In this limit, we may take the trace and keep nonvanishing terms. In particular, the expression for $\tilde{g}_{11}$ gives
\begin{align}
  & 1 = \lim_{\tilde{g}_{ii} \to 0} \frac{1}{\tilde{g}_{11}} \left\langle  \frac{\Gamma_{1}}{\mathbf{\Gamma}} \right\rangle = %\nonumber  
 %  &= \left\langle \frac{\sigma_{z}^{2} |\lambda \tau_{z} + 1|^{2} \left( |\lambda \tau_{r} + 1|^{2} r^{2} + (x_{2} + x_{3}) p^{2} Z\right)}{\mathbf{\Gamma}}\right\rangle.
 \gamma_{1} + \gamma_{2}(x_{2} + x_{3}). \label{eq:lin1}
\end{align}

Similarly for $\tilde{g}_{22}$ and $\tilde{g}_{33}$, we get
\begin{align}
    x_{2} & = \gamma_{3} + \gamma_{4}(x_{2} + x_{3}),\label{eq:lin2}\\%\left\langle \frac{\mathbb{D}^{2}\left( |\lambda \tau_{r} + 1|^{2} r^{2} + (x_{2} + x_{3}) p^{2} Z\right)}{\mathbf{\Gamma}}\right\rangle,\\
  x_{3} & = \gamma_{5}, \label{eq:lin3}%\left\langle \frac{|\lambda \tau_{z} + 1|^{2} |\lambda + \sigma_{z}|^{2} q^{2}  }{\mathbf{\Gamma}}\right\rangle .
\end{align}
where the coefficients $\gamma_{i}$, which are functions of $\lambda$, are given by 
\begin{align}
\gamma_{1} &= \left\langle \frac{\sigma_{z}^{2} r^{2}}{|\lambda + \sigma_{z}|^{2}} \right\rangle, \quad \gamma_{2}  = \left\langle \frac{p^{2} \sigma_{z}^{2} }{ |\lambda  + \sigma_{z}|^{2}} \right\rangle, \quad \gamma_{5} = \frac{\left\langle q^{2}\right\rangle}{|\lambda \tau_{r} + 1|^{2}},\nonumber\\
\gamma_{3} & = \left\langle \frac{\mathbb{D}^{2} r^{2} }{|\lambda \tau_{z} + 1|^{2} |\lambda + \sigma_{z}|^{2}}\right\rangle, \quad \gamma_{4} = \left\langle \frac{\mathbb{D}^{2} p^{2} }{|\lambda \tau_{z} + 1|^{2} |\lambda + \sigma_{z}|^{2}}\right\rangle,\nonumber
\end{align}

%and the denominator simplifies to
%\begin{align}
% \mathbf{\Gamma} = 
% \left|1+\tau_{r}  \lambda \right|^{2}
% \left|1+\tau_{z}  \lambda \right|^{2}
%\left|\sigma_{z} +  \lambda \right|^{2} .
%\end{align}
%Let us define now
The linear system of equations (\ref{eq:lin1}, \ref{eq:lin2}, \ref{eq:lin3}) is consistent iff %To recapitulate, we are now left with the linear system of equations %Then we must solve the system of equations
%\begin{align}
 %   1 &= \gamma_{1} + (x_{2} + x_{3})\gamma_{2},\\
 %   x_{2} &= \gamma_{3} + \gamma_{4}(x_{2} + x_{3}),\\
 %   x_{3} &= \gamma_{5},
%\end{align}
%only has a solution when
\begin{align}
    (1 - \gamma_{1})(1 - \gamma_{4}) = \gamma_{2} \left( \gamma_{3} + \gamma_{5}\right) \label{eq:gamma-eq}.
\end{align}
In other words, $\gamma_{i}$ must satisfy (\ref{eq:gamma-eq}) when $\tilde{g}_{ii} \to 0$. This expression depends on $\lambda$, and implicitly defines a curve in $\mathbbm{C}$, which is the boundary of the support of the spectral density. %. This is implicitly a . The values $\lambda \in \mathbbm{C}$ for which it is satisfied describes the boundary of the support of the eigenvalue spectrum. 

Plugging in the explicit expression for $\gamma_{i}$, we get the implicit equation for the spectral curve as all $\lambda \in \mathbbm{C}$ that satisfy
%\begin{widetext}
\begin{align}\label{eq:RMT_spectral_curve_general}
   & \left\{1 - \left\langle \frac{r^{2} \sigma_{z}^{2}}{|\lambda + \sigma_{z}|^{2}} \right\rangle \right\} \left\{ 1 - \left\langle \frac{\mathbb{D}^{2} p^{2} }{|\lambda \tau_{z} + 1|^{2} |\lambda + \sigma_{z}|^{2}} \right\rangle \right\} \nonumber\\
    &  = \left\langle \frac{\sigma_{z}^{2} p^{2} }{ |\lambda + \sigma_{z}|^{2}} \right\rangle \left\{ \left\langle \frac{\mathbb{D}^{2} r^{2} }{|\lambda \tau_{z} + 1|^{2} |\lambda + \sigma_{z}|^{2}} \right\rangle  + \frac{\langle q^{2} \rangle}{|\lambda \tau_{r} + 1|^{2}}\right\} .
\end{align}

For large systems we can replace the empirical traces of the state variable by their averages given by the DMFT variances. Then, the equation for the curve for a general steady-state is given by

\begin{align} \label{eq:RMT_MFT_spectral_curve}
\left( \langle \sigma_{r}^{2}\rangle +  \frac{\langle \phi^{2} \sigma_{r}'^{2}\rangle}{|1 + \tau_{r} \lambda|^{2}} \right)  \left\langle \frac{ \phi'^{2} \sigma_{z}^{2}}{|\lambda + \sigma_{z}|^{2}}\right\rangle  +  \frac{1}{|1 + \tau_{z}\lambda|^{2}} \left\langle \frac{\mathbb{D}^{2} \phi'^{2}}{|\lambda + \sigma_{z}|^{2}}\right\rangle = 1.
 \end{align}

\begin{comment}
\begin{align}
& \Bigg\{ 
\frac{1}{\left|1+\tau_{z}^{2}  \lambda \right|^{2}} 
 \textrm{Tr}\left[
 \frac{\mathbb{D}_{dyn}^{2}\mathbb{D}_{\phi_{}^{\prime} G_{r}}^2}{\left|\mathbb{D}_{G_z} +  \lambda \right|^{2}}
 \right]
 +  
 \frac{1}{\left|1+\tau_{r}^{2}  \lambda \right|^{2}} 
 \textrm{Tr}\left[ 
 \mathbb{D}_{\phi_{} G_{r}^{\prime}}^2\right]
\Bigg\} \times \nonumber \\
&
\textrm{Tr}\left[ \frac{\mathbb{D}_{G_z}^{2}
\mathbb{D}_{\phi_{}^{\prime} }^2}{\left|\mathbb{D}_{G_z} +  \lambda \right|^{2}}
\right] =  \nonumber \\
& \Bigg\{ 1- 
\textrm{Tr}\left[ \frac{\mathbb{D}_{G_z}^{2}
\mathbb{D}_{\phi_{}^{\prime} G_{r}}^2}{\left|\mathbb{D}_{G_z} +  \lambda \right|^{2}}
\right]
\Bigg\}  \Bigg\{ 1 - \frac{1}{\left|1+\tau_{z}^{2}  \lambda \right|^{2}}
\textrm{Tr}\left[ 
 \frac{\mathbb{D}_{dyn}^{2}\mathbb{D}_{\phi_{}^{\prime}}^{2} }{\left|\mathbb{D}_{G_z} +  \lambda \right|^{2}}
\right]
\Bigg\}
\end{align}
%\end{widetext}

\end{comment}

% ---------------- FIGURE SPEC CURVE FULL vs MFT -------------

\begin{figure}
\begin{centering}
\includegraphics[scale=0.55]{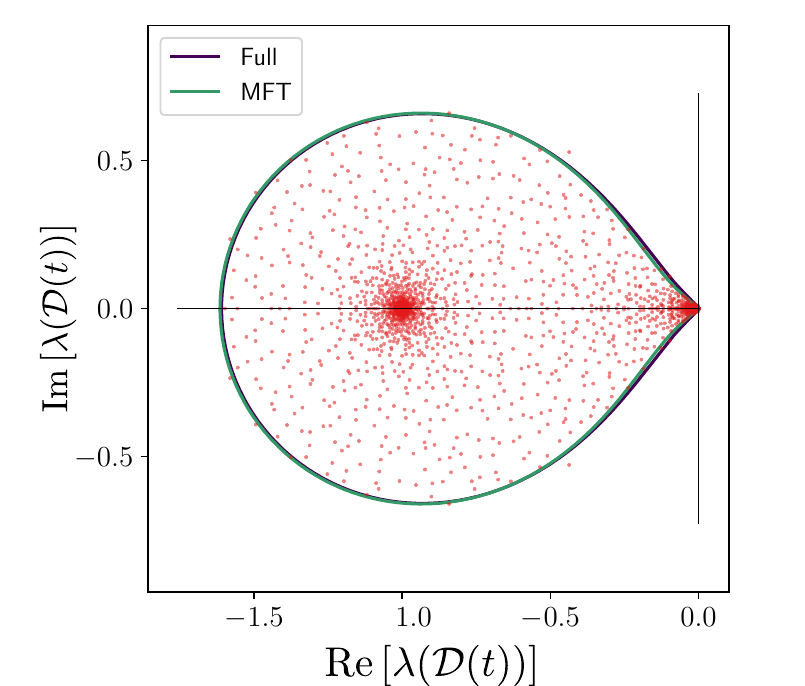}
\par\end{centering}
\caption{\label{fig:Jac_spec_full_v_MFT} {\it Jacobian spectrum at a time-varying state:}  Red dots are the Jacobian eigenvalues for the full network in a (time-varying) steady-state, and the spectral curve of the Jacobian is calculated using moments from (i) the full state vectors (blue curve) or using the variances from the fixed-point MFT (green). Surprisingly, the agreement is reasonably good.  For network simulations $N=1000$, $g_h=2.5, \alpha_r=1, \alpha_z=15$  and all biases are zero.
}
\end{figure}

For fixed-points, we have $\mathbb{D}=0$, which makes $\gamma_{3} = \gamma_{4} = 0$. The equation for the spectral curve simplifies to that which is quoted in the main text (eq. \ref{eq:RMT_FP_spectral_curve}):
\begin{align}
   & 1 =   \left\langle \frac{r^{2} \sigma_{z}^{2}}{|\lambda + \sigma_{z}|^{2}} \right\rangle  + \frac{\langle q^{2} \rangle}{|\lambda \tau_{r} + 1|^{2}}   \left\langle \frac{\sigma_{z}^{2} p^{2} }{ |\lambda + \sigma_{z}|^{2}} \right\rangle  .
\end{align}

\subsection{Jacobian spectrum for the case $\alpha_r=0$}
 
In the case when $\alpha_r=0$, it is possible to express the Green's function (eq. \ref{eq:RMT_Greens_function}) in a simpler form. Recall that,
\begin{align}
    G(\lambda, \bar{\lambda})=\lim _{\eta \rightarrow i 0^{+}} \frac{1}{3N} \operatorname{tr} \mathcal{G}_{21}(\eta, \lambda, \bar{\lambda}).
\end{align}
Let $\tilde{Y} = \mathbb{D}^{2} + \sigma_{r}^{2} \sigma_{z}^{2} |\lambda \tau_{z} + 1|^{2}$. Then the Green's 
function is given by  

\begin{align}\label{eq:app-GreensFcn-ar0-1}
    G(\lambda, \bar{\lambda}) =& \frac{1}{3} \left\langle \frac{| \lambda \tau_{z} +1|^{2} (\bar{\lambda} + \sigma_{z})}{|\lambda \tau_{z} + 1|^{2} |\lambda + \sigma_{z}|^{2} - \xi(\lambda, \bar{\lambda}) p^{2} \tilde{Y} } \right\rangle,\\
     + &\frac{1}{3} \left\langle \frac{ (\bar{\lambda}  + \tau_{z}^{-1}) \left( |\lambda + \sigma_{z}|^{2} - \xi(\lambda, \bar{\lambda})  p^{2} \sigma_{z}^{2} \right)}{|\lambda  + \tau_{z}^{-1}|^{2} |\lambda + \sigma_{z}|^{2} - \xi(\lambda, \bar{\lambda}) p^{2} \tilde{Y}} \right\rangle, \\
     &+ \frac{1}{3} \frac{1}{\lambda  + \tau_{r}^{-1}} ,
\end{align}

where $\xi(\lambda, \bar{\lambda})$ is defined implicitly to satisfy the equation
\begin{align}
    1 = \left\langle \frac{p^{2} \tilde{Y}}{|\lambda \tau_{z} + 1|^{2}|\lambda + \sigma_{z}|^{2} - \xi(\lambda, \bar{\lambda}) p^{2} \tilde{Y} } \right\rangle.
\end{align}

The function $\xi(\lambda, \bar{\lambda})$ acts as a sort of order parameter for the spectral density, indicating the transition on the complex plane between zero and finite density $\mu$. Outside the spectral support, $\lambda \in \Sigma^{c}$, this order parameter vanishes $\xi = 0$ and the Green's function is holomorphic 

\begin{align}\label{eq:app-GreensFcn-ar0-2}
    G(\lambda, \bar{\lambda}) = 
    \frac{1}{3}\left(
   \left\langle 
    \frac{1}{ \lambda + \sigma_{z}}
  \right\rangle 
    + \frac{1}{\lambda + \tau_{z}^{-1} } 
    + \frac{1}{\lambda + \tau_{r}^{-1}  } 
    \right) ,
\end{align}
which of course indicates that the density is zero since $
    \mu(\lambda) = \partial_{\bar{\lambda}} G(\lambda, \bar{\lambda})$. Inside the support $\lambda \in \Sigma$, the order parameter $\xi \ne 0$, and the Green's function consequently picks up non-analytic contributions, proportional to $\bar{\lambda}$. Since the Green's function is continuous on the complex plane, it must be continuous across the boundary of the spectral support. This must then occur precisely when the holomorphic solution meets the non-analytic solution, at $\xi = 0$. This is the condition used to find the boundary curve above.

% ---------- Details of clumping and pinching (with ar =0) 
\section{Spectral {\it clumping} and {\it pinching}
in the limit $\alpha_z \rightarrow \infty$} \label{app:clumping_and_pinching}

In this section we provide details on the accumulation of eigenvalues near zero and the {\it pinching} of the leading spectral curve (for certain values of $g_h$) as the update gate becomes switch-like ($\alpha_z \rightarrow \infty$). To focus on the 
key aspects of these phenomena, we  consider the case when the reset gate is off and there are no biases( $\alpha_r = 0, \beta_{r,h,z}=0$). Moreover, we consider 
a piece-wise linear approximation -- sometimes called ``hard'' $\tanh$ -- to the $\tanh$ function given by
\begin{align}
    \phi_{lin}(x) = \begin{dcases}
                1 & x > 1/g_h ,\\
                g_h x &  |x| \leq 1/g_h ,\\
                -1 & x < - 1/g_h.
    \end{dcases}
\end{align}
This approximation does not qualitatively change the nature of the clumping.

In the limit, $\alpha_z \rightarrow \infty$ the update gate $\sigma_z$ becomes binary with a distribution given by 
\begin{align}
    P(\sigma_z = x) = f_z \delta(x-1) + (1-f_z)\delta(x),
\end{align}

where  $f_z = \langle \sigma_{z} \rangle $ is the fraction of update gates that are open (i.e. equal to one). 
Using this, along with the assumption that $\mathbb{D} \approx 0$ -- which is valid in this regime -- we can simplify the expression for the Green's
function (eq. \ref{eq:app-GreensFcn-ar0-1} - \ref{eq:app-GreensFcn-ar0-2}) to yield:

\begin{align}\label{eq:app-GreensFcn-ar0-3}
    G(\lambda, \bar{\lambda}) = &
    \frac{1-f_{z}}{\lambda}  +
    f_{z}\left(1-f_{h}\right)\frac{1}{\lambda+1} 
     + \frac{1}{\lambda+ \tau_{z}^{-1}} 
     \nonumber \\
     & \: +
    \frac{(1+\bar{\lambda})}{g_{h}^{2} \sigma\left(\beta_{r}\right)^{2}} 
    \mathbb{I}_{\{ |\lambda| < g_{h}^{2} \sigma\left(\beta_{r}\right)^{2} \} },
\end{align}

where $f_h$ is the fraction of hard $\tanh$ activations that are not saturated.  In the limit of small $\tau_z$ and $\beta_r=0$, we get the expression for the density given in the text:
\begin{align}
    \mu(\lambda) 
    = & 
    (1-f_z) \delta(\lambda)+ f_z(1-f_h) \delta(\lambda+1) + \frac{4}{\pi g_h^2} 
     \mathbb{I}_{\{ |\lambda| \leq g_h^2/4 \} }.
\end{align}
Thus we see an extensive number of eigenvalues at zero.

Now, let us study the regime where $\alpha_z$ is large but not infinite. We would like to get the scaling behaviour of the leading edge of the spectrum and the density of eigenvalues contained in a radius $\delta$ around the origin. We make an ansatz for the spectral edge close to zero
$\lambda \sim e^{-c\alpha_z\sqrt{\Delta_h}}$, where $c$ is a positive constant. With this ansatz, the equation for the spectral curve reads
\begin{align}
  \int \mathcal{D}z \frac{\sigma_{z}(\sqrt{\Delta_{z}} \cdot z)^{2}}{\left| \lambda_0 e^{-c\alpha_z\sqrt{\Delta_h}} +\sigma_{z}(\sqrt{\Delta_{z}^{2}})\right|^{2}}=\frac{\sigma_{r}\left(\beta_{r}\right)^{-2}}{\left\langle\phi^{\prime}(\sqrt{\Delta_{h}} \cdot h)^{2}\right\rangle}.
\end{align}

In the limit of large $\alpha_z$ and $\beta_r=0$ this implies
\begin{align}
    \operatorname{erfc}\left( \frac{c}{\sqrt{2}}
    \right) \approx 
    \frac{4}{\left\langle\phi^{\prime}(\sqrt{\Delta_{h}} \cdot h)^{2}\right\rangle}.
\end{align}
If this has a positive solution for $c$, then the scaling of the spectral edge as $\lambda \sim e^{-c\alpha_z\sqrt{\Delta_h}}$ holds. Moreover, whenever there is a positive solution for $c$ we also expect pinching of the spectral curve and in the limit $\alpha_z \rightarrow \infty$ we will have marginal stability. 

Under the same approximation, we can approximate the
eigenvalue density in a radius $\delta$ around zero as
\begin{align}
    P\left(\left|\lambda\left(\mathcal{D}\right)\right|<\delta\right) = & 
    \frac{1}{2 \pi i} \oint_{\mathcal{C}} d z \:  G(z),
\end{align}
where we choose the contour along
$ z = e^{-c\alpha_z\sqrt{\Delta_h} + i \theta}$  for $\theta \in [0, 2\pi)$ 
    and  $\delta = e^{-c\alpha_z\sqrt{\Delta_h}}$. 
In the limit of large $\alpha_z$ (thus $\delta \ll 1$) we get the scaling form described in the main text :
\begin{align}
    P\left(\left|\lambda\left(\mathcal{D}\right)\right|<\delta\right) \approx \frac{1}{2} \operatorname{erfc}\left(-\frac{\log (\delta)}{\alpha_{z} \sqrt{2 \Delta_h}}\right).
\end{align}

%******************** DMFT APPENDIX *******************

\section{Details of the Dynamical Mean-Field Theory}
\label{app:DMFT_derivation}

The DMFT is a powerful analytical framework used to study the dynamics of disordered systems, and it traces its origins to the 
study of dynamical aspects of spin glasses \cite{sompolinsky1982relaxational,sompolinsky1981dynamic} and has been  later applied to the study of random neural networks \cite{sompolinsky1988chaos,chow2015path,schuecker2018optimal,kadmon2015transition}. In our case, the DMFT reduces the description of the full $3N$-dimensional (deterministic) ODEs describing $ (\mathbf{h},\mathbf{z},\mathbf{r})$ to a set of $3$ coupled stochastic differential equations for scalar variables $(h,z,r)$.

Here, we provide a detailed, self-contained description of the dynamical mean-field theory for the
gated RNN using the Martin-Siggia-Rose-DeDomnicis-Jansen formalism.
The starting point is a generating functional -- akin to the generating function of a random variable -- which takes an expectation over the paths generated by the dynamics. The generating functional is defined as 
\begin{align} \label{eq:MSRDJ_1}
    Z_{\mathcal{J}}[\mathbf{\hat{b}}, \mathbf{b}]=\mathbb{E}\left[\exp \left(i \sum_{j=1}^N \int \mathbf{\hat{b}}_{j}(t)^T \mathbf{x}_{j}(t) d t\right)\right]
\end{align}
where $\mathbf{x}_{j}(t) \equiv (h_j(t),z_j(t),r_j(t))$ is the trajectory and 
$\mathbf{\hat{b}}_j(t) = (\hat{b}_j^h,\hat{b}_j^z,\hat{b}_j^r) $ is the argument of the generating functional. We have also included external fields $\mathbf{b}_j = (b^h_j,b_j^z,b_j^r)$ which are used to calculate the response functions.
The measure in the expectation is a path integral over the dynamics. The generating functional is then used to calculate correlation and response functions using the appropriate (variational) derivatives. For instance, the two-point function for the $h$ field is given by:
\begin{align}\label{eq:MSRDJ_2}
\left\langle 
h_{i}(t) h_{i}\left(t^{\prime}\right)
\right\rangle = 
\left.\frac{\delta^{2}}
{\delta \hat{b}^h_{i}\left(t^{\prime}\right) \delta \hat{b}^h_{i}(t)} 
Z_{\mathcal{J}}[\mathbf{\hat{b}}, \mathbf{b}]
\right|_{\mathbf{b}=0}    
\end{align}

Up until this point, things are quite general, and do not rely on  the specific form of the dynamics.  However, for large random networks, we expect certain quantities such as the population averaged correlation function 
$C_h \equiv N^{-1}\sum_i \langle h_{i}(t)  h_{i}(t^{\prime}) \rangle$ to be self-averaging, and thus not vary much across realizations. Thus, we can study the disorder-averaged (over $\mathcal{J}$),  the generating functional $\bar{Z} = \langle Z_{\mathcal{J}} \rangle_{\mathcal{J}}$, and approximate $\bar{Z}$ with it's value evaluated at the saddle-point of the action. This approximation will give us the single-site DMFT picture of dynamics described in eqs.\ref{eq:DMFT_1}-\ref{eq:DMFT_3}.

To see how this all works, we start with the equations of motion (in vector form)
\begin{align}
\tau_{z}\dot{z}  = & -z+ J^{z}\phi_z(h),\\
\tau_{r}\dot{r}  = & -r+ J^{r}\phi_r(h),\\
\dot{h}  = & \sigma_z(z) \odot \left( -h  + 
			 \left[ J^h \left(  \sigma_r(r) \odot \phi_h(h) \right) \right] \right),
\end{align}
where $\odot$ stands for element-wise multiplication. 

To write down the MSRDJ generating functional, let us discretise the dynamics (in the It\^{o} convention).
Note that in this convention the Jacobian is unity.

\begin{align} \label{eq:eom_11}
h_{i}\left(t+1\right)-h_{i}\left(t\right)  = & \: \sigma_{z,i}(t) \Big\{-h_{i}\left(t\right)  
 + \nonumber \\
 &  \sum_{j}J_{ij}^{h}  \sigma_{r,j}(t)
   \phi_t \left(t\right)   + b_{i}^{h}\left(t\right)\Big\}\delta t ,\nonumber \\
\tau_{z}\left(z_{i}\left(t+1\right)-z_{i}\left(t\right)\right)  = & \: \Big\{-z_{i}\left(t\right)+\sum_{j}J_{ij}^{z}\phi\left(t\right)+b_{i}^{z}\left(t\right)\Big\}\delta t ,\nonumber \\
\tau_{r}\left(r_{i}\left(t+1\right)-r_{i}\left(t\right)\right) = & \:\Big\{-r_{i}\left(t\right)+\sum_{j}J_{ij}^{r}\phi \left(t\right)+b_{i}^{r}\left(t\right)\Big\}\delta t ,\nonumber
\end{align}

where we have introduced external fields in the dynamics $\{b^{h}_i(t)\}$,  $\{ b^{z}_i(t) \}$ and $\{b^{r}_i(t)\}$. 
The generating functional is given by

\begin{align} 
    Z_{\mathcal{J}}[\mathbf{\hat{b}}, \mathbf{b}]=\mathbb{E}\left[\exp \left(i \sum_{j=1}^N \sum_t \mathbf{\hat{b}}_{j}(t)^T \mathbf{x}_{j}(t) \delta t\right)\right],
\end{align}
where $\mathbf{\hat{b}}=(\hat{b}^h_j,\hat{b}^z_j,\hat{b}^r_j) $ ; 
$ \mathbf{b}=(b^h_j,b^z_j,b^r_j) $ and $\mathbf{x}_{j}(t) \equiv (h_j(t),z_j(t),r_j(t))$; also, the expectation is over the dynamics generated by the network. Writing this out explicitly, with 
$\delta-$functions enforcing the dynamics, we get the following integral for the generating functional

\begin{widetext}
\begin{align} \label{eq:MSRDJ_2}
Z_{\mathcal{J}}[\mathbf{\hat{b}}, 
\mathbf{b}] = &
  \int\prod_{i,t}\prod_{k,t'}\prod_{m,t''}dh_{i}\left(t\right)dz_{k}\left(t'\right) dr_{m}\left(t''\right)
    \cdot \exp\Big(i\Big\{ \sum_{i,t}\hat{b}_{i}^{h}\left(t\right)h_{i}\left(t\right)+\hat{b}_{i}^{z}\left(t\right)z_{i}\left(t\right) + \hat{b}_{i}^{r}\left(t\right)r_{i}\left(t\right) \Big\} \delta t\Big) ,\nonumber \\
 &  \times \delta\bigg(h_{i}\left(t+1\right)-h_{i}\left(t\right)+\Big\{ h_{i}\left(t\right) \sigma_{z,i}(t) 
  - \sigma_{z,i}(t) \Big[  \sum_{j}J_{ij}^{h}  \sigma_{r,j}(t) 
   \phi_{j}(t)  \Big] - b_{i}^{h}\left(t\right) \Big\} \delta t\bigg), \nonumber \\
 &  \times \delta\bigg(z_{k}\left(t'+1\right)-z_{k}\left(t'\right)+\frac{1}{\tau_z}\Big\{  z_{k}\left(t'\right)+\sum_{l}J_{kl}^{z}\phi_{l}\left(t'\right) +b_{k}^{z}\left(t'\right) \Big\} \delta t\bigg) ,\nonumber \\
 &  \times \delta\bigg(r_{m}\left(t''+1\right)-r_{m}\left(t''\right)+\frac{1}{\tau_r}\Big\{  r_{m}\left(t''\right)+\sum_{n}J_{mn}^{r}\phi_{n}\left(t''\right)+b_{m}^{r}\left(t'\right) \Big\} \delta t\bigg) .
\end{align}
\end{widetext}

Now, let us introduce the Fourier representation for the $\delta-$function; this introduces
an auxiliary field variable, which as we will see allows us to calculate the response function
in the MSRDJ formalism. The generating functional can then be expressed as
\begin{widetext}
\begin{eqnarray}
Z_{\mathcal{J}}[\mathbf{\hat{b}}, 
\mathbf{b}] & = 
   & \int\prod_{i,t}\prod_{k,t'}\prod_{m,t''}dh_{i}\left(t\right)  \frac{d\hat{h}_i(t)}{2\pi}  
   dz_{k}\left(t'\right)  \frac{d\hat{z}_k(t')}{2\pi}   
   dr_{m}\left(t''\right)  \frac{d\hat{r}_m(t'')}{2\pi}, \\
  & & \times  \exp\bigg[-i \sum_{i,t} \hat{h}_i(t) \Big( h_{i}\left(t+1\right)-h_{i}\left(t\right) - 
   f_h\big(h_i,z_i ,r_i \big)\delta t - b_{i}^{h}\left(t\right) \delta t \Big)  + 
  i \sum_{i,t}\hat{b}_{i}^{h}\left(t\right)h_{i}\left(t\right) \delta t  \bigg] ,\nonumber \\
  & & \times \exp\bigg[-i \sum_{k,t'} \hat{z}_k(t) \Big( z_{k}\left(t'+1\right)-z_{k}\left(t'\right) - 
   f_z\big(h_k,z_k \big)\frac{\delta t}{\tau_z} - b_{k}^{z}\left(t'\right) \frac{\delta t}{\tau_z} \Big)  + 
  i \sum_{k,t'}\hat{b}_{k}^{z}\left(t'\right)z_{k}\left(t'\right) \delta t  \bigg] ,\nonumber \\
  & & \times \exp\bigg[-i \sum_{m,t''} \hat{r}_m(t'') \Big( r_{m}\left(t''+1\right)-r_{m}\left(t''\right) - 
   f_r\big(h_m,r_m \big)\frac{\delta t}{\tau_r} - b_{m}^{r}\left(t''\right) \frac{\delta t}{\tau_r} \Big)   
 + \: i \sum_{m,t''}\hat{b}_{m}^{r}\left(t''\right)r_{m}\left(t''\right) \delta t  \bigg], \nonumber
\end{eqnarray} 
\end{widetext}
where the functions $f_{h,z,r}$ summarise the gated RNN dynamics
\begin{align}
f_{h}\left(h_{i}, z_{i}, r_{i}\right) &=
\sigma_{z,i}(t) \Big( -h_{i}(t) + \sum_{j} J_{i j}^{h} \sigma_{r,j}(t)  \phi_{j}(t) \Big) ,\nonumber \\
f_{z}\left(h_{k}, z_{k} \right) &=-z_{k}\left(t^{\prime}\right)+\sum_{l} J_{k l}^{z} \phi_{l}\left(t^{\prime}\right),
\nonumber \\
f_{r}\left(h_{m}, r_{m}\right) &=-r_{m}\left(t^{\prime \prime}\right)+\sum J_{m n}^{r} \phi_{n}\left(t^{\prime \prime}\right). \nonumber
\end{align}

Let us now take the continuum limit $\delta t \rightarrow 0$, and formally define the measures 
$\mathcal{D}h_i = \lim_{\delta t \rightarrow 0} \prod_{t} dh_{i}(t) $. We can then write the generating functional as a path integral 
\begin{align}
& Z_{\mathcal{J}}[\mathbf{\hat{b}}, 
\mathbf{b}]
=  \: \int \prod_i \mathcal{D}h_i \mathcal{D}\hat{h}_i   \mathcal{D}z_i \mathcal{D}\hat{z}_i  \mathcal{D}r_i \mathcal{D}\hat{r}_i  \exp \bigg\{ -S\left[ \hat{\mathbf{x}},\mathbf{x}\right] \nonumber \\
& \qquad +  i
 \int d t\left[\hat{\mathbf{b}}(t)^{T} \mathbf{x}(t)+\mathbf{b}(t)^{T} \hat{\mathbf{x}}(t)\right]
 \bigg\}  
\end{align}
Where $\hat{\mathbf{b}} = (\widehat{b}^h_i,\widehat{b}_i^z,\widehat{b}^r_i)$; $\mathbf{x}=(h_i,z_i,r_i)$ and $\mathbf{\hat{x}} = (\widehat{h}_i,\widehat{z}_i/\tau_z,\widehat{r}_i/\tau_r)$, and the action $S$ which gives weights to the paths is given by
\begin{align}
& S\left[ \hat{\mathbf{x}},\mathbf{x}\right]   =  i \sum_i \int dt \: \hat{h}_i(t) \Big[ \partial_th_i(t) 
-f_h\big(h_i,z_i ,r_i \big) \Big]   \nonumber
\\
 & \qquad + i \sum_k \int dt \: \hat{z}_k(t) \Big[ \partial_t z_k(t)
           - \frac{f_z\big(h_k,z_k \big)}{\tau_z} \Big]  \nonumber \\ 
  &\qquad + i \sum_k \int dt \: \hat{r}_m(t) \Big[ \partial_t r_m(t) - \frac{f_r\big(h_m,r_m \big)}{\tau_r} \Big] 
\end{align}
The functional is properly normalised, so $Z_{\mathcal{J}}[\mathbf{0}, 
\mathbf{b}] = 1$. We can calculate correlation functions and response
functions by taking appropriate variational derivatives of the generating functional $Z$, but first we address the role of the random couplings.

\textbf{Disorder Averaging:}

We are interested in the typical behaviour of ensembles of the networks, so we work with the disorder-averaged generating functional $\overline{Z}$;
 $Z_{\mathcal{J}}$ is properly normalised, so we are allowed to do this averaging on $Z_{\mathcal{J}}$. 
   Averaging over $J^{h}_{ij}$ involves
 the following integral
 \begin{align}
& \int dJ^{h}_{ij} \sqrt{\frac{N}{2\pi}} \exp \left\{ -\frac{N \left( J^{h}_{ij}\right)^2 }{2}  + \right. \nonumber \\
&  \left. \quad i\cdot J^{h}_{ij} \int dt \: \hat{h}_i(t)\sigma_{z,i}\left( t\right)\phi_{j}\left( t\right) \sigma_{r,j}\left( t\right) \right\}, \nonumber
 \end{align}
 which evaluates to

 $ \exp \left\{ -(1/2N)\cdot  \big( \int dt \: \hat{h}_i(t)\sigma_{z,i}\left( t\right)\phi_{j}\left( t\right) \sigma_{r,j}\left( t\right) \big)^2 \right\} $
 and similarly for $J^{z}$ and $J^{r}$  we get terms
 \begin{align}
 \exp \left\{ -(1/2N)\cdot  \big( \int dt \: \hat{z}_k(t)\phi_l\left( t\right) \big)^2  \tau_z^{-2}\right\}, \nonumber \\ 
  \exp \left\{ -(1/2N)\cdot  \big( \int dt \: \hat{r}_m(t)\phi_n\left( t\right) \big)^2  \tau_r^{-2}\right\} . \nonumber
\end{align}
The disorder-averaged generating functional is then given 
by 
\begin{align}
& \overline{Z}[\mathbf{\hat{b}}, 
\mathbf{b}]  =  \: \int \prod_i \mathcal{D}h_i \mathcal{D}\hat{h}_i   \mathcal{D}z_i \mathcal{D}\hat{z}_i  \mathcal{D}r_i \mathcal{D}\hat{r}_i  
\exp \bigg\{ -\overline{S} \left[\hat{\mathbf{x}}, \mathbf{x} \right]  \nonumber \\
&  \qquad +i \int d t\left[\hat{\mathbf{b}}(t)^{T} \mathbf{x}(t)+\mathbf{b}(t)^{T} \hat{\mathbf{x}}(t)\right]  \bigg\} 
\end{align}
where the disorder-averaged action $\overline{S}$ is given by

\begin{widetext}
\begin{align}
& \overline{S}\left[ \hat{\mathbf{x}},\mathbf{x}\right]    = 
i \sum_i \int dt \: \hat{h}_i(t) \Big(\partial_t h_i(t) + h_i(t)\sigma_{z,i}(t) \Big) + 
\frac{1}{2N} \sum_{i,j} \left( \int dt \: \hat{h}_i(t) \sigma_{z,i}(t) \phi_{j}(t) \sigma_{r,j}(t) \right)^2  \nonumber \\
  & \qquad + \: i \sum_k \int dt \: \hat{z}_k (t) \left( \partial_t z_k(t) +\frac{z_k(t)}{\tau_z}  \right) +
       \frac{1}{2N} \sum_{k,l} \left( \int dt \frac{\hat{z}_k(t)}{\tau_z} \cdot \phi_{l}(t) \right)^2 \nonumber \\
  &  \qquad + \: i \sum_m \int dt \: \hat{r}_m (t) \left( \partial_t r_m(t) +\frac{r_m(t)}{\tau_r}  \right) +
       \frac{1}{2N} \sum_{m,n} \left( \int dt \frac{\hat{r}_m(t)}{\tau_r} \cdot \phi_{n}(t) \right)^2
\end{align}
\end{widetext}

With some foresight, we see the action is extensive in the system size, and we can try to reduce it to a single-site description. However, the issue now is that we have non-local terms (e.g. involving both $i$ and $j$), and we can
introduce the following auxiliary fields to decouple these non-local terms

\begin{align}
C_{\phi \sigma_r}\left( t,t' \right)  := & \: \frac{1}{N} \sum_i \phi_{i}\left( t \right) \phi_{i}\left( t' \right)
\sigma_{r,i}\left( t \right) \sigma_{r,i}\left( t' \right),
\nonumber \\
C_{\phi}\left( t,t' \right)  := & \: \frac{1}{N} \sum_k \phi_{k}\left( t \right) \phi_{k}\left( t' \right) . \nonumber \\
\end{align}

To make the $C$'s free fields that we integrate over, we enforce these relations
using the Fourier representation of $\delta$ functions with additional auxiliary fields:
\begin{align}
& \delta \Big( N C_{\phi \sigma_r}\left( t,t' \right) - \sum_i \phi_{i}\left( t \right) \phi_{i}\left( t' \right)
\sigma_{r,i}\left( t \right) \sigma_{r,i}\left( t' \right) \Big)    = 
\nonumber \\
& \int \frac{N}{\pi} d\widehat{C}_{\phi \sigma_r}(t,t') 
  \exp \bigg[ -\frac{i}{2} \widehat{C}_{\phi \sigma_r}(t,t')  
   \Big(N\cdot C_{\phi \sigma_r}(t,t') \nonumber \\
   & \qquad  - \sum_i \phi_{i}\left( t \right) \phi_{i}\left( t' \right)
\sigma_{r,i}\left( t \right) \sigma_{r,i}\left( t' \right)  \Big)   \bigg] \nonumber
\\
&\delta \Big( N C_{\phi}\left( t,t' \right) - \sum_k \phi_{k}\left( t \right) \phi_{k}\left( t' \right) \Big)  =  
 \int \frac{N}{\pi} d\widehat{C}_{\phi}(t,t') \nonumber \\
&  \exp \bigg[ -\frac{i}{2} \widehat{C}_{\phi}(t,t') 
   \Big(N\cdot C_{\phi}(t,t') - \sum_k \phi_{k}\left( t \right) \phi_{k}\left( t' \right) \Big)   \bigg]
\nonumber
\end{align}

this allows us to make the following transformations to decouple the non-local terms in the action $\overline{S}$

\begin{align}
& \frac{1}{2N} \sum_{i,j} \left( \hat{h}_i(t)\sigma_{z,i}\left( t\right)\phi_{j}\left( t\right) \sigma_{r,j}\left( t\right) \right)^2  \longrightarrow \nonumber \\
& \frac{1}{2} \sum_{i} \int dt \: dt' \: \hat{h}_i\left( t \right) \sigma_{z,i}\left( t \right) C_{\phi \sigma_r}\left( t ,t' \right)
\hat{h}_i\left( t' \right) \sigma_{z,i}\left( t' \right) 
\nonumber \\
& \frac{1}{2N} \sum_{k,l} \left( \int dt \frac{\hat{z}_k(t)}{\tau_z} \cdot \phi_{l}(t) \right)^2  \longrightarrow \nonumber \\
& \qquad \qquad \frac{1}{2} \sum_{k} \int dt \: dt' \: \frac{\hat{z}_k(t)}{\tau_z}  C_{\phi}\left( t ,t' \right) \frac{\hat{z}_k(t')}{\tau_z} \nonumber 
\end{align}
\begin{align}
& \frac{1}{2N} \sum_{m,n} \left( \int dt  \frac{\hat{r}_m(t)}{\tau_r} \cdot \phi_{n}(t) \right)^2  \longrightarrow \nonumber \\
 & \qquad \qquad \frac{1}{2} \sum_{m} \int dt \: dt' \: \frac{\hat{r}_m(t)}{\tau_r}  C_{\phi}\left( t ,t' \right) \frac{\hat{r}_m(t')}{\tau_r} \nonumber
\end{align}

We see clearly that the $C_{\phi \sigma_r}$ and $C_{\phi}$  auxiliary fields which represent the (population averaged) $\phi \sigma_r-\phi \sigma_r$ and $\phi - \phi$ correlation functions have decoupled the sites 
by summarising  all the information present in the rest of the network in terms of two-point functions; two different sites interact only by means of the correlation functions. The disorder-averaged generating functional  can now be written as

\begin{align}
& \overline{Z}[\mathbf{\hat{b}}, 
\mathbf{b}] =  \: 
\int \mathcal{D}\widehat{\mathbf{C}} \: 
\mathbf{\mathcal{D}C}
 \exp  \Big( -N \cdot \mathcal{L}\big[\widehat{\mathbf{C}},
 \mathbf{C}; \hat{\mathbf{b}}, 
\mathbf{b}
\big] \Big) \label{eq:app_MSRDJ_1} \\
& \mathcal{L}
   =   \: \frac{i}{2} \int dt dt' \: \Big[ \mathbf{C}\left(t,t' \right)^T \widehat{\mathbf{C}} \left(t,t' \right) \Big]
 - W\big[\widehat{\mathbf{C}},
 \mathbf{C}; \widehat{\mathbf{b}}, 
\mathbf{b} \big] \nonumber \\
& \exp \left( N \cdot W \right)
 =  \:  \int \prod_{i} \mathcal{D}h_i \mathcal{D}\hat{h}_i   \mathcal{D}z_i \mathcal{D}\hat{z}_i \mathcal{D}r_i \mathcal{D}\hat{r}_i \times \nonumber \\
 & \exp \bigg\{   i  \int dt  \Big[
 \mathbf{b}(t)^T\hat{\mathbf{h}}(t) 
 +  \hat{\mathbf{b}}(t)^T\mathbf{h}(t)
 \Big]   - S_d\left[ \hat{\mathbf{h}},\mathbf{h}; \{ \mathbf{C},\widehat{\mathbf{C}} \} \right]  \bigg\}
 \nonumber
\end{align}
where $\mathbf{C} = (C_h,C_z,C_r)$ and $\widehat{\mathbf{C}} = (\widehat{C}_h,\widehat{C}_z,\widehat{C}_r)$.
The site-wise decoupled action, $S_d$ only contains terms involving a single site (and the $C$ fields). So, for a given value of $\widehat{\mathbf{C}}$ and $\mathbf{C}$, the different sites are decoupled and
driven by the site-wise action
\begin{align}
& S_d\left[ \hat{\mathbf{h}},\mathbf{h}; \{ \mathbf{C},\widehat{\mathbf{C}} \} \right]
= i \int dt \: \left[ \hat{\mathbf{h}}(t)^T 
\partial_t \mathbf{h}(t) +
\hat{\mathbf{h}}_{\tau}(t)^T\mathbf{h}(t) 
\right] \nonumber \\
& \qquad \qquad + \frac{1}{2} \int dt \: dt^{\prime} \:
\hat{\mathbf{h}}_{\tau}(t)^T \:
\mathbb{D}\mathbf{C}(t,t^{\prime})  \:
\hat{\mathbf{h}}_{\tau}(t^{\prime}) 
\nonumber \\
& \qquad \qquad - \frac{i}{2} \int dt \: dt^{\prime}
\: \mathbf{S}_x(t)^T \:
\mathbb{D}\widehat{\mathbf{C}}(t,t^{\prime}) \:
\mathbf{S}_x(t^{\prime})
\end{align}
where 
\begin{align}
 & \mathbf{\hat{h}}_{\tau}(t) =  (\hat{h}_i \sigma_{z,i},\hat{z}_i/\tau_z,\hat{r}_i/\tau_r) \nonumber  \\
&  \mathbf{\hat{h}}(t) =  (\hat{h}_i,\hat{z}_i,\hat{r}_i) \nonumber \\
&  \mathbf{S}_x =  (\phi_{i}\sigma_{r,i}, \phi_{i},\phi_{i})
\nonumber  \\
& \mathbb{D}\mathbf{C}(t,t') = 
\textrm{Diag}[C_{\phi \sigma_r}(t,t'),C_{\phi}(t,t'),C_{\phi}(t,t')] \nonumber \\
& \mathbb{D}\widehat{\mathbf{C}}(t,t') = 
\textrm{Diag}[\widehat{C}_{\phi \sigma_r}(t,t'),\widehat{C}_{\phi}(t,t'),\widehat{C}_{\phi}(t,t')] \nonumber
\end{align}

\textbf{Saddle-point approximation for} $N \rightarrow \infty$

So far, we have not made any use of the fact that we are considering large networks. However, noting that $N$ appears in the exponent in the expression for the disorder-averaged generating functional, we can approximate it using a saddle-point approximation: 
\begin{align}
\overline{Z}[\mathbf{\hat{b}}, 
\mathbf{b}] \simeq 
e^{N\cdot \mathcal{L}_0 \left[ 
\mathbf{\hat{b}}, 
\mathbf{b}; \mathbf{C}_0,\mathbf{\widehat{C}}_0
\right]} 
\int \mathcal{D} \mathbf{\widehat{Q}}
\: \mathcal{D} \mathbf{Q} 
e^{-N \cdot \mathcal{L}_{2}\left[
\mathbf{\widehat{Q}}, 
\mathbf{Q}, \mathbf{\widehat{b}}, \mathbf{b}\right]}
\nonumber
\end{align}

we have approximated the action $\mathcal{L}$ in eq. \ref{eq:app_MSRDJ_1} by its saddle-point value plus a Hessian term : $\mathcal{L} 
\simeq \mathcal{L}_0 + \mathcal{L}_2$ and the 
$\mathbf{Q}, \mathbf{\widehat{Q}}$ fields represent Gaussian 
fluctuations about the saddle-point values 
$\mathbf{C}^0, \widehat{\mathbf{C}}^0$, respectively. At the saddle-point
the action is stationary w.r.t variations, thus, the saddle-point values of $C$ fields satisfy  
\begin{align}
  C_{\phi\sigma_r}^{0}\left(t,t' \right) = & \: \frac{1}{N} \sum_{i=1}^{N} \left \langle \phi_{i} \left(t \right)\sigma_{r,i} \left(t \right) \phi_{i} \left(t' \right)\sigma_{r,i} \left(t' \right) \right \rangle_0 
  \nonumber
  \\
  \widehat{C}_{\phi\sigma_r}^{0}\left(t,t' \right) = & \: \frac{1}{N} \sum_{i=1}^{N} \left \langle 
  \widehat{h}_i\left( t \right) \sigma_{z,i}\left( t \right) \widehat{h}_i\left( t' \right) \sigma_{z,i}\left( t' \right) \right \rangle_0  \nonumber \\
   = & \: \frac{ \delta^2 \left \langle  \sigma_{z,i}\left( t \right) \sigma_{z,i}\left( t' \right) \right \rangle_0}{\delta b_i(t) \delta b_i(t')}  = 0 
   \nonumber \\
   C_{\phi}^{0}\left(t,t' \right) = & \: \frac{1}{N} \sum_{k=1}^{N} \left \langle \phi_{k} \left(t \right) \phi_{k} \left(t' \right) \right \rangle_0 \nonumber \\
  \widehat{C}_{\phi}^{0}\left(t,t' \right) = & \: 0
 \end{align}
In evaluating the saddle-point correlation function in the second line, we have used the fact that equal-time response functions in the It\^{o} convention are zero \cite{hertz2016path}. This is perhaps the first significant point of departure from previous studies of disordered neural networks, and forces us to confront the multiplicative nature of the $z-$gate. Here $\langle \cdots \rangle_0$ denotes averages w.r.t paths generated by the saddle-point action, thus these equations are a self-consistency constraint. With the correlation fields fixed at their saddle-point values, if we neglect the contribution of the 
fluctuations (i.e. ignore $\mathcal{L}_2$) then the generating functional is given by a product 
of identical site-wise generating functionals. 
\begin{align}
    \overline{Z}[\mathbf{\hat{b}}, 
\mathbf{b}] =  Z_0[\mathbf{\hat{b}}, 
\mathbf{b}]^N
\end{align}
where the site-wise functionals are given by
\begin{align}
    Z_0[\mathbf{\hat{b}},& \mathbf{b}] = 
    \:  \int \mathcal{D}h \mathcal{D}\hat{h}   \mathcal{D}z \mathcal{D}\hat{z} \mathcal{D}r \mathcal{D}\hat{r} \: \times \nonumber \\
   &  e^{ \big(   i  \int dt  \left[
 \mathbf{b}(t)^T\mathbf{\hat{h}}(t) 
 +  \mathbf{\hat{b}}(t)^T\mathbf{h}(t)
 \right]   - S_d\left[ \mathbf{\widehat{h}},\mathbf{h}; \{ \mathbf{C}^0,\mathbf{0} \} \right]  \big) }
\end{align}
where $\mathbf{C}^0 = (C_{\phi\sigma_r}^{0},C_{\phi}^{0})$.

The site-wise decoupled action is now quadratic with the correlation functions taking on their 
saddle-point values. This corresponds to an action for each site containing three scalar variables driven by  Gaussian processes. This can be seen explicitly by using a Hubbard-Stratonovich transform
which will make the action linear at the cost of introducing three auxiliary Gaussian fields $\eta_h$,
$\eta_z$ and $\eta_r$ with correlation functions $C_{\phi\sigma_r}^{0}\left( t ,t' \right)$, $C_{\phi}^{0}\left( t ,t' \right)$ and
$C_{\phi}^{0}\left( t ,t' \right)$, respectively. With this
transformation, the action for each site corresponds to stochastic dynamics for three scalar variables given by:
\begin{align}
\dot{h}(t) &=-\sigma_z(z) \cdot h(t)+\sigma_z(z) \cdot \eta_{h}(t) \label{eq:DMFT_3} \\
\tau_{z} \dot{z}(t) &=-z(t)+\eta_{z}(t) \label{eq:DMFT_1} \\
\tau_{r} \dot{r}(t) &=-r(t)+\eta_{r}(t) \label{eq:DMFT_2}
\end{align}    
where the Gaussian noise processes, $\eta_h$, $\eta_z$, and $\eta_r$ have 
correlation functions that must be determined self-consistently:
\begin{align}
\left\langle
\eta_{h}(t) \cdot \eta_{h}\left(t^{\prime}\right)
\right\rangle  
& =
\big \langle
\phi(t) \sigma_r(t) \cdot   
\phi\left(t^{\prime}\right) 
\sigma_r\left(t^{\prime}\right)
\big \rangle ,\nonumber \\
\left\langle
\eta_{z}(t) \cdot \eta_{z}\left(t^{\prime}\right)
\right\rangle  
&=
\big \langle
\phi(t)  \cdot   
\phi\left(t^{\prime}\right) 
\big \rangle ,\nonumber \\
\left\langle
\eta_{r}(t) \cdot \eta_{r}\left(t^{\prime}\right)
\right\rangle  
&=
\big \langle
\phi(t)  \cdot   
\phi\left(t^{\prime}\right) 
\big \rangle .\nonumber
\end{align}
The intuitive picture of the saddle-point approximation is as follows: the sites of the full network become decoupled, and they are each driven by a Gaussian processes whose correlation functions summarise the activity of the rest of the network `felt' by each site. It is possible to argue about the final result heuristically, but one does not have access to the systematic corrections that a field theory formulation affords. 

\begin{figure}
\begin{centering}
\includegraphics[scale=0.4
]{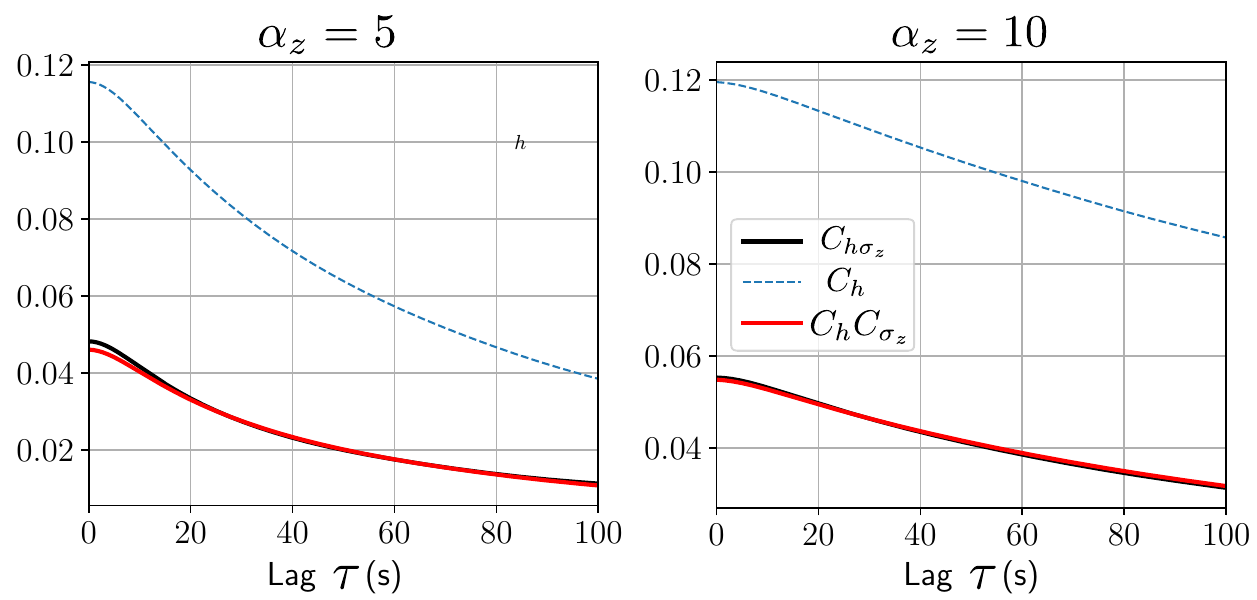}
\par\end{centering}
\caption{\label{fig:DMFT_hGz_separabiltiy} Figure showing the validity of the approximation $C_{h \sigma_z} = C_h C_{\sigma_z}$ for two values of $\alpha_z$. The correlation functions were calculated numerically in a network with $n=1000, g_h=3.5, \alpha_r=0$.  }
\end{figure}

We comment here on the unique difficulty that gating presents to an analysis of the DMFT. While $r(t)$ and $z(t)$ are both described by Gaussian processes in the DMFT, the multiplicative $\sigma_{z}(z)$ interaction in Eq.(\ref{eq:DMFT_3}) spoils the Gaussianity of $h(t)$. Note that $r(t)$ is always Gaussian and uncorrelated to $h(t)$. In order to try solving for the correlation functions, we need to make a factorization assumption, justified numerically in Fig.(\ref{fig:DMFT_hGz_separabiltiy}). The story simplifies at a fixed point, where $h = \eta^{h}$ (since $\sigma_{z}>0$) and is thus Gaussian and independent of $r$.

\begin{figure}
\begin{centering}
\includegraphics[scale=0.4]{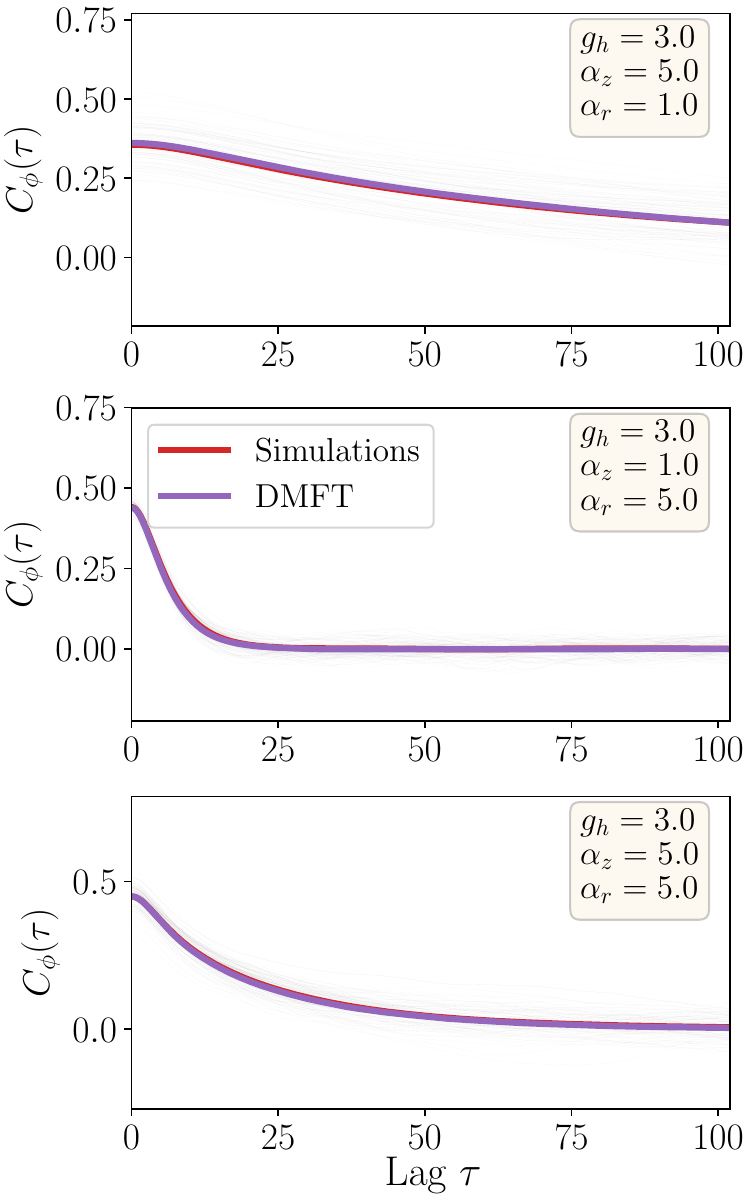}
\par\end{centering}
\caption{\label{fig:DMFT_fullNet_corr} {\it Validating the DMFT:}  panels show the comparison between the population averaged correlation functions $C_{\phi}(\tau) \equiv \langle \phi(t) \phi(t+\tau) \rangle$ obtained from the full network simulations of a single instantiation in steady-state (purple line) and from solving the DMFT equations (red line) for three distinct parameter values. The lag $\tau$  is relative to $\tau_h$ (taken to be unity). Thin black lines are the time-averaged correlation functions for individual neurons sampled from the network, to show  the scale of fluctuations around the population-averaged correlation functions. $N=5000$ for all the panels.}
\end{figure}

 In order to solve the DMFT equations, we use a numerical method described in \cite{roy2019numerical}. Specifically, we generate noise paths $\eta_{h,z,r}$ starting with an initial guess for the correlation functions, and then iteratively update the correlation functions using the mean-field equations till convergence.  The classical method of solving the DMFT by mapping the DMFT equations to a second-order ODE describing the motion of a particle in a potential cannot be used in the presence of multiplicative gates. In
Fig. \ref{fig:DMFT_fullNet_corr}, we see that the solution to the mean-field equations agrees well with the true population-averaged correlation function; Fig. \ref{fig:DMFT_fullNet_corr} also shows the scale of fluctuations around the mean-field solutions (Fig. \ref{fig:DMFT_fullNet_corr} thin black lines).

The correlation functions in the DMFT picture such as $C_h(t,t^{\prime}) =
\left\langle h(t)  h(t^{\prime}) \right\rangle  $ are the order parameters, and correspond to the population-averaged correlation functions in the full network. These will turn out to useful in our analysis of the RNN dynamics in some analyses. Qualitative changes in the correlation functions correspond to transitions between dynamical regimes of the RNN. 

In general, the non-Gaussian nature of $h$ makes it  impossible to get equations governing the correlation functions. However, when $\alpha_z$ is not too large , the eqns. \ref{eq:DMFT_1}-\ref{eq:DMFT_3} can be extended to get equations of motions for the correlation functions $C_h, C_z,C_r$, which will prove useful later on. This requires a separation assumption between the $h$ and $\sigma_z$ correlators, which seems reasonable for moderate $\alpha_z$ (see Fig. \ref{fig:DMFT_hGz_separabiltiy}). 
`Squaring' eqs. \ref{eq:DMFT_1}-\ref{eq:DMFT_3} we get

\begin{align}\label{eq:DMFT_5}
\left[ -\partial_{\tau}^2 + C_{\sigma_{z}}(\tau) \right]
C_h(\tau) = & \: C_{\sigma_{z}}(\tau)  C_{\sigma_{r}}(\tau) C_{\phi}(\tau), \\
\left[ -\tau_z^2 \partial_{\tau}^2 + 1 \right] C_z(\tau) 
= & \: C_{\phi}(\tau), \label{eq:DMFT_6} \\
\left[ -\tau_r^2 \partial_{\tau}^2 + 1 \right] C_r(\tau) 
= & \: C_{\phi}(\tau),  \label{eq:DMFT_7}
\end{align}

where we have used the shorthand $\sigma_{z}(t) \equiv \sigma_z(z(t)); \; \phi(t) \equiv \phi(h(t))$,  and denoted the two-time correlation functions as
\begin{align}
    C_{x}(t, t') = \langle x(t) x(t')\rangle,\label{eq:corr_fn}
\end{align}
where $x \in \{ h, z, r, \sigma_{z}, \sigma_{r}, \phi\}$, and the expectation here is over the random Gaussian fields in Eqs. (\ref{eq:DMFT_1},\ref{eq:DMFT_2},\ref{eq:DMFT_3}). We assume that the network has reached steady-state, so that the correlation functions are only a function of the time difference $\tau = t - t^{\prime}$. The role of the $z-$gate as an adaptive time constant is evident in eq. \ref{eq:DMFT_5}.

For time-independent solutions, i.e. fixed points, eqns. \ref{eq:DMFT_5} - \ref{eq:DMFT_7}  simplify to read
\begin{align}\label{eq:MFT_FP_variance_1}
\Delta_{z} \equiv \left\langle z^{2}\right\rangle = & \: \int D x \: \phi\left(\sqrt{\Delta_{h}} x\right)^{2}  = \: \Delta_{r}\\
\Delta_{h} \equiv  \left\langle h^{2}\right\rangle = & \: \int Dx Dy \: \phi\left( \sqrt{\Delta_{h}} x\right)^{2}\sigma_{r}(\sqrt{\Delta_{r}} y)^{2}\label{eq:MFT_FP_variance_2}
\end{align}
where we have used $\Delta$ instead of $C$ to indicate fixed-point variances, and  $Dx$ is the standard Gaussian measure.  It is interesting to note that these mean-field equations can be mapped to those obtained in \cite{can2020gating} for the discrete-time gated recurrent unit (GRU). 

We also make use of the MFT with static random inputs. For completeness, we include the resulting equations here. With $I_{i}^{h,z,r} \sim \mathcal{N}(0, \sigma_{h,z,r}^{2})$, the MFT time-independent solution satisfies

\begin{align}
    \Delta_{z} &= \int Dx \phi \left( \sqrt{\Delta_{h}} x\right)^{2} + \sigma_{z}^{2},\label{eq:mft_input1}\\
    \Delta_{r} & =  \int Dx \phi \left( \sqrt{\Delta_{h} } x\right)^{2} + \sigma_{r}^{2}, \label{eq:mft_input2}\\
    \Delta_{h} & =  \int Dx Dy \: \phi\left( \sqrt{\Delta_{h}} x\right)^{2}\sigma_{r}(\sqrt{\Delta_{r}} y)^{2} + \sigma_{h}^{2}.\label{eq:mft_input3}
\end{align}

%************* NUMERICAL LYAPUNOV APPENDIX *******
\section{Details of the numerics for the Lyapunov spectrum}
\label{app:Lyapunov_numerics}

The evolution of perturbations $\delta \mathbf{x}(t)$ along a trajectory 
follow the tangent-space dynamics governed by the Jacobian
\begin{align} \label{eq:tangent-space-dynamics}
    \partial_t \delta \mathbf{x}(t) = & \mathcal{D}(t) \delta \mathbf{x}(t)
\end{align}
So after a time $T$, the initial perturbation $\delta \mathbf{x}(0)$ will
be given by 
\begin{align} 
    \delta \mathbf{x}(t) = U(t,0) \delta {\bf x}(0), \quad U(t,0)  =  \mathcal{T}\left[ e^{\int_0^t ds \mathcal{D}(s)}
    \right] ,
\end{align}
 where $\mathcal{T}[..]$ is the time-ordering operator applied to the contents of the bracket. When the infinitesimal perturbations grow/shrink exponentially, the rate of this exponential growth/decay will be dictated by
the maximal Lyapunov exponent defined as \cite{eckmann1985ergodic}:
\begin{align}
    \lambda_{max} := & \lim_{T \rightarrow \infty}
        \frac{1}{T} \: \:
        \lim_{\| \delta \mathbf{x}(0) \| \rightarrow 0 }
        \ln \frac{\| \delta \mathbf{x}(T) \|}
        {\| \delta \mathbf{x}(0) \|}
\end{align}

For ergodic systems, this limit is independent of almost all initial conditions, as guaranteed by Oseledet's multiplicative ergodic theorem \cite{eckmann1985ergodic}. Positive values of $\lambda_{max}$ imply that the nearby trajectories diverge exponentially fast, and the
system is chaotic.
More generally, the set of all Lyapunov exponents -- the Lyapunov spectrum --  yields the 
rates at which perturbations along different directions shrink or diverge, and thus provide a fuller characterization of asymptotic behaviour. The first $k$ ordered Lyapunov exponents are 
given by the growth rates of $k$ linearly independent perturbations. 
These can be obtained as the logarithms of the eigenvalues 
of the Oseledet's matrix, defined as \cite{eckmann1985ergodic} 
\begin{align}
    \mathbf{M}(t) = \lim_{t \rightarrow \infty}
             \left( U(t, 0)^{T} U(t, 0)  \right)^{\frac{1}{2t}}
\end{align}
However, this expression cannot be directly used to calculate the Lyapunov spectra in practice
since $\mathbf{M}(t)$ rapidly becomes  ill-conditioned. 
We instead employ a method suggested by \cite{geist1990comparison} (also c.f. \cite{engelken2020lyapunov} for Lyapunov spectra of RNNs). We start with $k$ orthogonal vectors $Q^0 = [q_1,\ldots,q_k]$
and evolve them using the tangent-space dynamics, eq. \ref{eq:tangent-space-dynamics}, for a short time interval $t_0$. Therefore, the new set of vectors are given by 
\begin{align}
    \widehat{Q} =   U(t_{0}, 0) Q^{0} %\left[ e^{\int_0^{t_0} dt' \mathcal{D}(t')} \right] Q^0
\end{align}
We now decompose $\widehat{Q} = Q^1 R^1$ using a $QR-$decomposition, into an orthogonal matrix $Q^1$ and a upper-triangular matrix $R^1$ with positive diagonal elements, which give the rate of shrinkage/expansion of the volume element along the different directions. We iterate this procedure for a long time, $t_0 \times N_l$, and the first $k$ ordered Lyapunov exponents are given by 
\begin{align}
    \lambda_i = \lim_{N_l \rightarrow \infty} \frac{1}{N_l t_0}
              \sum_{j=1}^{N_l} \ln R_{ii}^j  \quad i \in \{1,\ldots,k\}
\end{align}

%************************** DMFT LYAPUNOV APPENDIX ***************

\section{Details of the DMFT prediction for $\lambda_{max}$}
\label{app:Lyapunov_DMFT}

 The starting point of the method to calculate the DMFT prediction for $\lambda_{max}$ is two replicas of the system $\mathbf{x}^1(t)$ and $\mathbf{x}^2(t)$
with the same coupling matrices $J^{h,z,r}$ and the same parameters. If the two systems are started with initial conditions which are close, then the rate of convergence/divergence of the trajectories reveals the maximal Lyapunov exponent. To this end, let's define $d(t,s) := N^{-1} \sum_{i} (x_i^1(t) - x_i^2(s))^2$, and study the growth rate of $d(t,t)$. In the large $N$ limit, we expect population averages like $ C^{12}(t,s) := N^{-1} \sum_{i} x_i^1(t)x_i^2(s)$ to be self-averaging (like in the DMFT for a single system) \footnote{The local chaos hypothesis employed by Cessac \cite{cessac1995increase} amounts to the same assumption}, and thus we can write 
\begin{align} \label{eq:replica-distance}
    d(t,s) = C^{11}(t,t) + C^{22}(s,s) - C^{12}(t,s) - C^{21}(t,s)
\end{align}

For trajectories that start nearby, the asymptotic growth rate of $d(t)$ is the maximal Lyapunov exponent. In order to calculate this using the DMFT, we need a way to calculate $C^{12}$ -- the correlation {\it between} replicas -- for a typical instantiation of systems in the large $N$ limit. As suggested by \cite{schuecker2018optimal}, this can be achieved by considering a joint generating functional for the replicated system:
\begin{align}
      \tilde{Z}_{\mathcal{J}}[\mathbf{\hat{b}}^1,
      \mathbf{\hat{b}}^2,\mathbf{b}^1,\mathbf{b}^2]
      =
      \mathbb{E}\left[\exp \left(i \sum_{\mu=1}^2\sum_{j=1}^N \int \mathbf{\hat{b}}^{\mu}_{j}(t)^T \mathbf{x}^{\mu}_{j}(t) d t\right)\right]
\end{align}

We then proceed to take the disorder average  of this generating functional -- in much the same way as a single system -- and this introduces correlations between the state vectors of the two replicas. A saddle-point approximation as in the single system case (c.f. Appendix \ref{app:DMFT_derivation}), yields a system of coupled SDEs (one for each replica), similar to eq. \ref{eq:DMFT_1}, but now the noise processes in the two replicas are coupled, so that terms like $\langle \eta_h^1(t) \eta_h^2(t') \rangle$    need to be considered. As before, the SDEs imply the equations of motion for the correlation functions
\begin{align}\label{eq:replica-DMFT-1}
\left[ -\partial_{\tau}^2 + C^{\mu \nu}_{\sigma_{z}}(\tau) \right]
C^{\mu \nu}_h(\tau) = & \: C^{\mu \nu}_{\sigma_{z}}(\tau) C^{\mu \nu}_{\phi}(\tau) C^{\mu \nu}_{\sigma_{r}}(\tau) \\
\left[ -\tau_z^2 \partial_{\tau}^2 + 1 \right] C^{\mu \nu}_z(\tau) 
= & \: C^{\mu \nu}_{\phi}(\tau) \\
\left[ -\tau_r^2 \partial_{\tau}^2 + 1 \right] C_r(\tau) 
= & \: C^{\mu \nu}_{\phi}(\tau) 
\end{align}
where $\mu,\nu \in \{1,2\}$ are the replica indices. 
Note that the single-replica solution will clearly be a solution to this system, reflecting the fact that marginal statistics of each replica is the same as before. When the replicas are started with initial conditions that are $\epsilon$-close, we expect the inter-replica correlation function to diverge from the single replica steady-state solution, so we expand $C^{12}$ to linear order as $C^{12}_{h,z,r}(t,s) \approx C_{h,z,r}(t-s) + \epsilon\tilde{\chi}_{h,z,r}(t,s)$. From eq. \ref{eq:replica-distance} we see that $d(t,t) \sim \epsilon \tilde{\chi}(t,t) $, and thus the growth rate of $\tilde{\chi}$ will yield the required Lyapunov exponent. To this end we make an ansatz $\tilde{\chi}_{h,z,r} = e^{\kappa T}\chi(\tau)$ where $2T = t+s$ and $2\tau = t-s$, and $\kappa$ is the DMFT prediction of the maximum Lyapunov exponent that needs to be solved for. Substituting this back in eq. \ref{eq:replica-DMFT-1}, we get a generalized eigenvalue problem for $\kappa$ as stated in the text (eqns. \ref{eq:Lyapunov-DMFT-eigenvalue_1} - \ref{eq:Lyapunov-DMFT-eigenvalue_3} ).

\section{Calculation of maximal Lyapunov exponent from RMT}
\label{app:RMT_Lyapunov}

The DMFT prediction for how gates shape $\lambda_{max}$ (via the correlation functions) is somewhat involved, thus we provide an alternate expression for the maximal Lyapunov exponent, $\lambda_{max}$, derived using RMT which relates it to the relaxation time of the dynamics. 
The starting point to get $\lambda_{max}$ is Oseledets' multiplicative ergodic theorem which guarantees that
 \footnote{Strictly speaking, Oseledets' theorem guarantees that $\lambda_{\max }=\lim _{t \rightarrow \infty} \frac{1}{2 t} \log \frac{\|\chi \mathbf{u}\|^{2}}{\|\mathbf{u}\|^{2}}$ for almost every $\mathbf{u}$. In particular, we can take $\mathbf{u}$ to be the all ones vector. The term inside the $\log$ then becomes $\frac{1}{N} \sum_{i, k} \chi_{i k}^{2}+\frac{1}{N} \sum_{i \neq j} \sum_{k} \chi_{i k} \chi_{j k}$, and the second term is subleading in $N$ since the susceptibilities are random functions. This justifies eq. \ref{eq:RMT_Lyapunov_0}
}

\begin{align} \label{eq:RMT_Lyapunov_0}
\lambda_{{\rm max}} = & \lim_{t \to \infty}   \frac{1}{2t} \log \frac{\| \chi(t)  \|^2}{N},  \\
= & \lim_{t \to \infty} \frac{1}{2t} \log \frac{1}{N}\operatorname{Tr} \left[ \chi(t) \chi(t)^T \right],
\end{align}
where  $\chi(t) = \mathcal{T} e^{\int_0^t dt' \mathcal{D}(t')}$ and $\mathcal{D}$ is the Jacobian.
For the vanilla RNN, the Jacobian is given by 
\begin{align}
    \mathcal{D} = -\mathbbm{1} + J \left[ \phi'(t) \right].
\end{align}
We expect the maximal Lyapunov exponent to be independent of the random network realization, and thus equal to its value after disorder-averaging. Furthermore, to make any progress, we use a short-time approximation for $\chi(t) \approx e^{ \int_{0}^{t} dt' \mathcal{D}(t')}$. Defining the diagonal matrix  $R(t) = \int^{t} \left[\phi'(t') \right]dt'$, these assumptions give
\begin{align} \label{eq:RMT_Lyapunov_1}
\frac{1}{N}\operatorname{Tr} \left[ \chi(t) \chi(t)^T \right]
& \approx  e^{ - 2t}  \left\langle \frac{1}{N} {\rm Tr} \, \,  e^{ J R(t)} e^{ R(t) J^{T}} \right\rangle, \\
& =  e^{ - 2 t} \sum_{n = 0}^{\infty} \frac{1}{(n!)^{2}} \left( \frac{1}{N} {\rm Tr} \, R(t)^{2} \right)^{n}  ,
\end{align}
where the second line in eq. \ref{eq:RMT_Lyapunov_1} follows after disorder averaging over $J$ and keeping only terms to leading order in $N$. Next, we may apply the DMFT to write

\begin{align}
\frac{1}{N} {\rm Tr} \, R(t)^{2} = & \int^{t} dt' dt'' \, \, \frac{1}{N}\sum_{i = 1}^{N}  \phi_{i}'(t'') \phi_{i}'(t') , \\ & \approx \int dt' dt'' C_{\phi'}(t', t'').
\end{align}
In steady-state, the correlation function depends only on the difference of the two times, and thus we can write
\begin{align}\label{eq:RMT_Lyapunov_2}
\int dt' dt'' C_{\phi'}(t', t'') \approx \int_{0}^{2 t} \frac{d u}{2} \int_{0}^{t} d\tau C_{\phi'}(\tau) \equiv t^{2} \tau_{R}, 
\end{align}
where we have defined the relaxation time for the $C_{\phi'}$ correlation function
\begin{align}
\tau_{R} \equiv \frac{1}{t} \int_{0}^{t} d\tau \, C_{\phi'}(\tau).
\end{align}
Substituting eq. \ref{eq:RMT_Lyapunov_2} in eq. \ref{eq:RMT_Lyapunov_1} we get
\begin{align}
\frac{1}{N}\operatorname{Tr} \left[ \chi(t) \chi(t)^T \right]
= e^{ - 2t} I_{0}(2 t \sqrt{\tau_{R}}) ,
\end{align}
which for long times behaves like $ \exp\left( 2 (\sqrt{\tau_{R}} - 1) t \right)$. By inserting this into eq. \ref{eq:RMT_Lyapunov_0}, we obtain a bound for the maximal Lyapunov exponent for the vanilla RNN 
\begin{align} \label{eq:RMT_Lyapunov_vanillaRNN_lambda}
\lambda_{max} \ge  & \sqrt{\tau_{R}} - 1 \\
\textrm{where} \quad \tau_{R} \equiv & \frac{1}{t} \int_{0}^{t} d\tau \, C_{\phi'}(\tau). 
\end{align}
This formula relates the asymptotic Lyapunov exponent to relaxation time of a local correlation function in steady state. It is interesting to note that the bound also follows by applying the variational theorem to the potential energy obtained from the Schr\"odinger equation that arises in computing the Lyapunov exponent using DMFT (e.g. see \cite{sompolinsky1988chaos,helias2019statistical}). Specifically, if one uses the potential obtained in these works $V(\tau) = 1 - C_{\phi'}(\tau)$, and assuming a uniform ``ground state wave function", the variational theorem implies that the true ground state energy $E_{0}$ is upper bounded $E_{0} \le \lim_{T \to \infty} \frac{1}{T}\int_{-T/2}^{T/2} V(\tau) d \tau  \equiv 1- \tau_{R}$, which consequently implies the bound (\ref{eq:RMT_Lyapunov_vanillaRNN_lambda}). 

Now we present the derivation for the mean squared singular value of the susceptibility matrix for the gated RNN with $\alpha_{z} = 0$, and $\beta_{z} = -\infty$. In this limit, $\sigma_{z}= 1$, and the instantaneous Jacobian becomes the $2N\times 2N$ matrix 
\begin{align}
\mathcal{D}_{t} &= - \mathbbm{1}_{2N} + \left( \begin{array}{cc}
 J^{r} & 0 \\
 0 & J^{h}	
 \end{array}\right) \left( \begin{array}{cc}
 0 & P_{t}\\
 Q_{t} & R_{t}	
 \end{array}\right)	 \equiv - 1_{2N} + \hat{J} S_{t},\\
  \quad Q_{t} &=\left[\phi({\bf h}(t)) \odot \sigma_{r}'({\bf r}(t) \right], \quad P_{t} = \left[\phi'({\bf h}(t))  \right],\\
   R_{t} &= \left[\phi'({\bf h}(t)) \odot \sigma_{r}({\bf r}(t) \right].
\end{align}

Let us define the quantity of interest

\begin{align}
\sigma_{\chi}^{2} &= \left\langle \frac{1}{2N} {\rm Tr} \left( \chi(t) \chi^{T}(t)\right) \right\rangle\\
& = e^{ - 2t} \left\langle, \frac{1}{2N} {\rm Tr} \,\,e^{ \hat{J}  \hat{S}_{t}} e^{ \hat{S}_{t}^{T} \hat{J}^{T}}  \right\rangle,
\end{align}

where we have additionally defined $\hat{S}_{t} = \int^{t} dt' S_{t}$, and the integration is performed elementwise. 
Expanding the exponentiated matrices and computing moments directly, one finds that the leading order in $N$ moments must have an equal number of $\hat{J}$ and $\hat{J}^{T}$. Thus, we must evaluate
\begin{align}
c_{n} = \left\langle \frac{1}{2N} {\rm Tr} \left[\left( \hat{J} \hat{S}_{t}\right)^{n} \left( \hat{S}_{t}^{T} \hat{J}^{T}\right)^{n}  \right]\right\rangle.
\end{align}

The ordering of the matrices is important in this expression. Since all of the $\hat{J}$ appear to the left of $\hat{J}^{T}$, the leading order contributions to the moment will come from Wick contractions that are ``non-crossing"- in the language of diagrams, the moment will be given by a ``rainbow" diagram. Consequently, we may evaluate $c_{n}$ by induction. First, the induction step. Define the expected value of the matrix moment

\begin{align}
\hat{c}_{n} &= \left\langle  \left( \hat{J} \hat{S}_{t}\right)^{n} \left( \hat{S}_{t}^{T} \hat{J}^{T}\right)^{n}  \right\rangle.\\
& = \left\langle \hat{J} \left( \hat{S}_{t} \left( \hat{J} \hat{S}_{t}\right)^{n-1} \left( \hat{S}_{t}^{T} \hat{J}^{T}\right)^{n-1} \hat{S}_{t}^{T}\right) \hat{J}^{T}\right\rangle\\
& = \left(\begin{array}{cc}
a_{n}\mathbbm{1} & 0\\
0 & b_{n} \mathbbm{1} \end{array}\right) + O(N^{-1}).
\end{align}
We wish to determine $a_{n}$ and $b_{n}$. Next,define
\begin{align}
g_{P} &= \frac{1}{N} {\rm Tr} \int^{t} dt' dt'' P_{t'} P_{t''},\\
g_{Q} &= \frac{1}{N} {\rm Tr} \int^{t} dt' dt'' Q_{t'} Q_{t''},\\
g_{R} &= \frac{1}{N} {\rm Tr} \int^{t} dt' dt'' R_{t'} R_{t''}.
\end{align}

Now we can directly determine the induction step at the level of matrix moments by Wick contraction of the rainbow diagram
\begin{align}
\hat{c}_{n} &= \left\langle  \hat{J} \hat{S}_{t} \left( \hat{J} \hat{S}_{t}\right)^{n-1} \left( \hat{S}_{t}^{T} \hat{J}^{T}\right)^{n-1} \hat{S}_{t}^{T} \hat{J}^{T}   \right\rangle,\\
& =  \left\langle  \hat{J} \hat{S}_{t} \hat{c}_{n-1} \hat{S}_{t}^{T} \hat{J}^{T}  \right\rangle + O(N^{-1}),\\
& = \left( \begin{array}{cc}
b_{n-1} g_{P} \mathbbm{1} & 0 \\
0 & \left(a_{n-1} g_{Q} + b_{n-1} g_{R} \right)\mathbbm{1}\end{array}\right)+ O(N^{-1}).
\end{align}
This implies the following recursion for the diagonal elements of $\hat{c}_{n}$
\begin{align}
a_{n} = g_{P} b_{n-1}, \quad b_{n} = g_{R} b_{n-1} + g_{Q} a_{n-1}.
\end{align}

The initial condition is given by observing that $\hat{c}_{0} = \mathbbm{1}$, which implies $a_{0} = b_{0} = 1$. The solution to this recursion relation can be written in terms of a transfer matrix

\begin{align}
\left( \begin{array}{c}
a_{n}  \\
b_{n}\end{array}\right)  & =  \left( \begin{array}{cc}
0 & g_{P} \\
g_{Q} &  g_{R}\end{array}\right)^{n} \left( \begin{array}{c}
1 \\
1\end{array}\right),
\end{align}

which implies the moment $c_{n} = \frac{1}{2} \left(a_{n} + b_{n}\right)$ is given by

\begin{align}
c_{n} = \frac{1}{2} \left(1 \, \quad 1\right)  \left( \begin{array}{cc}
0 & g_{P} \\
g_{Q} &  g_{R}\end{array}\right)^{n} \left( \begin{array}{c}
1 \\
1\end{array}\right).
\end{align}

To evaluate this, we use the fact that the eigenvalues of the transfer matrix are

\begin{align}
v_{\pm} = \frac{1}{2} \left( g_{R} \pm \sqrt{ g_{R}^{2} + 4 g_{P} g_{Q}}\right)	,
\end{align}

which are real valued. The eigenvectors are

\begin{align}
{\bf v}_{\pm} = 	\left( - \frac{v_{\mp}}{g_{Q}}, 1\right).
\end{align}

Then, defining ${\bf l} = (1, 1)$, the moment can be written 

\begin{align}
c_{n} &= \frac{1}{2} {\bf l}^{T} \left(  v_{+}^{n} {\bf v}_{+} {\bf v}_{+}^{T}  +  v_{-}^{n} {\bf v}_{-} {\bf v}_{-}^{T} \right) {\bf l}, \\
& = \frac{1}{2} \left( 1 - \frac{v_{-}}{g_{Q}}\right)^{2} v_{+}^{n} + \frac{1}{2} \left( 1 - \frac{v_{+}}{g_{Q}}\right)^{2}v_{-}^{n}.
\end{align}

The final expression for the mean squared singular value will then be

\begin{align}
\sigma_{\chi}^{2} = e^{ - 2t} \sum_{n = 0}^{\infty} \frac{c_{n}}{(n!)^{2}}.
\end{align}

After resumming this infinite series we wind up with an expression in terms of the modified Bessel function

\begin{align}
\sigma_{\chi}^{2} = \frac{1}{2} e^{ - 2 t} \left[ \left( 1 - \frac{v_{-}}{g_{Q}}\right)^{2}	 I_{0}( 2 \sqrt{v_{+}}) + \left( 1 - \frac{v_{+}}{g_{Q}}\right)^{2}	 I_{0}( 2 \sqrt{v_{-}}) \right].
\end{align}

In the steady-state, we approximate these expressions by assuming the correlation functions are time-translation invariant. Thus, we may write, for instance,

\begin{align}
g_{R} = \int dt dt' R_{t} R_{t'} \approx t^{2} \frac{1}{t}\int d\tau	C_{R}(\tau) = t^{2} \tau_{R},
\end{align}

and similarly for $g_{Q}$ and $g_{P}$. Then the eigenvalues of the transfer matrix become
\begin{align}
v_{\pm} = t^{2} \frac{1}{2} \left( \tau_{R} \pm \sqrt{\tau_{R}^{2} + 4 \tau_{P} \tau_{Q}}\right)	.
\end{align}

At late times, using the asymptotic behavior of the modified Bessel function, the moment becomes 

\begin{align}
\sigma_{\chi}^{2} \sim  \exp \left( - 2 t + 2 \sqrt{v_{+}}\right)	,
\end{align}

which gives the Lyapunov exponent

\begin{align}
\lambda_{L} \ge  \left( \frac{\tau_{R} + \sqrt{\tau_{R}^{2} + 4 \tau_{P}\tau_{Q}}}{2} \right)^{1/2} - 1	.
\end{align}

where the relaxation times $\tau_A,\tau_R,\tau_Q$ are defined as
\begin{align}
\tau_{R} &= \lim_{t \to \infty} \frac{1}{t}\int_{0}^{t} d\tau C_{\phi'}(\tau) C_{\sigma_{r}}(\tau),\\
\tau_{A} & =  \lim_{t \to \infty} \frac{1}{t}\int_{0}^{t} d\tau C_{\phi'}(\tau) ,\\
\tau_{Q} &= \lim_{t \to \infty} \frac{1}{t}\int_{0}^{t} d\tau C_{\phi}(\tau) C_{\sigma_{r}'}(\tau).
\end{align}

%The relations we find in this section relate timescales associated with two important characteristic functions: the two-point correlation function, and the response function. In equilibrium statistical mechanics, one expects these timescales to be related via the fluctuation-dissipation theorem. However, for the random neural networks with asymmetric coupling, it is not clear a priori that such a relation should hold. This results 

% ******************* Discontinuous chaotic transition appendix******************

\section{Details of the discontinuous chaotic transition} \label{app:DMFT_BF}

In this section, we provide the details for the calculations involved in the discontinuous chaotic transition.

\subsection{Spontaneous emergence of fixed-points}
\label{app:subsec:FP_BF_transition}

For $g_h < 2.0$ and small $\alpha_r$, the zero fixed-point is the globally stable state for the dynamics and the only solution to the fixed-point equations, eq. \ref{eq:MFT_FP_variance_1}, for $\Delta_h$. However, as we increase $\alpha_r$ for a fixed $g_h$, two additional non-zero solutions to
$\Delta_h$ spontaneously appear at a critical value $\alpha^{*}_{FP}(g_h)$ as shown in Fig. \ref{fig:DMFT_bifurcation}a. Numerical solutions to the fixed-point equations reveal the form
of the bifurcation curve $\alpha^{*}_{r,FP}(g_h)$ and the associated
value of $\Delta_h^{*}(g_h)$. We see that $\alpha^{*}_{r,FP}(g_h)$ increases rapidly with decreasing $g_h$, dividing the parameter space into regions with either 1 or 3 solutions for $\Delta_h$. The corresponding $\Delta_h^{*}(g_h)$  vanishes at two boundary values of $g_h$ -- one at 2.0 and another, $g_c$, below 1.5 where $\alpha_r^{*} \rightarrow \infty$. 
This naturally leads to the question of whether the fixed-point bifurcation exists for all values of $g_h$ below 2.0.

To answer this, we perturbatively solve the fixed-point equations in two asymptotic regimes: i)  $g_h \rightarrow 2^-$ and  ii) $g_h \rightarrow g_c^{+}$. Details of the perturbative treatment are in Appendix \ref{app:FP_perturbative_BF}. For $g_h=2-\epsilon$, we see that the perturbative problem undergoes a bifurcation from one solution ($\Delta_h=0$) to three when $\alpha_r$ crosses the bifurcation threshold $\alpha_r^{*}(2.0) = \sqrt{8}$, and this is the left limit of the bifurcation curve in Fig. \ref{fig:DMFT_bifurcation}b. The larger non-zero solution for the variance at the bifurcation point scales as
\begin{align}
    \Delta_h^{*} \approx (\alpha_r^2 - 8)\cdot \xi_0 + \xi_1 \epsilon \quad \textrm{ for } \alpha_r \rightarrow \alpha_{r,FP}^*(2) = \sqrt{8},
\end{align}
where $\xi_0$ and $\xi_0$ are positive constants (see Appendix \ref{app:FP_perturbative_BF}).

At the other extreme, to determine the smallest value of $g_h$ for which a bifurcation is possible, we note from Fig. \ref{fig:DMFT_bifurcation}b that in this limit $\alpha_r \rightarrow \infty$, and thus we can look for solutions to $\Delta_h$ in the limit: $\Delta_h \ll 1$ and $\alpha_r \rightarrow \infty$ and $\alpha_r \sqrt{\Delta_h} \gg 1$. In this limit, there is a 
bifurcation in the perturbative solution when $g_h > g_h^* =  \sqrt{2}$, and close to the critical point, the fixed-point solution is given by (see Appendix \ref{app:FP_perturbative_BF}):
\begin{align}
    \Delta_h^*(\sqrt{2}^{\:+}) \sim \frac{g_h^2 - 2}{2g_h^4} \qquad \textrm{ for } g_h \rightarrow  \sqrt{2}^{\: +}.
\end{align}

Thus in the region $g_h \in (\sqrt{2},2)$ there exist non-zero solutions to the fixed-point equations once $\alpha_r$ is above a critical value $\alpha_r^{*}(g_h)$. These solutions correspond to unstable fixed-points appearing in the phase space.

%----------------------- TRANSIENT TIME IGURE

\begin{figure*}
\begin{centering}
\includegraphics[scale=0.4
]{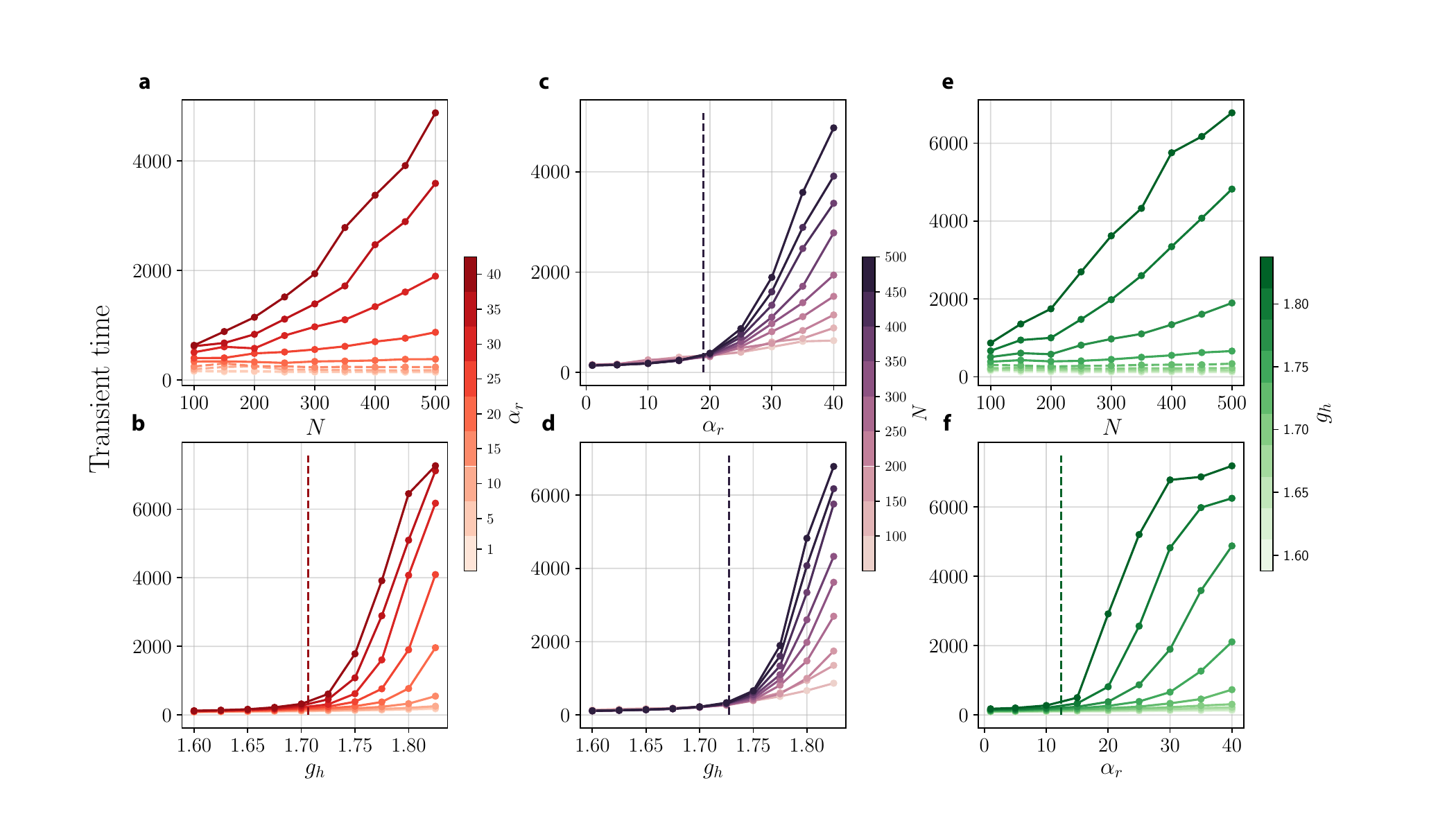}
\par\end{centering}
\caption{\label{fig:BF_transient_times} {\it Transient times at the bifurcation transition} Transient times($\tau_T$, relative to $\tau_h$) as a function of $g_h$, $\alpha_r$ and  system size $N$ Dashed plot lines correspond to situations where $\alpha_r < \alpha_{r,DMFT}^{*}(g_h)$. Dashed vertical lines are critical values of $\alpha_r$ or $g_h$ (a) $\tau_T$ vs. $N$ for $g_h=1.775$ (b) $\tau_T$ vs. $g_h$ for $N=500$; dashed line indicated $g_h$ such that $\alpha_{r,DMFT}^{*}(g_h) = 40$. (c) $\tau_T$ vs. $\alpha_r$ for $g_h=1.775$; dashed vertical line is $\alpha_{r,DMFT}^{*}(1.775)$. (d) $\tau_T$ vs. $g_h$ for $\alpha_r=30$; dashed line is $g_h$ such that $\alpha_{r,DMFT}^{*}(g_h) = 30$. (e) $\tau_T$ vs. $N$ for $\alpha_r=30$; dashed plot lines correspond to situations where $30 < \alpha_{r,DMFT}^{*}(g_h)$. (f) $\tau_T$ vs. $\alpha_r$ for $N=500$; dashed vertical line is $\alpha_{r,DMFT}^{*}(1.85)$
Transient times are averaged over 2000 instances of random networks. 
}
\end{figure*}

\subsection{Delayed dynamical transition shows a decoupling between topological and dynamical complexity}
\label{app:subsec:DMFT_transition}

The picture from the fixed-point transition above is that when $g_h$ is in the interval $(\sqrt{2},2)$, there is a proliferation of unstable fixed-points in the phase space provided $\alpha_r > \alpha_{r,FP}^{*}(g_h)$. However, 
it turns out that the spontaneous appearance of these unstable fixed-points is not accompanied by any  asymptotic dynamical signatures -- as measured by the Lyapunov exponents (see Fig. \ref{fig:DMFT_bifurcation}) or by the transient times (see Fig. \ref{fig:BF_transient_times}). It is  only when $\alpha_r$ is increased further beyond a second critical value $\alpha_{r,DMFT}^{*}(g_h)$, that we see the appearance of 
chaotic and long-lived transients. 
This is significant in regard to a result by Wainrib \& Touboul \cite{wainrib2013topological}, where they showed that the transition to chaotic dynamics (dynamical complexity) in random RNNs is tightly linked to the proliferation of critical points (topological complexity), and in their case, the exponential rate of growth of critical points (a topological property) was the same as the maximal Lyapunov exponent (a dynamical property).  

Let us characterize the second dynamical transition curve given by 
$\alpha_{r,DMFT}^{*}(g_h)$ (Fig. \ref{fig:DMFT_bifurcation}c, red curve).
For ease of discussion, we turn off the update gate $(\alpha_z=0)$ and introduce a functional $F_{\psi}$ for a 2-D Gaussian average of a given function $\psi(x)$:
\begin{align} \label{eq:F_Gaussian_int}
    & F_{\psi}\left(
    C_h(0), C_h(\tau)
    \right) =  \mathbb{E}\left[  
    \psi(z_1) \psi(z_2)  
    \right] ,  \\
    & \textrm{  where  }
    \begin{pmatrix}
    z_1\\
    z_2
    \end{pmatrix}
    \sim
    \mathcal{N}\left(
    \mathbf{0},{\bf C}_{h}
    \right),
    \quad 
    {\bf C}_{h} =
    \begin{pmatrix}
    C_h(0) & C_h(\tau)&\\
    C_h(\tau) & C_h(0)& 
    \end{pmatrix}.
\end{align}
The DMFT equations for the correlation functions then become
\begin{align}
  \frac{1}{4}C_h(\tau) -  \partial^2_{\tau} C_h(\tau) = & \:
    \frac{1}{4}F_{\phi}(C_h(0), C_h(\tau)) 
    F_{\sigma_r}(C_r(0), C_r(\tau)), \nonumber \\
    C_r(\tau) -  \tau_r^2\partial^2_{\tau} C_r(\tau) = & \:
    F_{\phi}(C_{\phi}(0), C_{\phi}(\tau)).
\end{align}
We further make an approximation that $\tau_r \ll 1$, which in turn  implies $C_r(\tau) \approx C_{\phi}(\tau)$. This approximation turns out to hold even for moderately 
large $\tau_r$. With these approximations, we can integrate the equations for $C_h(\tau)$ to arrive at an equation for the variance $C_h^0 \equiv C_h(0)$. We do this by 
multiplying by $\partial_{\tau}C_h(\tau)$ and integrating from $\tau$ to $\infty$, we get
\begin{align}
    & \frac{1}{2}\dot{C}_h(\tau)^2  = \frac{1}{4} \frac{(C_h)^2}{2} 
    - \frac{1}{4}\int_{0}^{C_{h}^0} d C_{h}  F_{\phi}\left(C_{h}, C_{h}^{0}\right) F_{\sigma_{r}}\left( C_{\phi}, C_{\phi}^{0}\right) .
\end{align}
Using the boundary condition that $\dot{C}_h(0) =0$, we get the equation for the variance
\begin{align}\label{eq:DMFT_BF_root}
\frac{1}{8} C_{h}(0)^{2}  - \frac{1}{4}\int_{0}^{C_{h}^0} d C_{h}  F_{\phi}\left(C_{h}, C_{h}^{0}\right) F_{\sigma_{r}}\left( C_{\phi}, C_{\phi}^{0}\right)  = 0.
\end{align}

Solving this equation will give the DMFT prediction for the variance for any $g_h$ and $\alpha_r$. Beyond the critical value of $\alpha_r$, two non-zero solutions for $C_h^0$ spontaneously emerge. In order to use eq. \ref{eq:DMFT_BF_root} to find a prediction for the DMFT bifurcation curve $\alpha_{r,DMFT}^{*}(g_h)$, we need to use the additional fact that at the bifurcation point the two solutions coincide, and there is only one non-zero solution. To proceed, 
we can view the L.H.S of eq. \ref{eq:DMFT_BF_root}, as a function of $\alpha_r$, $g_h$ and $C_h^0$: $\mathcal{F}(g_h,\alpha_r,C_h^0)$. Then the equation
for the bifurcation curve is obtained by solving 
the following two equations for $C_h^{0,*}$ and $\alpha_r^*$
\begin{align}\label{eq:DMFT_BF_eqn_1}
   \mathcal{F}(g_h,\alpha_r^*,C_h^{0,*}) = & \: \: 0 ,\\
   \frac{\partial \mathcal{F}(g_h,\alpha_r,C_h^{0}) }{\partial C_h^0} \Biggr|_{\alpha_r^*,C_h^{0,*}} = & \: \: 0 .\label{eq:DMFT_BF_eqn_2}
\end{align}

To get the condition for the dynamical bifurcation 
transition, we need to differentiate the L.H.S of 
eq. \ref{eq:DMFT_BF_root} ($\mathcal{F}\left(g_{h}, \alpha_{r}, C_{h}^{0}\right)$ ) w.r.t $C_h^0$ and set it to 0. This involves terms like
\begin{align}
    \frac{\partial F_{\psi}(C_h^0,C_h^0) }
    {\partial C_h^0 }\quad ; 
    \frac{\partial F_{\psi}(C_h^0,0) }
    {\partial C_h^0 }.
\end{align}
We give a brief outline of calculating the first term. It's easier to work in the Fourier domain:
\begin{align}
     & F_{\psi}(C_h^0,C_h)  =  \mathbb{E} \left[ \int \frac{dk}{2\pi} \int \frac{dk'}{2\pi}
    \widetilde{\psi}(k)e^{-kz_1} 
    \widetilde{\psi}(k')e^{-k'z_2}
    \right] ,\nonumber \\
    & = \int \frac{dk}{2\pi} \int \frac{dk'}{2\pi}
    \widetilde{\psi}(k)\widetilde{\psi}(k')
    \exp\left[ -\frac{C_h^0}{2}(k^2+k'^2) - C_h(\tau)k k' \right].
\end{align}
This immediately gives us,
\begin{align}
    \frac{\partial F_{\psi}(C_h^0,C_h^0) }
    {\partial C_h^0 } = & \:
    \int \mathcal{D}x \: \psi(\sqrt{c_h^0} x)
    \psi^{\prime \prime}(\sqrt{c_h^0} x) + \nonumber \\
    & \quad \int \mathcal{D}x \: \psi^{\prime}(\sqrt{c_h^0} x)^2 , \nonumber \\
    \frac{\partial F_{\psi}(C_h^0,0) }
    {\partial C_h^0 } = & \:
    \int \mathcal{D}x \: \psi(\sqrt{c_h^0} x)
    \int \mathcal{D}x \: 
    \psi^{\prime \prime}(\sqrt{c_h^0} x).
\end{align}
Using this fact, we can calculate the derivative of $\mathcal{F}\left(g_{h}, \alpha_{r}, C_{h}^{0}\right)$ as a straightforward (but long!) sum of Gaussian integrals. We then numerically solve eqns. \ref{eq:DMFT_BF_eqn_1},\ref{eq:DMFT_BF_eqn_2} to get the 
bifurcation curve shown in Fig. \ref{fig:DMFT_bifurcation}c. Fig. \ref{fig:DMFT_bifurcation}d shows the corresponding
variance at the bifurcation point $C_h^{0,*}$ (red curves). We note two salient points: i) the DMFT bifurcation curve is 
always above the fixed-point bifurcation curve (black, in Fig. \ref{fig:DMFT_bifurcation}a) and ii) 
the lower critical value of $g_h$ which permits a dynamical transition (dashed green curve in 
Fig. \ref{fig:DMFT_bifurcation}a,b) is smaller than the corresponding fixed-point critical value of $\sqrt{2}$.

We now calculate the lower critical value of $g_h$ and provide an analytical description of the asymptotic behaviour near the lower and higher critical values of $g_h$. From the red curve in Fig. \ref{fig:DMFT_bifurcation}c we know that as $g_h$ tends towards the lower critical value, $\alpha_{r,DMFT}^{*} \rightarrow \infty$ and $C_h^0 \rightarrow 0$. So, we can approximate $\sigma_r$ as a step function in this limit, and $F_{\sigma_r}$ is approximated as
\begin{align}
F_{\sigma_r}(C_{\phi}^0,C_{\phi}) \approx &
\frac{1}{4}+\frac{1}{2 \pi} \tan ^{-1}\left(\frac{x}{\sqrt{1-x^{2}}}\right), \\
\textrm{where    } x := & \frac{C_h(\tau)}{C_h(0)}  \approx \frac{C_{\phi}(\tau)}{C_{\phi}(0)}.
\end{align}
The DMFT equation then reads:
\begin{align}\nonumber
    4 \ddot{x}=x-g_{h}^{2} x \left(\frac{1}{4}+\frac{1}{2 \pi} \tan ^{-1}\left(\frac{x}{\sqrt{1-x^{2}}}\right)\right) + O(C_{h}(0)^{2}).
\end{align}
Integrating, this equation we get
\begin{align}
    2 \dot{x}^{2}=\frac{x^{2}}{2}\left(1-\frac{g_{h}^{2}}{4}\right)+\frac{g_{h}^{2}}{8 \pi}\left[\left(1-2 x^{2}\right) \sin ^{-1}(x) - x \sqrt{1-x^{2}}\right],\nonumber
\end{align}
which will have $O(C_{h}(0)^{2})$ corrections. From the boundary condition $\dot{C}_h(0)=0$, we know that as $x \rightarrow 1$ then $\dot{x} \rightarrow 0$. We thus find that these boundary conditions are only consistent to leading order in $C_{h}(0)$ when $g_{h}$ is equal to its critical value:
\begin{align}
    g_h^* = \sqrt{ \frac{8}{3} }.
\end{align}

Which indicates that $C_{h}(0)$ must vanish as $g_{h} \to \sqrt{8/3}^{+}$. 

In the other limit when $g_h \rightarrow 2^-$, we see that $\alpha_r^*$ remains finite and $C_h^{0,*} \rightarrow 0$. We assume that for $g_h = 2- \epsilon$, $C_h^0$ has a power-series expansion
\begin{align}
    C_h^0 = c_0 + c_1 \epsilon + c_2 \epsilon^2  + ....
\end{align}
We also expand $F_{\phi}$ and $F_{\sigma_r}$ to $O(C_h(0)^2)$ 
\begin{align}
    F_{\phi} \approx g_{h}^{2} C_{h}(\tau)-2 g_{h}^{4} C_{h}^{0} \cdot C_{h}(\tau)+5 g_{h}^{6}\left(C_{h}^{0}\right)^{2} \cdot C_{h}(\tau), 
\end{align}

 and look for values of $\alpha_r$ which permit a non-zero value for $c_0$ in the leading order solutions to the DMFT. We find that critical value of $\alpha_r$ from the perturbative solution is given by 
 \begin{align}
     \alpha_{r,DMFT}^{*}(2) = \sqrt{12}.
 \end{align}

The DMFT prediction for the dynamical bifurcation agrees well with the full network simulations. In Fig. \ref{fig:DMFT_bifurcation}e we see that the maximum Lyapunov exponent experiences a discontinuous transition from a negative value (network activity decays to fixed-point) to a positive value (activity is chaotic) at the critical value of $\alpha_r$ predicted by the DMFT (dashed vertical lines).

\subsection{Influence of update gate on the discontinuous transition}\label{app:z-gate-BF-line}

Here we comment briefly on the possible influence of the z-gate on the discontinuous dynamical phase transition given by the curve $\alpha_{r,DMFT}^{*}$. Assuming Eq.\ref{eq:DMFT_5} is valid (discussed in more detail toward the end of Appendix \ref{app:DMFT_derivation}), we may rewrite the DMFT equation for the two-point correlation functions as

\begin{align}
     \frac{1}{2} \dot{C}_{h}(\tau)^{2} % =& \int_{0}^{C_{h}} d C_{h}  C_{h} F_{\sigma_{z}}(C_{\phi}^{0}, C_{\phi})\\
    % & - \int_{0}^{C_{h}}  dC_{h} F_{\phi}(C_{h}, C_{h}^{0})  F_{\sigma_{r}}(C_{\phi}, C_{\phi}^{0}) F_{\sigma_{z}} (C_{\phi},C_{\phi}^{0})\\
     & = \int_{0}^{C_{h}(\tau)} F_{\sigma_{z}}(C_{\phi}^{0}, C_{\phi}) Q(C_{h}, C_{h}^{0})
\end{align}

where
\begin{align}
    Q(C_{h}, C_{h}^{0}) = C_{h} - F_{\phi}(C_{h}, C_{h}^{0}) F_{\sigma_{r}}(C_{\phi}, C_{\phi}^{0}).
\end{align}

Noting that a time-dependent solution corresponds to a non-zero solution for $C_h(0)$ and  satisfies the boundary condition $\dot{C}_h(0) = 0$ then requires

\begin{align}
    \mathcal{F}_{\alpha_{z}} \equiv  \int_{0}^{C_{h}^{0}} dC_{h} F_{\sigma_{z}}(C_{\phi}, C_{\phi}^{0}) Q(C_{h}, C_{h}^{0}) = 0.
\end{align}

Were we have defined a new ``potential" function which is related to that defined above by

\begin{align}
    \mathcal{F}_{\alpha_{z}}\Big|_{\alpha_{z} = 0} =\mathcal{F} =  \frac{1}{4}\int_{0}^{C_{h}^{0}} dC_{h}  Q(C_{h}, C_{h}^{0}).
\end{align}
We have left the arguments $(g_{h}, \alpha_{r}, C_{h}^{0})$ implicit, for ease of presentation. We proceed to bound the new potential by establishing bounds on $F_{\sigma_{z}}$. To be explicit, we have

\begin{align}
    F_{\sigma_{z}} = \langle \sigma_{z}(\tau) \sigma_{z}(0) \rangle = \langle \sigma_{z}(\tau) \sigma_{z}(0)\rangle_{c} + \langle \sigma_{z}\rangle^{2},
\end{align}
which we express as the sum of a connected component (indicated by a subscript $c$) and a disconnected component. We can consider two limiting behaviors. When the correlation time tends to zero, the connected component will vanish and (at zero bias $\beta_{z} = 0$)
\begin{align}
    F_{\sigma_{z}} \approx \langle \sigma_{z} \rangle^{2}  = \frac{1}{4}. 
\end{align}
Increasing the correlation time can only serve to increase the two-point function, since $\sigma \ge 0$. In the extreme limit of very long correlation time, we have that

\begin{align}
    F_{\sigma_{z}} \approx \langle \sigma_{z}^{2}(0)\rangle \le \frac{1}{2}.
\end{align}
The inequality is saturated at $\alpha_{z} = \infty$, when $\sigma_{z}$ becomes a step function of its argument. Therefore, the two-point correlation function of the update gate is bounded above and below

\begin{align}
    \frac{1}{4} \le F_{\sigma_{z}} \le \frac{1}{2},
\end{align}
and this bound is uniform in the sense that it holds for all values of the argument $0\le C_{h} \le C_{h}^{0}< \infty$. Consequently, we are able to bound the potential 
\begin{align}
    \frac{1}{4} \mathcal{F} \le \mathcal{F}_{\alpha_{z}} \le \frac{1}{2} \mathcal{F}.
\end{align}
It follows immediately that the derivative is similarly bounded. Consequently, the zeros of $\mathcal{F}_{\alpha_{z}}$  and $\partial \mathcal{F}_{\alpha_{z}}/\partial C_{h}^{0}$ will coincide with the zeros of $\mathcal{F}$ and $\partial \mathcal{F}/\partial C_{h}^{0}$, respectively. As a result, the discontinuous transition, determined by (\ref{eq:DMFT_BF_eqn_1}) and (\ref{eq:DMFT_BF_eqn_2}), will remain unchanged for values of $\alpha_{z}$ for which (\ref{eq:DMFT_5}) is valid. Thus, for moderately large $\alpha_z$ ($\sim 10)$, where \ref{eq:DMFT_5} is valid, the critical line for the discontinuous transition will remain unchanged.

%*************** ROLE OF  BIASES ********************

\section{The role of biases} \label{app:role_of_biases}

We have thus far described the salient dynamical aspects for the gated RNN in the absence of biases. Here we describe
the role of the biases $\beta_h$ (bias of the activation $\phi$) and $\beta_r$ (bias of the output gate $\sigma_r$). We first
note that when $\beta_h = 0$, zero is always a fixed-point 
of the dynamics, and  the zero fixed-point is stable provided

\begin{align} \label{eq:stability-zeroFP}
   -1+\phi^{\prime}(0) \sigma_{r}(0) < 0  
\end{align}
where $\phi(x)=\tanh(g_hx + \beta_h)$. This gives the familiar $g_h < 2$ condtion when $\beta_r=0$ \footnote{in previous work, $g = 1$ sets the critical value. The difference is simply due to the factor $\sigma_{r}(0) = 1/2$. The vanilla RNN result is recovered by sending $\beta_{r} \to \infty$}.
Thus, in this case, there is an interplay between $g_h$ and $\beta_r$ in determining the leading edge of the Jacobian around the zero fixed-point, and thus its stability. In the limit $\beta_r \rightarrow -\infty$ the  leading edge retreats to $-\tau_r^{-1}$. 
When $\beta_h > 0$, zero cannot be a fixed-point of 
the dynamics. Therefore, $\beta_h$ facilitates the appearance of non-zero fixed-points, and both $\beta_r$ and $\beta_h$
will determine the stability of these non-zero fixed-points.

% ---------------- ROLE OF BIASES --------------------
\begin{figure*}
\begin{centering}
\includegraphics[scale=0.45
]{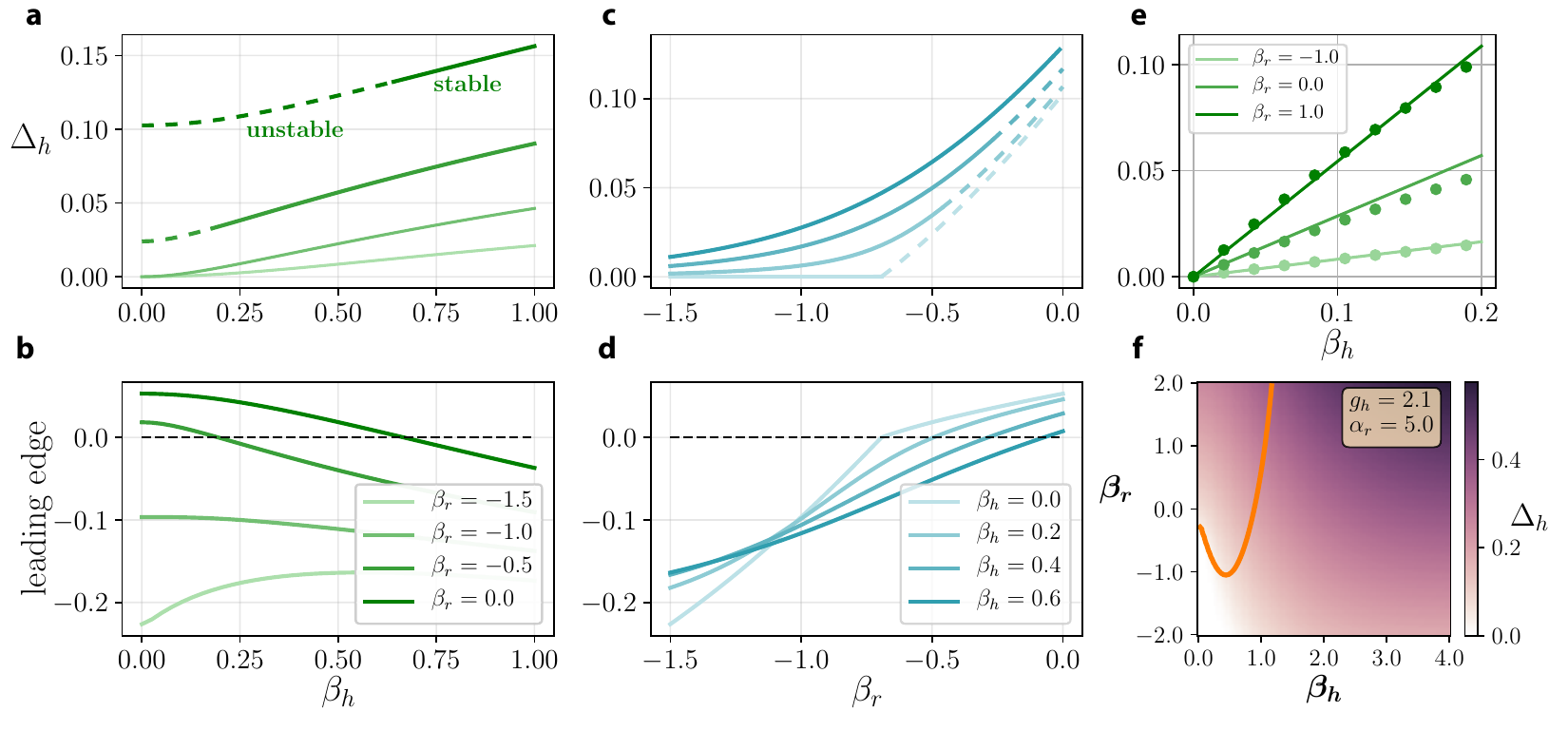}
\par\end{centering}
\caption{\label{fig:role_of_biases} { \it The role of biases}
(a) FP solutions as a function of increasing $\beta_h$; different
shades of green correspond to different values of $\beta_r$. 
Dashed lines correspond to FP solutions that are unstable (time-varying states). (b) The leading edge of the spectrum corresponding to the FP solutions calculated in (a); the FP solution is unstable when the leading edge is positive. 
(c) similar to (a) but for $\beta_r$; different shades of blue
correspond to different values of $\beta_h$. (d) similar to (b) but for $\beta_r$. (e) FP solutions near critical $g_c$ where the zero FP becomes unstable (circles) compared with the perturbative solution predicted by 
eq. \ref{eq:FP-perturbative-bias} (solid lines).
(f) FP solution as a function of $\beta_r$ and $\beta_h$. Orange line indicates stability line -- i.e. regions on top of the orange line correspond to unstable/time-varying states. 
}
\end{figure*}

To gain some insight into the role of $\beta_h$ in generating
fixed-points, we treat the mean-field FP equations (eq. \ref{eq:MFT_FP_variance_1}) perturbatively around the 
operating point $g_c$ where the zero fixed-point becomes unstable (eq. \ref{eq:stability-zeroFP}). For small $\beta_h$ and $\epsilon = g_h - g_c$, we can express the solution $\Delta_h$ as a power series in $\epsilon$, and we see that to leading order
the fixed-point variance behaves as (details in Appendix \ref{app:FP_perturbative_bias}):
\begin{align} \label{eq:FP-perturbative-bias}
    \Delta_h \approx & \begin{dcases}
            \frac{ \beta_h + \epsilon}
            {g_c^2(2 - g_c^2 a_1)} & \quad
            g_c^2 a_1 < 2 \\
            & \\
            \left(
            g_c^2 a_1 - 2
            \right)f_1
             +
            \epsilon \cdot f_2
            & \quad
            g_c^2 a_1 > 2
    \end{dcases} \\
    \textrm{ where  } a_1 = & \: \frac{\alpha_r^2}{16}\left[ 
    \phi_0^{(1)}(\beta_r/2)^2
    + 
    \phi_0(\beta_r/2)
    \phi_0^{(2)}(\beta_r/2)
    \right]
\end{align}
where $\phi_0 \equiv \tanh$ and $f_2(\alpha_r,\beta_r)$ and $f_2(\alpha_r,\beta_r)$ are constant functions w.r.t $\epsilon$. Therefore, we see that the bias $\beta_h$ gives rise to non-zero fixed-points near the critical point which scale linearly with the bias. In Fig. \ref{fig:role_of_biases}e, 
we show this linear scaling of the solution for the case when
$\beta_h = \epsilon$, and we see that the prediction (lines) matches the true solution (circles) over a reasonably wide range. 

More generally, away from the critical $g_c$, an increasing
$\beta_h$ gives rise to fixed-point solutions with increasing variance, and this can arise continuously from zero, or it can arise by stabilizing an unstable, time-varying state depending on the value of $\beta_r$. In Fig. \ref{fig:role_of_biases}a we see how the $\Delta_h$ behaves for increasing $\beta_h$ for different $\beta_r$, and we can see the stabilizing effect of $\beta_h$ on unstable solutions by looking at its effect on the leading spectral edge (Fig. \ref{fig:role_of_biases}b). In Fig. \ref{fig:role_of_biases}c, we see that an increasing $\beta_r$ also gives rise to increasing $\Delta_h$. However in this case, it has a destabilizing effect by shifting the leading spectral edge to the right. In particular, when $\beta_h=0$, increasing $\beta_r$ destabilizes the zero fixed-point and give rise to a time-varying solution. We note that when $\beta_h=0$, varying $\beta_r$ cannot yield stable non-zero FPs. The combined effect of $\beta_h$ and $\beta_r$ can been seen in Fig. \ref{fig:role_of_biases}f where the  non-zero solutions to the left of the orange line indicate unstable (time-varying) solutions. We have chosen the parameters to illustrate an interesting aspect of the biases: in some cases, increasing $\beta_h$ can have a non-monotonic effect on the stability, wherein the solution becomes unstable with increasing $\beta_h$ and is then eventually stabilized for sufficiently large $\beta_h$.

\subsection{Effect of biases on the phase boundaries}

%----------------------- PHASE DIAGRAM arVgh with BIASES -------

\begin{figure}
\begin{centering}
\includegraphics[scale=0.4
]{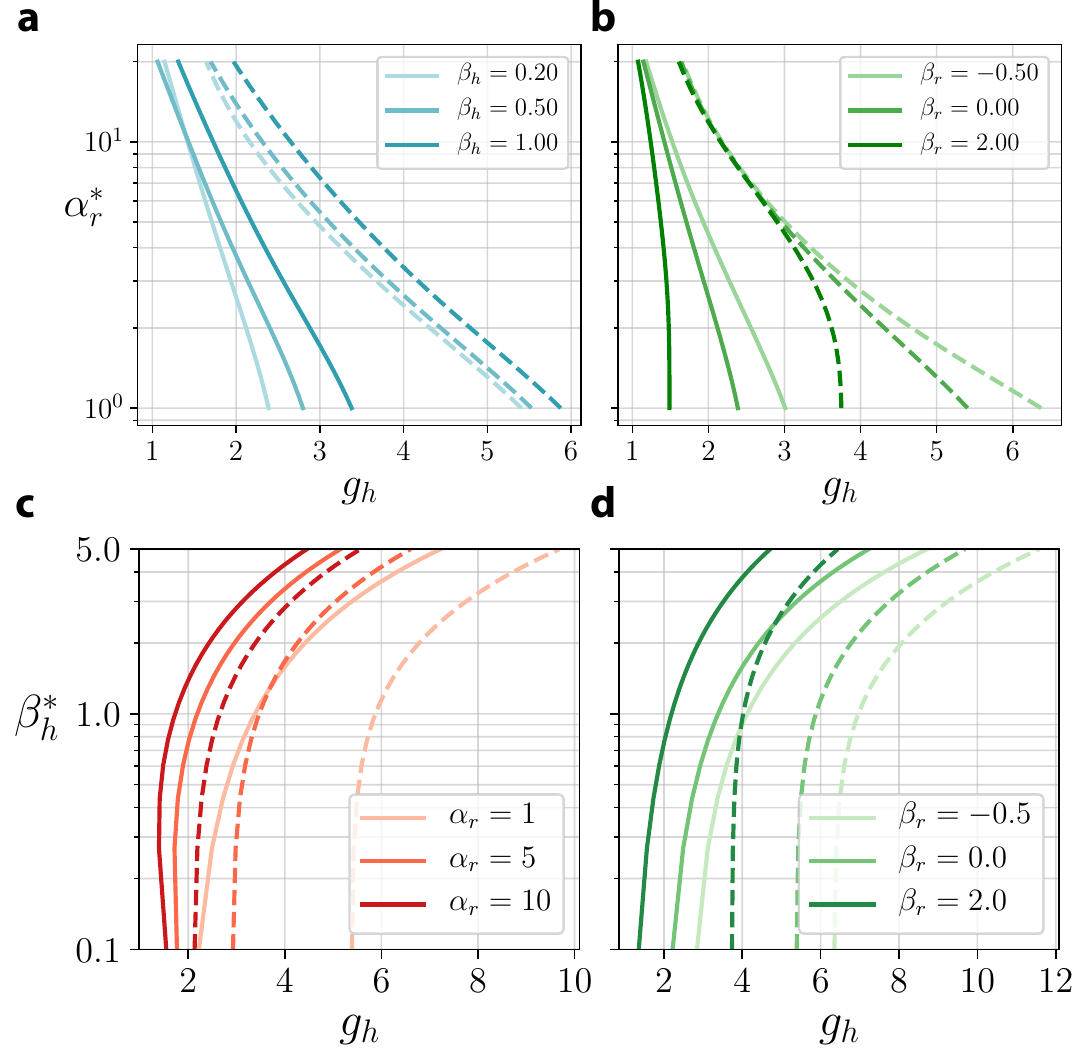}
\par\end{centering}
\caption{\label{fig:phaseDiag_arVgh_withBias} {\it How the
biases alter the transition between stability and chaos}
(a) Critical lines indicating boundaries for stability (solid lines) or marginal stability (dashed lines) for different values of $\beta_h$. (b) Similar to the (a) panel but for different values of $\beta_r$. (c,d): how the boundaries of stability (solid lines) or marginal stability (dashed lines) change as we vary $\alpha_r$ (c) or $\beta_r$ (d). 
}
\end{figure}

In Fig. \ref{fig:phaseDiag_arVgh_withBias},a,b we look at how the critical line for the chaotic transition, in the $\alpha_r - g_h$ plane, changes as we vary $\beta_h$ (a) or $\beta_r$ (b).
Positive values of $\beta_r$ (``open'' output gate) tend to make the transition line less dependent on $\alpha_r$ (Fig. \ref{fig:phaseDiag_arVgh_withBias}, b), and negative values of $\beta_r$ have a stabilizing effect by requiring larger values of $g_h$ and $\alpha_r$ to transition to chaos.
As we have seen above,  higher values of $\beta_h$ have a stabilizing effect, requiring higher $g_h$ and $\alpha_r$ to make the (non-zero) stable fixed-point unstable. In both cases, the critical lines for marginal stability (Fig. \ref{fig:phaseDiag_arVgh_withBias},a,b dashed lines) are also influenced in a similar way.  In Fig. \ref{fig:phaseDiag_arVgh_withBias}, (c,d), we see how the stability-to-chaos transition is affected by $\alpha_r$ (c) and $\beta_r$ (d). Consistent with the discussion above, larger $\alpha_r$ and $\beta_r$ have a destabilizing effect, requiring a larger $\beta_h$ to make the system stable.

% ***************** PERTURBATIVE SOLNS ************

\section{Details of the perturbative solutions to the mean-field equations}

%---------- perturbative solns. to FP equations ----

\subsection{Perturbative solutions for the fixed-point variance $\Delta_h$ with biases} \label{app:FP_perturbative_bias}

In this section, we derive the  perturbative solutions for the fixed-point variance $\Delta_h$
with finite biases, near the critical point where the zero fixed-point becomes unstable. 
Recall, that fixed-point variances are obtained by solving
\begin{align}\label{eq:MFT_FP_variance}
\Delta_{z} \equiv \left\langle z^{2}\right\rangle = & \: \int \mathcal{D} x \: \phi\left( \sqrt{\Delta_{h}} x\right)^{2}  = \: \Delta_{r}\\
\Delta_{h} \equiv  \left\langle h^{2}\right\rangle = & \: \int \mathcal{D}x \mathcal{D}y \: \phi\left( \sqrt{\Delta_{h}} x\right)^{2}\sigma_{r}(\sqrt{\Delta_{r}} y)^{2}
\end{align}
The expansion we seek is perturbative in $\Delta_h$. So, expanding the gating and activating functions about their biases under the assumption  $\Delta_r \approx g_h^2 \Delta_h$, we have a series expansion to $O(\Delta_h^2)$

\begin{align}
   &  \left \langle \sigma_{r}(\sqrt{\Delta_{r}} x)^{2}
    \right \rangle_x =  \: \quad a_0 + a_1 g_h^2 \Delta_h + a_2 g_h^4 \Delta_h^2 \nonumber \\
    & a_0 =  \: \frac{1}{4} \left[ 
    1 + \phi_0(\beta_r/2)
    \right]^2 \\
    & a_1 =  \: \frac{\alpha_r^2}{16}\left[ 
    \phi_0^{(1)}(\beta_r/2)^2
    + \right. \nonumber \\
    & \quad \left.
    \phi_0(\beta_r/2)
    \phi_0^{(2)}(\beta_r/2)
    + \phi_0^{(2)}(\beta_r/2)
    \right] \\
   &  a_2 =  \: \frac{\alpha_r^4}{256} 
    \left[ 
    12 \phi_0^{(2)}(\beta_r/2)^2 + 
    4 \phi_0(\beta_r/2) 
    \phi_0^{(4)}(\beta_r/2) \right. \nonumber \\
    & \quad  \left.
    + 16 \phi_0^{(1)}(\beta_r/2)
    \phi_0^{(3)}(\beta_r/2)
    +  \phi_0^{(4)}(\beta_r/2)
    \right]
\end{align}

where we have used the following identities involving 
the derivatives of $\tanh$:
\begin{align}
    \phi_0(x) = & \: \tanh(x) \\
    \phi_0^{(1)}(x) = & \: 1 - \phi_0(x)^2 \\
    \phi_0^{(2)}(x) = & \: 
    -2\phi_0(x) \left( 1- \phi_0(x)^2 \right)  \\
    \phi_0^{(3)}(x) = & \:
    2\left( 1- \phi_0(x)^2 \right)
    \left( 3\phi_0(x)^2 - 1\right) \\
     \phi_0^{(4)}(x) = & \:
     -8 \phi_0(x) \left( 1- \phi_0(x)^2 \right)
     \left( 3\phi_0(x)^2 - 2\right)
\end{align}

This gives us to $O(\Delta_h^2)$
\begin{align}
  &  \Delta_{h} \approx   \:
    \left[c_0 + c_1 \Delta_h + c_2 \Delta_h^2 \right]
    \left \langle \sigma_{r}(\sqrt{\Delta_{r}} x)^{2}
    \right \rangle_x \\
    & c_0 =  \: \phi_0(\beta_{h})^2 \\
     c_1 = & \: g_h^2
    \left[
    \phi_0^{(1)}(\beta_{h})^2
    + \phi_0^{(2)}(\beta_{h})
    \phi_0(\beta_{h}) 
    \right] \\
   & c_2 =  \:      
    g_h^4 \left[
    \frac{1}{4}\phi_0(\beta_{h})\phi_0^{(4)}(\beta_{h})
    + \right. \nonumber \\
    & \left. 
    \phi_0^{(1)}(\beta_{h})\phi_0^{(3)}(\beta_{h})
    +
    \frac{3}{4}\phi_0^{(2)}(\beta_{h})^2
    \right]
\end{align}
and therefore, 
\begin{align}\label{eq:FP_perturb_1}
    \Delta_{h} \approx  & \: 
    \left( c_0 + c_1 \Delta_h + c_2 \Delta_h^2  \right) 
    \left( a_0 + a_1 g_h^2 \Delta_h + a_2 g_h^4 \Delta_h^2 \right) 
\end{align} 
To proceed further, we study the solutions to this equation for small deviations for a critical value of $g_h$. Which critical value should we use? 
Recall, that the zero fixed-point becomes unstable when 
\begin{align}
    -1 + \phi^{\prime}(0)\sigma_r(0) = 0
\end{align}

Therefore, we expand around this operating point and our small parameter $\epsilon = g_h - g_c$ where  
$g_c = \sigma_r(0)^{-1}$. We make an ansatz that we can express $\Delta_h$ as a power series in $\epsilon$,
\begin{align}\label{eq:FP_perturb_2}
    \Delta_h = \epsilon^{\eta}\left(
    d_0 + d_1 \epsilon + d_2 \epsilon^2
    \right)
\end{align}
where $\eta$ is the exponent for the prefactor scaling, and needs to be determined self-consistently. To get the scaling relations for $\Delta_h$ we need to expand the coefficients in the Taylor series for $\Delta_h$ in terms of $\epsilon$. We note that $c_0 = \tanh(\beta_h)^2$, and therefore, these approximations only make sense for small $\beta_h$. How small should $\beta_h$ be relative to $\epsilon$? We make the following ansatz:
\begin{align}\label{eq:FP_perturb_3}
    \beta_h = \beta_{0} \epsilon^{\delta}
\end{align}
and thus if $\delta > 1/2$ then $c_0 \sim \beta_{0}^2 \epsilon^{2\delta}$ will increase slower than $\epsilon$. 

We now express the coefficients for small $\beta_h$ :
\begin{align}\label{eq:FP_perturb_4}
    c_0 \approx & \: \beta_{0}^2 \epsilon^{2\delta} \\
    c_1 \approx & \: g_h^2 \left( 
    1 - 2 \beta_h^2
    \right) \\
    c_2 \approx & \: g_h^4 \left( 
    -2 + 17 \beta_h^2
    \right)
\end{align}

After solving Eqns. [\ref{eq:FP_perturb_1}-\ref{eq:FP_perturb_4}] self-consistently in terms of the expansion parameter $\epsilon$, we get the following perturbative solution for $\delta \leq 1$ : 
\begin{align}
    \Delta_h \approx & \begin{dcases}
            \frac{ 2\beta_0 \epsilon^{\delta} }
            {g_c^2(2 - g_c^2 a_1)} & \quad
            g_c^2 a_1 < 2 \\
            & \\
            \left(
            g_c^2 a_1 - 2
            \right)f_1
             +
            \epsilon \cdot f_2
            & \quad
            g_c^2 a_1 > 2
    \end{dcases} \\
    \textrm{ where  } a_1 = & \: \frac{\alpha_r^2}{16}\left[ 
    \phi_0^{(1)}(\beta_r/2)^2
    + 
    \phi_0(\beta_r/2)
    \phi_0^{(2)}(\beta_r/2)
    \right]
\end{align}
$f_2(\alpha_r,\beta_r)$ and $f_2(\alpha_r,\beta_r)$ are constant functions (w.r.t $\epsilon$). Therefore, we see a linear scaling with the bias $\beta_h$.

\subsection{Perturbative solutions for the fixed-point variance $\Delta_h$ in the bifurcation region with no biases}\label{app:FP_perturbative_BF}

The perturbative treatment of the fixed-point solutions in this case closely follows that described above. For $g_h = 2 -\epsilon$, we can express $\Delta_h$ as a power-series in $\epsilon$ : $\Delta_h = c_0 + c_1 \epsilon + c_2 \epsilon^2 $, and look for a condition that allows for a non-zero $c_0$ corresponding to the bifurcation point. Since we expect, $\Delta_h$ to be small in this regime, we can expand $\Delta_r$ as :
\begin{align}
    \Delta_r \approx & \: g_{h}^{2}\Delta_h - 2 g_{h}^{4} \Delta_h^2+ \frac{17}{3} g_{h}^{6} \Delta_h^3 + O(\Delta_h^4)
\end{align}
and similarly, we can also approximate
\begin{align}
    \left \langle \sigma_{r}(\sqrt{\Delta_{r}} x)^{2}
    \right \rangle_x \approx & \:
    \frac{1}{4} \left[ 1 + \frac{\alpha_{r}^{2}}{4} \Delta_r-\frac{\alpha_{r}^{4}}{8} \Delta_r^2
    \right]
\end{align}
Now, equating coefficient of powers of $\epsilon$, we get
that either $c_0= 0$ or 
\begin{align}
    c_0 = \frac{ 3(\alpha_r^2 - 8)}{2 (-136 + 24 \alpha_r^2 + 3\alpha_r^4)}
\end{align}
which is a valid solution when $\alpha_r \geq \sqrt{8}$. This is the bifurcation curve limit near $g_h=2^-$. 

In the other limit, $\alpha_r^* \rightarrow \infty$ and $\Delta_h^* \rightarrow 0$. We can work in the regime where $\alpha_r \sqrt{\Delta_h} \gg 1$ to see what values of $g_h$ admit a bifurcation in the perturbative solutions. The equation (to $O(\Delta_h^2)$ is given by :
\begin{align}
    \Delta_h \approx & \: \frac{1}{2} \left[ g_{h}^{2}\Delta_h - 2 g_{h}^{4} \Delta_h^2
    \right]
\end{align}
Thus, we get a positive solution for $\Delta_h$, when $g_h > \sqrt{2}$ and to the leading order, the solution scales as
\begin{align}
    \Delta_h^*(\sqrt{2}^{\:+}) \sim \frac{g_h^2 - 2}{2g_h^4} \qquad \textrm{ for } g_h \rightarrow \sqrt{2}^{\: +}
\end{align}

%--------------------- Perturbative Ch(tau) -----------------

\subsection{$C_h(\tau)$ near critical point}
\label{app:DMFT_perturbative}
Here we study the asymptotic behaviour of $C_h(\tau)$ near the critical point $g_h=2.0$  for small $\alpha_z$. For simplicity, we set the biases to be zero. In this limit we can assume that $C_h(\tau)$ and 
$C_{\phi}(\tau)$ are small. 
Let us begin by approximating $C_{\sigma_z}(\tau)$.

%\begin{align}
%    \sigma_z(t) = & \: \left[ 1 + \exp(-\alpha_z z(t) \right]^{-1} \nonumber \\
%   = & \:  \frac{1}{2}\left[ 1+ \tanh\left(   \frac{\alpha_z z(t)}{2}   \right)\right]
%\end{align}

We get up to $O(C_z^3)$,
\begin{align}
    C_{\sigma_z}(\tau) = & \: 
    g_0 + 
    g_1 C_z(\tau) + g_3 C_z(\tau)^3, \\
    \textrm{where } \quad  g_0 = & \:
    \frac{1}{4} ,\\
    g_1 = & \:
    \frac{\alpha_z^2}{16}
    -\frac{\alpha_z^4}{32}C_z(0)
    + \frac{5 \alpha_z^6}{256}C_z(0)^2, \\
    g_3 = & \: \frac{ \alpha_z^6}{384}
    - \frac{ \alpha_z^8}{192}C_z(0).
\end{align}
 This can be obtained, for instance, by expanding $\sigma_{z}(z(t))$ and taking the Gaussian averages over the argument $z(t)$ in the steady-state. The relation between $C_{\phi}(\tau)$ and $C_z(\tau)$
, in general, does not have a simple form; however, 
when $g_h \sim 2$, we expect the relaxation time $\tau_R \gg 1$, and therefore, we can approximate $C_z(\tau) \approx C_{\phi}(\tau)$. We can then approximate $C_{\phi}$ as 
\begin{align}
    C_{\phi}(\tau) = & \: 
    g_0 + g_1 C_h(\tau) + g_3 C_h(\tau)^3 ,\\
    \textrm{where } \quad  g_0 = & \:
    0 , \quad ({\rm for\, \beta_{h} = 0})\\
    g_1 = & \:
    g_h^2 - 2g_h^4 C_h(0) + 5g_h^6C_h(0)^2,
     \\
    g_3 = & \: \frac{ 2}{3}g_h^6
    - \frac{ 16}{3}g_h^8 C_h(0).
\end{align}

Note that this also gives us an approximation for $C_{\phi}(0)$. 
Putting all this together, the equation governing $C_h(\tau)$:
\begin{align}
    \left[ -\partial_{\tau}^2  + C_{\sigma_z}(\tau)\right]
    C_h(\tau) = \frac{1}{4}C_{\sigma_z}(\tau)C_{\phi}(\tau),
\end{align}
becomes (up to $O(C_h^3)$)
\begin{align}
  \partial_{\tau}^2 C_h(\tau) \simeq & \: 
  a_1 C_h(\tau)  + a_2 C_h(\tau)^2 + a_3 C_h(\tau)^3, \\
  \textrm{where } \quad  a_1 = & \:
  \frac{1}{16}\left(4 -  \Gamma    
  \right), \\
  a_2 = & \: \frac{\alpha_z^2}{64} 
  \left(
  4 -  \Gamma
  \right)
  \Gamma ,\\
  a_3 = & \: -\frac{g_h^6}{24} \\
  \Gamma = & \: g_h^2 - 2g_h^4 C_h(0)  + 5g_h^6 C_h(0)^2.
\end{align}

Integrating w.r.t $\tau$ gives
\begin{align}
\left( \partial_{\tau} C_h(\tau) \right)^2 = 
& \: 2 \Big(
\frac{a_1}{2} C_h(\tau)^2 + 
\frac{a_2}{3} C_h(\tau)^3 + \nonumber \\
& \quad \frac{a_3}{4} C_h(\tau)^4 +
\textrm{const.} 
\Big).
\end{align}

The boundary conditions are 
\begin{align}
    \partial_{\tau}C_h(0) = 0 \qquad  \lim_{\tau \rightarrow \infty }\partial_{\tau}C_h(\tau) = 0.
\end{align}
The second condition implies the constant is 0. And, the first condition implies
\begin{align}
\frac{a_1}{2} + 
\frac{a_2}{3} C_h(0) + 
\frac{a_3}{4} C_h(0)^2  = 0.
\end{align}
From this, we can solve for $C_h(0)$ (neglecting terms higher than quadratic) to get a solution that is perturbative in the deviation $\epsilon$ from the critical point ($g_h = 2 + \epsilon$). To the leading order the variance grows as
\begin{align}
    C_h(0) 
    \approx & \: \frac{1}{8}\epsilon + O(\epsilon^2),
\end{align}
and the $\alpha_z$ only enters the timescale-governing term $a_1$ at $O(\epsilon^2)$ . At first, it might seem counter-intuitive that $\alpha_{z}$, which effectively controls the dynamical time constant in the equations of motion, should not influence the relaxation rate to leading order. However, this result is for the dynamical behavior close to the critical point, where the relaxation time is a scaling function of $\epsilon$. Moving away from this critical point, the relaxation time becomes finite, and the $z-$ gate, and thus $\alpha_{z}$, should have a more visible effect.

%********************** Kac-Rice **********************

\section{Topological Complexity via Kac-Rice formula}
\label{app:Kac-Rice}

The arguments here are similar to those presented in \cite{ipsen}, which use a self-averaging assumption to express the topological complexity (defined below) in terms of a spectral integral. Let's begin.

 The goal is to estimate the total number of fixed points for a dynamical system $\dot{{\bf x}} = G({\bf x})$. The Kac-Rice analysis proceeds by constructing the integral over the state space ${\bf x}$ whose integrand has delta-functional support only on the fixed points:

\begin{align}
\mathcal{N} = \int d{\bf x} \mathbbm{E} \left[ \delta(G({\bf x})) | {\rm det} \mathcal{D} |  \right] ,
\end{align}
where $\mathcal{D} = \partial G/\partial {\bf x}$ is the instantaneous Jacobian. The expectation value here is over the random coupling matrices. The average number of fixed points is related to the so-called topological complexity $\mathcal{C}$ via the definition $$\mathcal{N} = \exp \left( N \mathcal{C}\right).$$ We seek a saddle-point approximation of this quantity below.

For the gated RNN, the state space ${\bf x} = ({\bf h}, {\bf z}, {\bf r})$, and the fixed points satisfy
\begin{align}
    \sigma_{z}(z_{i}) \left( - h_{i} + \eta^{h}_{i} \right) &= 0,\\
    -z_{i} + \eta^{z}_{i}& = 0,\\
    -r_{i} + \eta^{r}_{i} & = 0,
\end{align}
where for notational shorthand we have introduced $\eta_{i}^{h} = \sum_{j} J_{ij}^{h} \phi(h_{j}) \sigma_{r}(r_{j})$, and $\eta_{i}^{r/z} = \sum_{j}J_{ij}^{r/z} \phi(h_{j})$, anticipating the mean-field approximation to come. Notice that only the first equation for $h$ provides a nontrivial constraint. Once $h$ is found, the second and third equations can be used to determine $z$ and $r$, respectively. Notice furthermore that since $\sigma(z_{i})>0$, the solutions $h_{i}$ to the first equation do not depend on $z_i$. Indeed, the dependence on $\sigma(z)$ can be factorized out of the Kac-Rice integral. This requires noting first that for the fixed point Jacobian, Eq (\ref{eq:JacRMT_linearisation}) implies that the Jacobian can be written (setting $\tau_{r} = \tau_{z} = 1$ for simplicity)
\begin{align}
    \mathcal{D} = A \left( - 1 + \mathcal{J} R \right),
\end{align}
and that the determinant can be factorized
\begin{align}
  \det | \mathcal{D}| &= \det | A| \times \det \left| - 1 + \mathcal{J} R \right| \\
  &= \left(\prod_{i} \sigma(z_{i}) \right)\det \left| - 1 + \mathcal{J} R \right| .
\end{align}
The product of $\sigma(z_{i})$ produced by the determinant will be canceled by the product of delta functions, using the fact that $\sigma(z_{i})>0$ and the transformation law
\begin{align}
   \prod_{i} \delta\left[\sigma(z_{i}) ( - h_{i} + \eta_{i}^{h} ) \right] =  \frac{1}{\prod_{i}\sigma(z_{i})} \prod_{i}\delta \left[ - h_{i} + \eta_{i}^{h}\right],
\end{align}

So we see that what evidently matters for the topological complexity is the {\it fixed-point Jacobian}

\begin{align}
    \mathcal{D}^{fp} = - 1 + \mathcal{J} R,
\end{align}
whose eigenvalues we denote by $\lambda_{i}$ for $i = 1, ... ,N$, and with the spectral density 
\begin{align}
    \hat{\mu}(z) = \frac{1}{N} \sum_{i} \delta^{(2)}(z - \lambda_{i}).
\end{align}

The preceding analysis was all basically to show that we could easily have set $\alpha_{z} = 0$ and gotten the same answer, i.e. the z-gate does not influence the topological properties of the dynamics. For $\alpha_{z} = \infty$, the situation changes drastically, and the analysis will likely need to be significantly reworked. Indeed, in this limit, we most likely do not have discrete fixed points anymore, so the very notion of counting fixed points no longer makes sense. 

Having introduced the spectral density,  we can rewrite the Kac-Rice integral as 

\begin{align}
\mathcal{N} = \int \prod_{x \in \{ h, r\}} d{\bf x}  \mathbbm{E} \left[ \delta^{(N)}({\bf x} - \boldsymbol{\eta}^{x})  e^{  N \int d^{2} z \hat{\mu}(z) \log |z| } \right] 
\end{align}

Note that since the spectral density of $\mathcal{D}^{fp}$ is independent of $z$, the integral over $z$ is trivial to perform, and leaves only $h$ and $r$ in the integrand.

So far, everything has been exact. We begin now to make some approximations. The first crucial approximation is that the spectral density is self-averaging. The RMT analysis in the previous sections showed us furthermore that the spectral density depends only on macroscopic correlation functions of the state variables. Let us denote the spectral integral factor

\begin{align}
    I({\bf x}, \mathcal{J}) = \exp \left( N \int d^{2} z \hat{\mu}(z) \log |z|\right),
\end{align}
by which we mean that it will depend on the particular realization of the random coupling $\mathcal{J}$, and the state vector ${\bf x}$. The self-averaging assumption implies that

\begin{align}
    I({\bf x}, \mathcal{J}) \approx \mathbbm{E}\left[ I( {\bf x},\mathcal{J})\right] \equiv \bar{I}({\bf x})
\end{align}

i.e. this factor does not depend on the particular realization of $\mathcal{J}$, but just on the state vector. Equivalently, we are assuming that the spectral density $\hat{\mu}_{{\bf h}, {\bf r}}(z)$ only depends on the configurations ${\bf h}, {\bf r}$, and not the particular realization $J^{h,r}$. This allows us to pull this factor outside of the expectation value

\begin{align}
   \mathcal{N} \approx \int \prod_{{\bf x} \in \{ h, r\}} \bar{I}({\bf x}) \mathbbm{E} \left[ \delta^{(N)}({\bf x} - \boldsymbol{\eta}^{x}) \right].
\end{align}

Now we give some non-rigorous arguments for how one might evaluate the remaining expectation value. In order to carry out the average over $J^{h}$ and $J^{r}$, we utilize the Fourier representation of the delta function to write

\begin{align}
\mathbbm{E}\left[ \int d \hat{ {\bf x}} \, e^{ i \hat{{\bf x}}( {\bf x} - \boldsymbol{\eta}^{x}) } \right]&\\
 = \int d {\bf x} d \hat{{\bf x}} \mathbbm{E}\Big[ \exp \sum_{i,j} \Big\{  &i \hat{h}_{i} \left(h_{i} - J_{ij}^{h} \phi(h_{j}) \sigma_{r}(r_{j}) \right) \\
&+ i \hat{r}_{i} ( r_{i} -  J^{r}_{ij} \phi(h_{j})) \Big\}\Big], 
\end{align}

which upon disorder averaging yields

\begin{align}
   \int d{\bf x} d \hat{{\bf x}} \exp \left\{\sum_{i} \left( i \hat{h}_{i} h_{i} + i \hat{r}_{i} r_{i} - \frac{1}{2 }  \hat{h}_{i}^{2} \hat{C}_{\phi \sigma_{r}} - \frac{1}{2} \hat{r}_{i}^{2} \hat{C}_{\phi} \right) \right\} \label{eq:integral}
\end{align}

where we have defined

\begin{align}
    \hat{C}_{\phi \sigma_{r}} = \frac{1}{N} \sum_{i} \phi(h_{i})^{2} \sigma_{r}(r_{i})^{2}, \quad \hat{C}_{\phi} = \frac{1}{N} \sum_{i} \phi(h_{i})^{2} \label{eq:emp-ave}
\end{align}

This is where we make our second crucial assumption: that the empirical averages appearing in (\ref{eq:emp-ave}) converge to their average value

\begin{align}
    \hat{C}_{\phi \sigma_{r}} &\to C_{\phi \sigma_{r}}  =  \mathbbm{E}_{{\bf h},{\bf r}} \left[ \frac{1}{N} \sum_{i} \phi(h_{i})^{2} \sigma_{r}(r_{i})^{2}\right], \\
    \hat{C}_{\phi} &\to   C_{\phi} \equiv \mathbbm{E}_{{\bf h}} \left[ \frac{1}{N} \sum_{i} \phi(h_{i})^{2} \right].
\end{align}

This means we are assuming the strong law of large numbers. With this essential step, the integral in (\ref{eq:integral}) evaluates to

\begin{align}
  &\frac{1}{\sqrt{2\pi \Delta_{h}}}\frac{1}{\sqrt{2\pi \Delta_{r}}}  \exp\left( - || {\bf h}||^{2}/2 \Delta_{h} - || {\bf r}||^{2} / 2 \Delta_{r}\right) \\
  & = \prod_{i = 1}^{N} P_{h}(h_{i}) P_{r}(r_{i})
\end{align}

where $\Delta_{h} = C_{\phi \sigma_{r}}$ and $\Delta_{r} = C_{\phi}$ - which are just the time-independent (fixed point) MFT equations (\ref{eq:MFT_FP_variance_1}).

Returning to the expression for the complexity, these series of approximations give us 

\begin{align}
    \mathcal{N} \approx  \int \prod_{i = 1}^{N} dh_{i} d r_{i} P_{h}(h_{i}) P_{r}(r_{i}) \bar{I}({\bf r}, {\bf h})
\end{align}

Let us now describe more intuitively what has happened. We started with the formal expression for the Kac-Rice formula, which uses the delta functional integrand to find fixed-points, and counts them with the weighting factor related to the Jacobian. Our first assumption allows us to simplify the calculation involving the Jacobian, since we argued that this term is self-averaging. The second assumption allows us to deal with the remaining expectation value of the delta functions. The expectation value adds a number of delta functions (however many there may be for that $J^{h/r}$) for each configuration of the connectivity. For continuously distributed connectivity, this implies that the expectation value will smear out the delta functions, and result in a smooth distribution. What should this distribution be? Well, we know from the mean-field analysis that the state vectors are distributed as Gaussians at a fixed point. Furthermore, the mean-field theory becomes exact for large $N$. Therefore, we should expect that in this limit, the delta functions are smeared out into the Gaussian distributions determined by the MFT. This is what our derivation shows. 

The final step is to recall that the spectral density depends on the state vectors only via empirical averages. For instance, in the absence of an $r-gate$, the spectral density would depend on the empirical average $\hat{C}_{\phi'}$. Again invoking the strong law of large numbers, we may argue that the self-averaging goes a step further, and that 

\begin{align}
    \mathcal{N} &\approx \int \prod_{i} dP_{h}(h_{i}) dP_{r}(r_{i}) \exp \left( N \int d^{2} z  \hat{\mu}_{{\bf h}, {\bf r}}(z)  \log |z| \right),\\
    & \approx \exp \left( N \int d^{2}z \bar{\mu}(z) \log |z| \right),
\end{align}
where 

\begin{align}
 \bar{\mu}(z) =  \mathbb{E}_{{\bf h}, {\bf r}} \left[ \hat{\mu}_{{\bf h},{\bf r}}(z) \right] =   \int d{\bf h} d{\bf r} P({\bf h}) P({\bf r}) \hat{\mu}_{{\bf h}, {\bf r}}(z) 
\end{align}

This precisely the spectral density we study in a preceding appendix, and the one for which we have obtained an explicit expression for the spectral curve. These approximations give us the topological complexity

\begin{align}
\mathcal{C} & = 	\int d^{2} z \bar{\mu}(z) \log |z| .
\end{align}

Now we take a closer look at the spectral density. The eigenvalues of $\mathcal{D}^{fp}$ form a circular droplet of finite radius $\rho$, and centered on $- 1$. Therefore, the eigenvalues have the form $\lambda = -1 + r e^{ i \theta}$, and the spectral density is a function only of $r$. The value of the radius is found from Eq.  (\ref{eq:RMT_FP_spectral_curve}) by removing the z-gate (i.e. setting $\alpha_{z} = 0$). After some algebraic steps, we find for the radius
\begin{align}
    \rho^{2} &=\frac{1}{2}\left( C_{1} + 	\sqrt{C_{1}^{2} + 4 C_{2}}\right),\\
    C_{1} &= C_{\phi'} C_{\sigma_{r}}, \quad C_{2} = C_{\phi'} C_{\sigma_{r}'} C_{\phi}.
\end{align}
Using these facts, we can write the topological complexity as
\begin{align}
    \mathcal{C} &=  \int \, r dr  d\theta \, \bar{\mu}(r) \mathbbm{I}_{\{r<\rho\}} \log|r e^{ i \theta} - 1|,\\
    &= \begin{cases}
   2\pi  \int_{1}^{\rho} r dr \log r \, \bar{\mu}(r)  \ge  0 , \quad &{\rm for} \, \rho > 1,\\
    0 , \quad & {\rm for } \, \rho < 1,
    \end{cases}
\end{align}
where $\mathbbm{I}_{\{r<\rho\}}$ is the indicator function which is one for $r < \rho$ and vanishes for $r > \rho$. Thus we see that the topological complexity is zero for $\rho<1$. This is {\it precisely the fixed-point stability condition derived in the main text Eq. (\ref{eq:transition-to-chaos})}. Conversely, the topological complexity will be nonzero for $\rho >1$, which corresponds to unstable fixed points. Thus we see, under our set of reasonable approximations, unstable MFT fixed-points correspond to a finite topological complexity, and consequently to a number of ``microscopic" fixed-points that grows exponentially with $N$. 

The final missing ingredient, necessary to show that Region 2 in the phase diagram has an exponentially growing number of fixed points, is to show that the MFT fixed points which appear after the bifurcation, are indeed unstable. At the moment, we lack any analytical handle on this. However, we confirmed numerically that along the bifurcation curve, the fixed points are unstable, and that increasing the variance $\Delta_{h}$ only serves to increase $\rho$. However, is it possible for the lower branch, on which $\Delta_{h}$ decreases with $\alpha_{r}$? Evidently not, since $\Delta_{h}$ scales with $\alpha_{r}$ in such a way that $C_{\sigma_{r}'}$ ends up growing like $\alpha_{r}^{2}$, thus once again increasing $\rho$. Therefore, we conclude that the MFT fixed-points appearing after the bifurcation are always unstable, with $\rho >1$. This concludes our informal proof of the transition in topological complexity between regions 1 and 2 in the phase diagram Fig. (\ref{fig:phaseDiag_combined}).

\bibliography{gated_RNN}% Produces the bibliography via BibTeX.

\end{document}